\documentclass[aps,prd,groupedaddress, tightenlines,
preprint,eqsecnum,nofootinbib,showpacs%
]{revtex4}
\usepackage{graphicx,epsf,amssymb,amsbsy,amsfonts,amssymb,amsmath}
\usepackage[usenames, dvipsnames]{color}
\newif\ifdraft
\drafttrue
\newif\ifpreprint
\preprinttrue

\def\sect#1{section~{\ref{#1}}}
\def\fig#1{fig.~{\ref{#1}}}

\def\app#1{appendix~{\ref{#1}}}
\def\eqn#1{eq.~(\ref{#1})}
\def\Eqn#1{Equation~(\ref{#1})}
\def\eqns#1#2{eqs.~(\ref{#1}) and~(\ref{#2})}

\def\tab#1{table~{\ref{#1}}}

\def\spa#1.#2{\left\langle#1\,#2\right\rangle}
\def\spb#1.#2{\left[#1\,#2\right]}
\def\spash#1.#2{\spa{\smash{#1}}.{\smash{#2}}}
\def\spbsh#1.#2{\spb{\smash{#1}}.{\smash{#2}}}
\def\sand#1.#2.#3{%
\left\langle\smash{#1}{\vphantom1}^{-}\right|{#2}%
\left|\smash{#3}{\vphantom1}^{-}\right\rangle}
\def\sandpp#1.#2.#3{%
\left\langle\smash{#1}{\vphantom1}^{+}\right|{#2}%
\left|\smash{#3}{\vphantom1}^{+}\right\rangle}
\def\sandpm#1.#2.#3{%
\left\langle\smash{#1}{\vphantom1}^{+}\right|{#2}%
\left|\smash{#3}{\vphantom1}^{-}\right\rangle}
\def\sandmp#1.#2.#3{%
\left\langle\smash{#1}{\vphantom1}^{-}\right|{#2}%
\left|\smash{#3}{\vphantom1}^{+}\right\rangle}

\def\colorf#1{\tilde f^{#1}}
\def\colorc#1{c_{(#1)}}
\def\c#1{c_{(#1)}}

\def\t{\tau}
\def\tree{{\rm tree}}
\def\Loop{{(L)}}

\def\Tr{\, {\rm Tr}}

\def\eps{\epsilon}
\def\e{\epsilon}

\def\nn{\nonumber}

\def\n{{\tilde n}}
\def\f{\tilde f}
\def\P{{\rm (P)}}
\def\NP{{\rm (NP)}}
\def\A{{\rm (A)}}
\def\B{{\rm (B)}}

\def\NeqFoursYM{{${\cal N}=4$~sYM}}
\def\NeqFour{{{\cal N}=4}}

\def\NeqEight{{{\cal N}=8}}

\def\be{\begin{equation}}
\def\ee{\end{equation}}
\def\bea{\begin{eqnarray}}
\def\eea{\end{eqnarray}}
\def\ba{\begin{eqnarray}}
\def\ea{\end{eqnarray}}

\def\Ord{{\cal O}}
\def\Frac#1#2{{\textstyle \frac{#1}{#2}}}

\def\fourloop{{(4)}}
\def\ss{s}
\def\tt{t}
\def\uu{u}

\newbox\charbox
\newbox\slabox
\def\s#1{{      
        \setbox\charbox=\hbox{$#1$}
        \setbox\slabox=\hbox{$/$}
        \dimen\charbox=\ht\slabox
        \advance\dimen\charbox by -\dp\slabox
        \advance\dimen\charbox by -\ht\charbox
        \advance\dimen\charbox by \dp\charbox
        \divide\dimen\charbox by 2
        \raise-\dimen\charbox\hbox to \wd\charbox{\hss/\hss}
        \llap{$#1$} }}

\begin{document}
\hfuzz 20pt

\ifpreprint
\noindent
UCLA/11/TEP/110 $\null\hskip0.35cm\null$ \hfill 
SU-ITP-11/34$\null\hskip0.35cm\null$ \hfill 
Saclay--IPhT--T11/175\\
SLAC--PUB--14529 $\null\hskip0.35cm\null$ \hfill
NSF--KITP--11--236  $\null\hskip0.35cm\null$ \hfill
CERN--PH--TH/2011/190
\fi

\vskip0.8cm
\title{Simplifying Multiloop Integrands and Ultraviolet Divergences\\
of Gauge Theory and Gravity Amplitudes}

\vskip0.8cm
\author{Z.~Bern${}^{a,b,c}$, J.~J.~M.~Carrasco${}^{b,d}$,
L.~J.~Dixon${}^{c,e}$, H.~Johansson${}^{b,f}$
and R.~Roiban${}^{b,g}$}

\affiliation{
${}^a$\hbox{Department of Physics and Astronomy, UCLA, Los Angeles, 
CA 90095, USA}\\
${}^b$\hbox{Kavli Institute for Theoretical Physics,
University\! of\! California,\! Santa\! Barbara,\! CA\! 93106,\! USA}
${}^c$\hbox{Theory Group, Physics Department, CERN, CH--1211 Geneva 23, 
Switzerland}\\
${}^d$\hbox{Stanford Institute for Theoretical Physics and 
Department of Physics,}\\ 
\hbox{Stanford University, Stanford, CA 94305, USA}\\
${}^e$\hbox{SLAC National Accelerator Laboratory, Stanford University, 
Stanford, CA 94309, USA}\\
${}^f$\hbox{Institut de Physique Th\'eorique, CEA--Saclay,  F--91191
Gif-sur-Yvette cedex, France}\\
${}^g$\hbox{Department of Physics, Pennsylvania State University,
University Park, PA 16802, USA}
\\
}


\begin{abstract}
We use the duality between color and kinematics to simplify the
construction of the complete four-loop four-point amplitude of  
$\NeqFour$ super-Yang-Mills theory, including the nonplanar
contributions. The duality completely determines the amplitude's
integrand in terms of just two planar graphs.  The existence of a
manifestly dual gauge-theory amplitude trivializes the construction of
the corresponding $\NeqEight$ supergravity integrand, whose graph
numerators are double copies (squares) of the $\NeqFour$
super-Yang-Mills numerators.  The success of this procedure provides 
further nontrivial evidence that the duality and
double-copy properties hold at loop level. The new form of the
four-loop four-point supergravity amplitude makes manifest the same
ultraviolet power counting as the corresponding $\NeqFour$
super-Yang-Mills amplitude.  We determine the amplitude's ultraviolet
pole in the critical dimension of $D=11/2$, the same dimension as for
$\NeqFour$ super-Yang-Mills theory.  Strikingly, exactly the same
combination of vacuum integrals (after simplification) describes
the ultraviolet divergence of $\NeqEight$ supergravity as the
subleading-in-$1/N_c^2$ single-trace divergence in $\NeqFour$
super-Yang-Mills theory.
\end{abstract}

\pacs{04.65.+e, 11.15.Bt, 11.30.Pb, 11.55.Bq \hspace{1cm}}

\maketitle


\section{Introduction}

The past few years have brought remarkable advances in understanding
scattering amplitudes in the maximally supersymmetric $\NeqFour$
super-Yang-Mills (sYM)
theory~\cite{N4YM} in the planar limit of a large number of colors.
It may soon be possible to completely determine all planar
scattering amplitudes in this theory, for all values of the coupling,
going far beyond the (now thoroughly understood) cases of four and five
external gluons~\cite{BDS}.  Much of this progress has been surveyed
recently~\cite{SolveAmplReview}. Planar scattering
amplitudes  exhibit a new symmetry known as dual conformal
symmetry~\cite{DualConformal,BCDKS}, which severely restricts their
structure.  Together with supersymmetry and (position space) conformal
symmetry, dual conformal invariance gives rise to a
Yangian~\cite{Yangian}---an algebraic structure common in integrable
models. Indeed, it is widely believed that several aspects of the
planar sector of $\NeqFour$ sYM theory are controlled by an integrable
model (see {\it e.g.}~ref.~\cite{IntegrableReview}).

In contrast, much less is known about the nonplanar sector---the
subject of the present paper.   Consider $\NeqFour$ sYM theory for
the gauge group $SU(N_c)$.   In the limit $N_c \rightarrow\infty$,
the nonplanar, or subleading-color, contributions are suppressed by
powers of $1/N_c$.  Once one takes into account these corrections,
for finite $N_c$, the scattering amplitudes no longer appear to possess
dual conformal symmetry, nor do they demonstrate any obvious integrability
properties.  

Understanding the subleading-color terms is critical to a complete
description of the behavior of gauge theories.  For example, many types of
color correlations are suppressed in the large-$N_c$ limit.
Furthermore,  the information provided by the full-color expression
for $\NeqFour$ sYM amplitudes, expressed in terms of their loop-momentum
integrands, can be used to construct corresponding
amplitudes~\cite{BDDPR,GravityThree,GravityFour,Neq44np} in
$\NeqEight$ supergravity~\cite{CremmerJulia}.  From each set of amplitudes
one can extract information about ultraviolet divergences in the respective 
theory.

The ultraviolet (UV) properties of $\NeqEight$ supergravity
have been the focus of intense investigation.  There have been
several recent reviews of the situation~\cite{GravityUVReview}.
Long ago, an $\NeqEight$ supersymmetric local
counterterm at three loops in $D=4$ was
proposed~\cite{DeserKayStelle,Ferrara1977mv,Deser1978br,Howe1980th,%
Kallosh1980fi}.  An explicit computation of the three-loop four-graviton
amplitude first revealed that the counterterm has a vanishing
coefficient~\cite{GravityThree}.  Subsequently it was
realized~\cite{EKR4} that this counterterm is forbidden
in $D=4$ by the $E_{7(7)}$ duality symmetry~\cite{CremmerJulia}.
Other analyses have extended the finiteness constraints from $E_{7(7)}$
and linearized supersymmetry, such that the first potential
divergence in $D=4$ is now at seven
loops~\cite{BHN,BHS2010,Beisertetal,BHSV}.  Finiteness until this
loop order happens to agree with an earlier naive power-counting, based
on the assumption of an off-shell $\NeqEight$ superspace~\cite{Siegel7l}.
A potential seven-loop divergence
is also suggested by other approaches, including an analysis of
string theory dualities~\cite{GRV2010}, a first-quantized
world-line approach~\cite{FirstQuantized}, and light-cone
supergraphs~\cite{KalloshRamond}.  However, it has also
been argued that the theory may remain finite beyond seven
loops~\cite{FiniteArgue}.

In this paper, we will show how a conjectured duality between color and
kinematics~\cite{BCJ,BCJLoop} provides a powerful method for
computing subleading-color terms in $\NeqFour$ sYM amplitudes,
in a way that makes the construction of the corresponding
$\NeqEight$ supergravity amplitudes extremely simple.  Also, the
$\NeqEight$ result is expressed in a form that makes manifest 
the true ultraviolet behavior of the amplitude (when continued to higher
space-time dimension $D$).  Thus this method provides unprecedented
access to the precise coefficients of potential counterterms in
$\NeqEight$ supergravity, as well as in its higher-dimensional versions.
It may eventually offer a means for settling the question of whether
additional UV cancellations exist in $\NeqEight$ supergravity,
beyond the known or expected ones.  Perhaps even more importantly,
the method gives a means for constructing complete amplitudes,
allowing for detailed studies of their symmetries and properties.

A key point is that when the color-kinematics duality holds
manifestly, it locks the nonplanar contributions to the planar ones.
The nonplanar contributions are essential for evaluating gravity
amplitudes, because in gravity theories no separation exists between
planar and nonplanar
contributions.  This duality allows one to efficiently
export information from the planar sector, {\it e.g.}~that provided by
dual conformal symmetry, to the much more intricate nonplanar sector. 

A second key point is the claim~\cite{BCJLoop} that if a
duality-respecting representation of $\NeqFour$ sYM amplitudes can be
found, then the loop-momentum integrands of the corresponding
$\NeqEight$ supergravity amplitudes can be obtained simply by taking
the graphs of $\NeqFour$ sYM theory, dropping the color factors and
squaring their kinematic numerators.  This double-copy property is a
loop-level generalization of the corresponding tree-level
property~\cite{BCJ}, equivalent to the Kawai-Lewellen-Tye (KLT)
relations between gravity and gauge-theory amplitudes~\cite{KLT}.
Using the color-kinematics duality, followed by the double-copy
property, advances in constructing integrands for the planar sector of
gauge theory can be carried over to the nonplanar sector, and then on
to gravity.  The color-kinematic duality and the gravity double-copy
property do not appear to require supersymmetry, although amplitudes in
supersymmetric theories are generally much simpler to work with than
non-supersymmetric amplitudes.  Another important aspect of the
duality is that it appears to hold in any dimension, thus making it
compatible with dimensional regularization.

In this paper we will exploit the color-kinematic duality to construct
the complete four-loop four-point amplitudes of $\NeqFour$ sYM theory
and $\NeqEight$ supergravity.  Both amplitudes were constructed
previously by us~\cite{GravityFour,Neq44np}; however, the present
construction is considerably more efficient, and makes various
properties of the amplitudes more manifest.  The color-kinematic
duality relations allow us to express the four-loop loop-momentum
integrands, for 83 different cubic (trivalent) graphs, as functionals
of the integrands of just two planar graphs.  For the \mbox{one-,} two- and
three-loop four-point~\cite{BCJLoop,JJHenrikReview}, and the one- and
two-loop five-point cases~\cite{loop5ptBCJ}, the duality is
even more restrictive: a single planar graph suffices to determine
all the others.  As it is becoming
increasingly simple to construct planar amplitudes, a particularly
attractive aspect of using the color-kinematic duality is that it
determines nonplanar contributions from planar ones.  Perhaps even
more remarkable, in terms of measuring the redundancy found in local
gauge-theory scattering amplitudes, we shall find that the entire
non-trivial dynamical information in the four-loop four-point
amplitude is contained in a single nonplanar graph; all other graphs
are related to this one by the duality.  

While a general proof of the duality
conjecture is yet to be given beyond tree level, 
the four-loop construction we offer in this paper provides further
evidence in favor of it,
in the form of a highly nontrivial example.  In this work, we have
confirmed the duality-based construction by verifying that the integrand
matches a complete (spanning) set of generalized unitarity cuts.

Based on the double-copy structure of supergravity amplitudes, we will
give a new representation of the four-loop four-point $\NeqEight$
supergravity amplitude.  This construction provides a direct multiloop
confirmation of the double-copy property, because we verify
the generalized unitarity cuts for the new form of the
supergravity amplitude, against the cuts of the known 
expression~\cite{GravityFour}, originally constructed using the KLT
relations.  We also explore the ultraviolet properties
of the amplitude in $D>4$ dimensions.
An essential feature of the new representation is that
the UV behavior is manifest:  Individual integrals diverge logarithmically
in precisely the same critical dimension $D_c$ as their sum.  
This property did not hold for the previous form of the
amplitude~\cite{GravityFour}.
The critical dimension is also the same as that for the planar and
single-trace sectors of $\NeqFour$ sYM theory.  In a previous
paper~\cite{GravityFour}, we showed that the supergravity amplitude is
finite for $D<11/2$, which is also the bound obeyed by $\NeqFour$ sYM
theory.  However, the previous form of the amplitude did not display
this bound manifestly.  To see the cancellation of potential UV
divergences, the integrals had to be expanded a few orders in
powers of the external momenta.  The lack of manifest UV behavior
in that representation made it difficult to carry out the required
integration in $D=11/2$ and to determine whether the amplitude actually
does diverge in this dimension.  With the new form,
this task is greatly simplified, allowing us to carry it out here.

Due to the double-copy construction, the numerators of the integrands
for the $\NeqEight$ supergravity amplitude are perfect squares.
However, they multiply propagator denominators that do not have
definite signs.  Therefore, individual integrals contributing to the
amplitude can have different signs.  To probe
whether or not the four-loop amplitude diverges in $D=11/2$, it is
necessary to actually evaluate the UV divergences in the loop
integrals in this dimension.  Using the double-copy form of the
four-loop four-point amplitude we do so, finding that the $\NeqEight$
finiteness bound is in fact saturated at $D=11/2$ at
four loops, which matches the behavior of $\NeqFour$ sYM theory.
Moreover, we calculate the precise coefficient of the $\NeqEight$
supergravity divergence.  We find that it exactly matches the coefficient
of the divergence of the $1/N_c^2$-suppressed single-trace term in the
four-loop four-point 
amplitude of $\NeqFour$ sYM theory, up to an overall rational factor.
Although this property is most striking at four loops, only emerging
after a number of simplifications, it is consistent with
lower-loop behavior.  Presumably this consistent connection is a
clue for unraveling the
general UV properties of $\NeqEight$ supergravity.

Regularization is a crucial point in the construction of
loop-level amplitudes in massless theories, because such amplitudes are 
usually either infrared or UV
divergent.  The issue of regularization has been studied in some
detail in the context of unitarity cuts in ref.~\cite{SixD}, where the
six-dimensional helicity formalism~\cite{SixDHel} was suggested as a
general means for implementing either dimensional
regularization~\cite{DimReg} or a massive infrared regulator
equivalent to the one in ref.~\cite{HiggsReg}.  In the present paper
we take advantage of an earlier construction of the four-loop
four-point amplitude of $\NeqFour$ sYM theory, which provides
expressions with demonstrated validity for $D\le 6$~\cite{Neq44np, SixD}.
In this paper, we compare the $D$-dimensional unitarity cuts of the new
results with the cuts of the earlier results.  We find exact agreement,
confirming the new representations.

This paper is organized as follows.  In \sect{OverviewSection} we
explain our strategy for constructing multiloop integrands,
illustrating it with the three-loop four-point $\NeqFour$ sYM
amplitude.  In \sect{FourLoopSection} (and
appendix~\ref{NumeratorAppendix}) we present the new forms of the
four-loop integrands of $\NeqFour$ sYM theory and $\NeqEight$
supergravity. We also outline their construction.  In \sect{UVSection}, we
obtain the explicit value of the UV divergence of the
$\NeqEight$ supergravity amplitude in $D=11/2$ and discuss its
properties.  We also determine the UV divergence of the color
double-trace terms in the four-loop $\NeqFour$ sYM amplitude
in $D=6$.  (We had found earlier~\cite{Neq44np} that the double-trace
divergence canceled in the next possible lower dimension, $D=11/2$.)
We give our conclusions and outlook in
\sect{ConclusionSection}.  Several appendices are included.
The first one gives functional defining relations between the numerators
in the four-loop four-point amplitude, which are derived from the
Jacobi relations after imposing some auxiliary conditions
valid for $\NeqFour$ sYM theory.
Appendix~\ref{NumeratorAppendix} presents the analytic expressions for
the numerators.
Appendix~\ref{VacuumNumeratorsAppendix} gives the values of the
vacuum integrals entering the UV divergence for the supergravity
amplitude in the critical dimension, as well as expressions for the
vacuum integrals' numerators. 
Explicit expressions for the color factors for each contribution
to the full four-loop amplitude may be found online~\cite{Online}, 
where we also provide plain-text, computer-readable versions of the
numerator factors and the kinematic dual Jacobi relations that they obey.


\section{Constructing multiloop integrands}
\label{OverviewSection}

The unitarity
method~\cite{UnitarityMethod,BCFUnitarity,FiveLoop,Neq44np} has become
a general-purpose tool for constructing multiloop amplitudes in gauge
and gravity theories.  In this section we demonstrate how one can
dramatically reduce the complexity of unitarity-based calculations for
gauge theories by assuming the conjectured duality between color and
kinematics~\cite{BCJ,BCJLoop}.  This duality
reduces the construction of an amplitude at the integrand level to the
determination of the numerator factors for a small set of graphs,
which we call {\it ``master graphs''}.  For the four-loop four-point
$\NeqFour$ sYM amplitude, it suffices to use just two planar master graphs.
Alternatively, a single nonplanar master graph is sufficient. 

 Given
the duality-satisfying form of the $\NeqFour$ sYM amplitude, the
$\NeqEight$ supergravity amplitude can be written down immediately by
squaring the $\NeqFour$ sYM numerator factors.  We confirm the
correctness of the derived gauge and gravity amplitudes using a
spanning set of generalized unitarity cuts, showing that they
agree with our previous forms~\cite{GravityFour, Neq44np} on these
cuts.

\subsection{Duality between color and kinematics}

\begin{figure*}[tb]
\begin{center}
\vskip .7 cm 
\includegraphics[scale=.76]{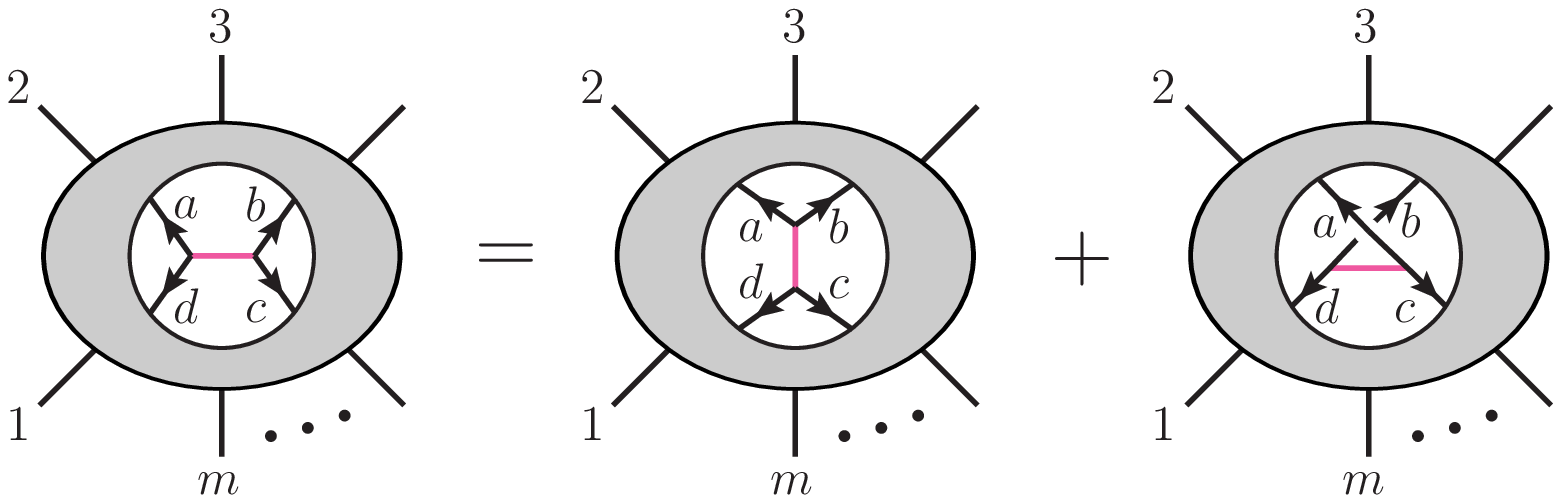}
\end{center}
\vskip -.7 cm 
\caption[a]{\small Pictorial Jacobi relation for a group of three
  graphs. The graphs can represent color factors or numerator factors.
  Except for the connections to the central (pink) lines, the graphs
  are identical in the three cases, as indicated by the common
  (momentum or color) labels $a,b,c,d$.  The gray area represents some
  unspecified subgraph which is identical in all three graphs.  
\label{GeneralJacobiFigure}
}
\end{figure*}

In general, a massless $m$-point $L$-loop gauge-theory 
amplitude ${\cal A}^{\Loop}_m$ in $D$
space-time dimensions, with all particles in the adjoint
representation, may be written as
\begin{equation}
{\cal A}^{\Loop}_m\ =\ 
i^L \, g^{m-2 +2L } \,
\sum_{i\in \Gamma}{\int{\prod_{l = 1}^L \frac{d^D p_l}{(2 \pi)^D}
\frac{1}{S_i} \frac {n_i C_i}{\prod_{\alpha_i}{p^2_{\alpha_i}}}}}\,, 
\label{LoopGauge} 
\end{equation}
where $g$ is the gauge coupling constant.
The sum runs over the complete set $\Gamma$ of $m$-point
$L$-loop graphs with only cubic (trivalent) vertices, including all permutations of
external legs.  In each term, the product in the denominator runs over
all propagators of the corresponding cubic graph.  The integrations
are over the independent loop momenta $p_l$.  The coefficients $C_i$ are the
color factors obtained from the gauge-group structure constants by
dressing every three-vertex in the graph with a factor
\be
\f^{abc} = i \sqrt{2} f^{abc}=\Tr([T^{a},T^{b}]T^{c})\,,
\ee
where the hermitian generators of the gauge group are normalized via
$\Tr(T^a T^b) = \delta^{ab}$.
The coefficients $n_i$ are kinematic numerator factors depending on
momenta, polarization vectors and spinors. For supersymmetric
amplitudes in an on-shell superspace, the numerators will also contain
Grassmann parameters.
The symmetry factors $S_i$ of each graph remove any overcount
introduced by summing over all permutations of external legs, as well
as any internal automorphisms---symmetries of a graph with external
legs fixed.  
This form of the amplitude may be
obtained from representations involving higher-point contact interactions 
(as long as they are built out of $\colorf{abc}$'s).  One
reexpresses all contact terms as the product between a propagator and
its inverse; {\it i.e.}~one inserts $1 = p_{\alpha_i}^2/p_{\alpha_i}^2$,
and assigns the inverse propagator $p_{\alpha_i}^2$ to be part of a
numerator factor $n_i$.
The duality conjecture~\cite{BCJ,BCJLoop} states that there should exist a
representation of the amplitude where the numerator factors $n_i$
satisfy equations in one-to-one correspondence with the Jacobi
identities of the color factors.  Explicitly, it requires that
\begin{equation}
C_i = C_j + C_k \qquad  \Rightarrow \qquad  n_i = n_j + n_k \,,
\label{BCJDuality}
\end{equation}
where the first relation holds, thanks to the usual color Jacobi
identity, for any triplet $(i,j,k)$ of graphs
which are identical within the gray region in \fig{GeneralJacobiFigure}.
Moreover, the numerator factors carry the same antisymmetry 
properties as color factors, {\it i.e.}~if a color factor changes sign 
under the interchange of two legs, then so does the corresponding kinematic
numerator factor:
\begin{equation}
C_i \rightarrow -C_i \qquad  \Rightarrow \qquad  n_i \rightarrow -n_i \,.
\label{antisymmetry}
\end{equation}
These relations are conjectured to hold to all multiplicities, to all
loop orders in a weak-coupling expansion, and in any dimension in both
supersymmetric and non-supersymmetric Yang-Mills theory. Such
relations were noticed long ago for four-point tree
amplitudes~\cite{Halzen}. Beyond the four-point level, the relations 
are rather nontrivial and only work after appropriate rearrangements of
the amplitudes.

While the sign of the complete numerator factor $C_i n_i$ of a graph
is unambiguous, the sign of each factor, as well as the signs in
\eqn{BCJDuality}, may be changed as a consequence of the relations
(\ref{antisymmetry}) by simply interchanging two lines at any vertex.
This ambiguity reflects the different possible sign conventions in Jacobi
identities.  In this paper, the sign of each color factor $C_i$ (and
implicitly of $n_i$) is fixed by the corresponding graph figure: for
each integral, $C_i$ is built out of the structure constants
$\f^{abc}$ corresponding to each trivalent vertex, with legs $a,b,c$
ordered clockwise in the plane of the figure.

The kinematic version of the Jacobi identity (\ref{BCJDuality}) is the
key equation for the duality.  At loop level, this equation relates
graph numerators at the integrand level, as illustrated in
\fig{GeneralJacobiFigure}.  Therefore it is important to properly line
up both external and internal momenta. There is one such equation for
every propagator of every graph.  Of course, many of the equations are
simply related to each other by automorphic symmetries of graphs, and the
fact that any given equation can be obtained starting from each of the 
three contributing graphs.   Simultaneous consideration of all relations
 gives a  system of linear
functional equations that the amplitude's numerators should obey.  
As we will see, only a tiny subset of the possible equations needs to be
used when solving the system.  Once a tentative solution is found,
one must verify that the full set of equations is satisfied, in order to
have a duality-satisfying representation of the amplitude.
The existence of at least one
solution consistent with the unitarity cuts is the critical assumption
of the conjecture. Indeed, as we will see, the system of equations at
four loops is quite nontrivial and the emergence of a 
solution is striking.

There is by now substantial evidence in favor of the duality,
especially at tree level
($L=0$)~\cite{OtherTreeBCJ,virtuousTrees,Square,Oconnell},
where explicit representations of the numerators in terms of partial
amplitudes are known for any number of external
legs~\cite{ExplicitForms}.  A consequence of this duality is the
existence of nontrivial relations between the color-ordered partial
tree amplitudes of gauge theory~\cite{BCJ}, which have been proven both
from field theory~\cite{Feng} and string theory~\cite{Bjerrum1}
perspectives.  These relations were important in the recent
construction of all open-string tree
amplitudes~\cite{MSSallStringAmpl}. A partial Lagrangian understanding
of the duality has also been given~\cite{Square}.  An alternative
trace-based presentation of the duality relation~(\ref{BCJDuality}),
which emphasizes its group-theoretic structure, was described
recently~\cite{Trace}.

While less is known at loop level, several nontrivial tests have been
carried out. In particular, it has been confirmed that the duality
holds for the three-loop four-point amplitude
of $\NeqFour$ sYM theory~\cite{BCJLoop}.
(The one- and two-loop four-point amplitudes~\cite{GSB,BRY,BDDPR}
in this theory also manifestly satisfy the duality).  Similarly, the
duality-satisfying five-point one-, two- and three-loop amplitudes of
$\NeqFour$ sYM theory have recently been constructed~\cite{loop5ptBCJ}.
The color-kinematic duality is also known to hold~\cite{BCJLoop} for
the two-loop four-point identical-helicity amplitude of pure
Yang-Mills theory~\cite{AllPlus2}.  At present there is no proof that
the system of equations generated by the duality~(\ref{BCJDuality})
always has a solution consistent with the unitarity cuts of a given
theory, so it needs to be checked case by case.
In \sect{FourLoopSection} we will find a solution for the
four-loop four-point amplitude of $\NeqFour$ sYM theory.


Perhaps more surprising than the duality itself is a consequent
relation between gauge and gravity amplitudes.  Once the 
gauge-theory amplitudes are arranged into a
form satisfying eq.~(\ref{BCJDuality}), the numerator factors of the
corresponding $L$-loop gravity amplitudes, ${\cal M}^{\Loop}_m$,
can be obtained simply by multiplying together two copies of gauge-theory
numerator factors~\cite{BCJ,BCJLoop},
\begin{equation}
 {\cal M}^{\Loop}_m = i^{L+1} \, \left(\frac{\kappa}{2}\right)^{m-2+2L} \,
\sum_{i\in\Gamma} {\int{ \prod_{l = 1}^L \frac{d^D p_l}{(2 \pi)^D}
 \frac{1}{S_i}
   \frac{n_i \n_i}{\prod_{\alpha_i}{p^2_{\alpha_i}}}}} \,, 
\hskip .7 cm 
\label{DoubleCopy}
\end{equation}
where $\kappa$ is the gravitational coupling.  The
$\n_i$ represent numerator factors of a second gauge-theory amplitude
and the sum runs over the same set of graphs as in \eqn{LoopGauge}.
At least one family of numerators ($n_i$ or $\n_i$) must satisfy the
duality~(\ref{BCJDuality}). The construction (\ref{DoubleCopy}) is
expected to hold in a large class of gravity theories, including all
theories that are the low-energy limits of string theories. At tree
level, this double-copy property encodes the KLT
relations between gravity and gauge-theory amplitudes~\cite{KLT}.  For
$\NeqEight$ supergravity both $n_i$ and  $\n_i$ are numerators of
$\NeqFour$ sYM theory.

The double-copy formula (\ref{DoubleCopy}) has been
proven~\cite{Square} for both pure gravity and for $\NeqEight$
supergravity tree amplitudes, under the assumption that the
duality~(\ref{BCJDuality}) holds in the corresponding gauge theories,
pure Yang-Mills and $\NeqFour$ sYM theory, respectively.  The
nontrivial part of the loop-level conjecture is the existence of a
representation of gauge-theory amplitudes that satisfies the duality
constraints.  The double-copy property was explicitly checked for the
three-loop four-point amplitude of $\NeqEight$ supergravity in
ref.~\cite{BCJLoop} by comparing \eqn{DoubleCopy} against a spanning
set of unitarity cuts of the previously-calculated
amplitude~\cite{GravityThree,CompactThree}.  Here we perform a similar
check for the four-loop four-point amplitude of $\NeqEight$
supergravity, using the maximal cut
method~\cite{FiveLoop,CompactThree}.
The duality and double-copy property have
also been confirmed in one- and two-loop five-point amplitudes in
$\NeqEight$ supergravity~\cite{loop5ptBCJ}.  For less-than-maximal
supergravities, the double-copy property has been checked explicitly
for the one-loop four- and five-graviton amplitudes of ${\cal N} =
4,5,6$ supergravity~\cite{N46Sugra} by showing it matches known
results~\cite{DunbarEttle}.  In the two-loop four-graviton amplitudes
in these theories, it has been verified to be consistent
with the known infrared divergences and other properties~\cite{N46Sugra2}.
The double-copy property also
leads to some interesting relations between certain ${\cal N} \ge 4$
supergravity and subleading-color sYM amplitudes~\cite{SchnitzerBCJ}.

\subsection{Calculational setup}
\label{StrategySubsection}

We now demonstrate how the conjectured duality between color and
kinematics streamlines the construction of integrands of multiloop
gauge-theory amplitudes.  We first give an overview of the procedure
and illustrate it with the three-loop four-point amplitude, before
turning to the four-loop case in the following section.

To start the construction we enumerate the graphs with only cubic
vertices that can appear in a particular amplitude. Although this step
can be carried out in many different ways, we describe one 
that conveniently also generates the needed duality relations: We
assume we have a given set of known cubic graphs ({\it e.g.}~at four
loops we can start with the planar cubic graphs ones given in
ref.~\cite{BCDKS}).  Any missing graphs can then be generated by
applying the Jacobi relations~(\ref{BCJDuality}) to the set of known
graphs.  New graphs generated in this way are then added to the list
of known ones.  This process continues recursively, until no further
graphs or relations are found. At the end of the process all cubic
graphs related via the duality are known, and all duality
relations~(\ref{BCJDuality}) have been written down.  

The next step is to solve the relations thus generated. This is the
most complicated part of the construction. We can, however, simplify
the step by dividing it up into two separate parts,  the first of which
is straightforward.  First use a subset of the duality relations to express all
numerator factors in terms of the numerators of a judiciously chosen
small set of graphs, which we call ``master graphs''.  We identify
master graphs by systematically eliminating numerator factors from the
duality relations, via a functional analog of the standard row
reduction of systems of linear equations.  
This problem is analogous to the reduction of loop integrals
to a set of master integrals using the Laporta algorithm~\cite{Laporta}.
In both cases, there is freedom to change the order in which the
linear equations are solved.  Here, there is a freedom in the
choice of master graphs, which is equivalent to a choice of path in
solving the system of duality relations.  In all cases we have
examined, it is convenient to choose the master graphs to be planar
(although such a restriction does not necessarily yield the smallest
set).  This choice has the advantage that the planar contributions are
relatively simple and well studied in the literature.  In
particular, the planar contributions to the four-loop
four-point amplitude have a fairly simple form~\cite{BCDKS}.  For the
three-loop four-point $\NeqFour$ sYM amplitude we only need a single
master graph~\cite{BCJLoop,JJHenrikReview}.  In
\sect{FourLoopSection}, we will find that at four loops we
can express all numerators in terms of the numerators of only two
planar master graphs (or a single nonplanar master graph).

After the reduction of the system of duality constraints, our task is
to find explicit expressions for the master numerators.  As with any
functional equations, a good strategy is to write down Ans{\"a}tze for
the master numerators.  The Ans{\"a}tze are then constrained using input
from unitarity cuts, as well as symmetry requirements on both the master
numerators and on the numerators derived from them through the
duality relations.

In addition to the duality relations~(\ref{BCJDuality}) and unitarity
cuts, we may add extra constraints on numerator factors, motivated
by our prejudices about the structure and properties of the amplitude.
Although not necessary, such constraints, when well chosen, can greatly
facilitate the construction.  To find the four-loop four-point
$\NeqFour$ sYM amplitude we  use the following
{\it auxiliary constraints},
which are known to be valid for the duality-satisfying numerators at
three loops~\cite{BCJLoop}:
\begin{enumerate}
\item One-loop tadpole, bubble and triangle subgraphs do not appear
in any graph.
\item A one-loop $n$-gon subgraph carries no more than $n-4$ powers
  of loop momentum for that loop.
\item After extracting an overall factor of $s t A^\tree_4$, the 
numerators are polynomials in $D$-dimensional Lorentz products 
of the independent loop and external momenta.
\item Numerators carry the same relabeling symmetries as the graphs.
\end{enumerate}
We will assume that these observations carry over to the four-loop
four-point amplitude. If one of these auxiliary conditions had turned out
to be
too restrictive, it would have led to an inconsistency with either
the unitarity cuts or the duality relations. We would then have removed
conditions one by one until a consistent solution were found.
As we shall see in \sect{FourLoopSection}, these auxiliary constraints
are quite helpful for quickly finding a duality-satisfying representation
for the four-loop four-point amplitude.  A surprisingly small
subset of generalized unitarity cuts is then sufficient to completely determine
this amplitude.

The specific auxiliary constraints that should be imposed depend on
the problem at hand.  The third constraint is clearly specific to the
four-point amplitude, and should be modified for higher-point
amplitudes because they have a more complicated structure.  For the
five-point case, however, a simple generalization has been
found~\cite{loop5ptBCJ}, involving pre-factors that are
proportional~\cite{virtuousTrees} to linear combinations of five-point
tree-amplitudes.  For amplitudes in less supersymmetric theories, the
first and second conditions should be relaxed (in addition to the
third one), because one-loop triangle and bubble subgraphs are known
to appear in such theories.

\begin{figure}[t]
\includegraphics[scale=0.58]{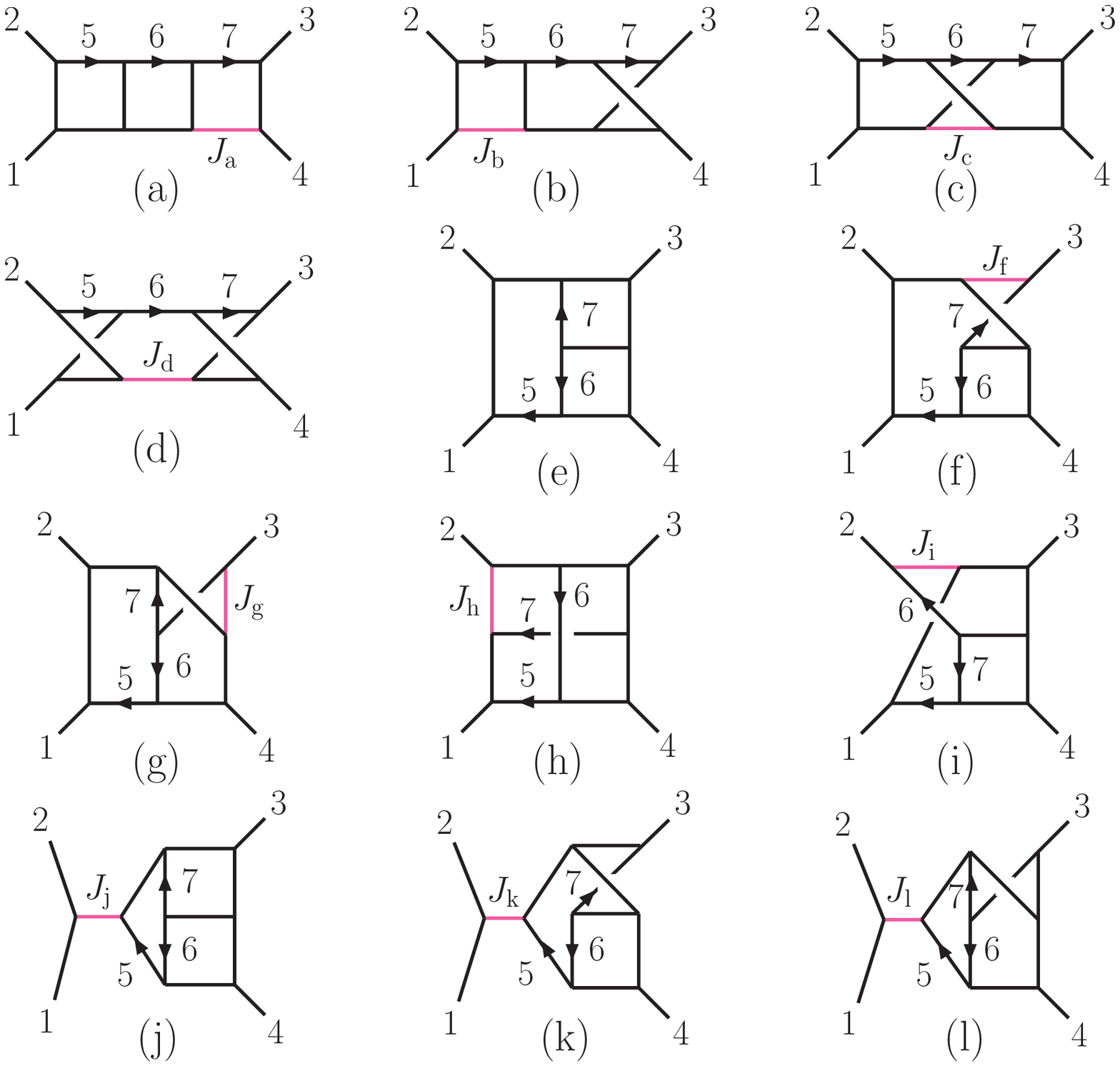}
\caption{The 12 nonvanishing graphs used in the construction of the
  $\NeqFour$ sYM and $\NeqEight$ supergravity three-loop four-point
  amplitude.  The shaded (pink) lines mark the application of the
  duality relation used to determine the numerator of the graph.  The
  external momenta are outgoing and the arrows mark the directions of
  the labeled loop momenta.  }
\label{ThreeLoopDiagramsFigure}
\end{figure}

\subsection{Three-loop warmup}

To illustrate the above procedure in some detail, we reconstruct the
well-studied three-loop four-point amplitude of $\NeqFour$ sYM theory.
This amplitude was originally constructed in
refs.~\cite{GravityThree,CompactThree}.  A form compatible with the
duality (\ref{BCJDuality}) was then found~\cite{BCJLoop}.
Here we describe how to streamline the construction of the latter form,
before following a similar procedure at four loops.

A straightforward enumeration shows that there are 17 distinct cubic
graphs with three loops and four external legs, which do not have
one-loop triangle or bubble (or tadpole) subgraphs. Only 12 contribute
to the amplitude, as shown in ref.~\cite{BCJLoop}. These 12
nonvanishing graphs, shown in \fig{ThreeLoopDiagramsFigure}, are
sufficient for explaining the construction.  (Had we kept all 17
graphs, the construction would be only slightly more involved, with
the result that the numerators of the additional five graphs vanish.)

Each numerator depends on three independent external momenta,
labeled by $k_1, k_2, k_3$, and on (at most) three independent loop
momenta, labeled by $l_5, l_6, l_7$, as well as on the external states.
The Mandelstam variables are $s = (k_1+k_2)^2$, $t=(k_1+k_4)^2$
and $u=(k_1+k_3)^2$.   We denote the color-ordered
tree-level amplitude by $A_4^\tree\equiv A_4^\tree(1,2,3,4)$.
The four-point amplitudes of $\NeqFour$ sYM theory are special.
Supersymmetry Ward identities fix the external state dependence,
and imply that an overall prefactor of $A_4^\tree$,
or equivalently the crossing-symmetric
prefactor $st A_4^\tree$~\cite{GravityThree,CompactThree}, 
can be extracted from every numerator factor $n^{(x)}$,
leaving behind new numerator factors $N^{(x)}$ that depend only on
the momenta,
\bea
n^{(x)} &=& st A_4^\tree(1,2,3,4) \, N^{(x)}\, , \nn \\
N^{(x)}&\equiv& N^{(x)}(k_1,k_2,k_3,l_5,l_6,l_7)\,.
\eea
Here $(x)$ refers to the label for each graph in
\fig{ThreeLoopDiagramsFigure}.  This result has been
argued to be valid in any dimension $D\le 10$ \cite{Neq44np},
justifying the third assumption above for these amplitudes.
The homogeneity of the Jacobi relations implies that they
hold for $N^{(x)}$ just as for $n^{(x)}$.  The crossing symmetry of
$st A_4^\tree$ implies that the symmetry properties of $N^{(x)}$ are
the same as those of $n^{(x)}$.

Next, we write down a subset of the duality relations that allows us
to identify the master graphs~\cite{JJHenrikReview}. 
For the three-loop four-point amplitude, one such
restricted set of duality relations is:
\begin{eqnarray}
 N^{(\rm a)}&=& N^{(\rm b)}(k_1,k_2,k_3,l_5,l_6,l_7)\,, \nn \\
 N^{(\rm b)}&=& N^{(\rm d)}(k_1,k_2,k_3,l_5,l_6,l_7)\,, \nn \\
 N^{(\rm c)}&=& N^{(\rm a)}(k_1,k_2,k_3,l_5,l_6,l_7)\,, \nn \\
 N^{(\rm d)}&=& N^{(\rm h)}(k_3,k_1,k_2, l_7, l_6, k_{1,3}-l_5+l_6-l_7)
   + N^{(\rm h)}(k_3, k_2, k_1,l_7,l_6,  k_{2,3}+l_5-l_7)\,,\nn \\
 N^{(\rm f)}&=& N^{(\rm e)}(k_1,k_2,k_3,l_5,l_6,l_7)\,, \nn \\
 N^{(\rm g)}&=& N^{(\rm e)}(k_1,k_2,k_3,l_5,l_6,l_7)\,, \nn \\
 N^{(\rm h)}&=& - N^{(\rm g)}(k_1, k_2, k_3, l_5, l_6, k_{1,2} - l_5 - l_7)
- N^{(\rm i)}(k_4, k_3, k_2,  l_6-l_5,  l_5 - l_6 + l_7-k_{1 ,2}, l_6)\,,\nn \\
 N^{(\rm i)}\,&=& N^{(\rm e)}(k_1,k_2,k_3,l_5,l_7,l_6)
   - N^{(\rm e)}(k_3, k_2, k_1, -k_4 - l_5 - l_6, - l_6 - l_7, l_6)\,, \nn\\
 N^{(\rm j)}\,&=& N^{(\rm e)}(k_1,k_2,k_3,l_5,l_6,l_7)
             - N^{(\rm e)}(k_2,k_1,k_3,l_5,l_6,l_7)\,,\nn \\
 N^{(\rm k)}&=& N^{(\rm f)}(k_1,k_2,k_3,l_5,l_6,l_7)
             - N^{(\rm f)}(k_2,k_1,k_3,l_5,l_6,l_7)\,,\nn \\
 N^{(\rm l)}&=& N^{(\rm g)}(k_1,k_2,k_3,l_5,l_6,l_7)
             - N^{(\rm g)}(k_2,k_1,k_3,l_5,l_6,l_7)\,,
 \label{BCJjacobi}
\end{eqnarray}
where $k_{i,j}\equiv k_i + k_j$.  For convenience we have suppressed
the canonical arguments $(k_1,k_2,k_3,l_5,l_6,l_7)$ of the numerators
on the left-hand side of the equations (\ref{BCJjacobi}), as we will
often do in the remainder of the paper. Each relation specifying an
$N^{(x)}$ is generated by considering the dual Jacobi
relations focusing around the lightly colored (pink) line labeled
$J_x$ in \fig{ThreeLoopDiagramsFigure}.  In general, the duality
condition relates triplets of numerators; sometimes, however, one or
two of the numerators vanish because the associated graph has a
one-loop triangle subgraph forbidden by our auxiliary
constraints.  Specifically, for five of the above equations, the
duality sets pairs of numerators equal; this occurs because the third
term in the triplet of numerators of \eqn{BCJDuality} vanishes due to
the presence of a triangle subgraph.
The above system can be used to express any numerator factor in terms
of combinations of the numerator $N^{\rm (e)}$, with various different
arguments. Thus, graph (e) can be taken as the sole master graph. This
is a convenient choice, but not the only possible one; for example,
either graph (f) or (g) can also be used as a single master
graph. None of the remaining nine graphs, however, can act alone as a
master graph.

One valid numerator factor (consistent with unitarity cuts) for graph
(e) is the ``rung-rule'' numerator~\cite{BRY},
\begin{equation}
N_{\rm rr}^{\rm (e)} = s (l_5 + k_4)^2\,.
\label{RungRuleNumerator}
\end{equation}
With this numerator, the graph possesses dual conformal symmetry.
However, it turns out that this numerator is incompatible
with the duality between color and kinematics~(\ref{BCJDuality}).

We are therefore looking for a modification of $N^{\rm (e)}$
consistent with both the maximal cut of the graph and with the
duality constraints (\ref{BCJjacobi}). We start by requiring that the
maximal cut of graph (e) is correct, and that the auxiliary
constraints in \sect{StrategySubsection} are satisfied. That is, the
numerator $N^{\rm (e)}$ has mass dimension four and possesses the
symmetry of the graph; no loop momentum for 
any box subgraph in (e) appears in it (ruling out $l_6$ and $l_7$);
and $N^{\rm (e)}$ is at most quadratic in the pentagon loop
momenta $l_5$.  (This last condition is looser than the second
auxiliary constraint in \sect{StrategySubsection}; we will
tighten it shortly.)  The symmetry condition implies that $N^{\rm (e)}$
should be invariant under
\begin{equation}
\{ k_1 \leftrightarrow k_2, k_3 \leftrightarrow k_4, 
     l_5 \rightarrow k_1 + k_2 - l_5 \}\,.
\end{equation}
The most general polynomial consistent with these constraints is
of the form,
\begin{equation}
N^{\rm (e)} =  s (l_5 + k_4)^2 +
         (\alpha s + \beta t) l_5^2 +
         (\gamma s + \delta t)(l_5-k_1)^2 + 
         (\alpha s + \beta t) (l_5-k_1-k_2)^2 \,,
\label{ThreeLoopAnsatz}
\end{equation}
where the four parameters $\alpha, \beta, \gamma, \delta$ are to be
determined by further constraints.  All added terms are
proportional to inverse propagators and therefore vanish on the maximal cut.
Thus, since \eqn{RungRuleNumerator} is consistent with the maximal cuts so is
\eqn{ThreeLoopAnsatz}.

According to the second auxiliary constraint in
\sect{StrategySubsection}, the numerator of a pentagon subgraph should
be at most linear in the corresponding loop momentum, not quadratic
as assumed above. Therefore the coefficient of $l_5^2$ in
\eqn{ThreeLoopAnsatz} should vanish, yielding the relations
$\gamma = -1- 2 \alpha $ and $\delta = - 2 \beta$. Consequently,
the Ansatz for $N^{\rm (e)}$ is reduced to
\begin{equation}
N^{\rm (e)} = s (\tau_{45}  +\tau_{15}) 
+  ( \alpha s +   \beta t) (s+\tau_{15} - \tau_{25})  \,,
\label{CleanAnsatz}
\end{equation}
where we use the notation,
\begin{equation}
\tau_{ij} \equiv 2 k_i \cdot l_j  \hskip .3 cm 
 (i \le 4, j \ge 5) \,, \hskip 2 cm 
\tau_{ij} \equiv 2 l_i \cdot l_j  \hskip .3 cm   (i, j \ge 5) \,.
\label{tauDef}
\end{equation}
Now there are just two undetermined parameters, $\alpha$ and $\beta$.

To determine one of the remaining parameters we use the properties of 
graph (j), and the expression for its numerator in terms of the numerator
of graph (e), which is given by the 9th duality constraint in
\eqn{BCJjacobi}. Inserting \eqn{CleanAnsatz} into this relation leads to
\begin{equation}
 N^{(\rm j)}
 = s (1+2\alpha -\beta) (\tau_{15} - \tau_{25}) 
               + \beta s (t-u) \,.
\label{Njtemp}
\end{equation}
Because the smallest loop in graph (j) carrying $l_5$ is a box
subgraph, our auxiliary constraints require that this momentum be
absent from $N^{\rm (e)}$.  Setting the first term in \eqn{Njtemp}
to zero implies that $\beta = 1+2 \alpha$, which in turn leads to
\begin{eqnarray}
N^{\rm (e)} &=& s (\tau_{45}  +\tau_{15}) +  
( \alpha (t-u)  + t) (s+\tau_{15} - \tau_{25})  \,, 
\label{Simplified_e}\\
N^{\rm (j)} &=& (1+2 \alpha) ( t-u ) s\,,
\label{Simplified_j}
\end{eqnarray}
leaving undetermined a single parameter $\alpha$. 

There are a variety of ways to determine the final parameter.
For example, one can use planar cuts to
enforce that the planar part of the amplitude is correctly reproduced.
A particularly instructive method is to use the duality relations to
obtain the numerator for planar graph (a) in terms of master numerator
$N^{\rm (e)}$.
The numerator $N^{\rm (a)}$ is quite simple once we impose the condition
that a one-loop box subgraph cannot carry loop momentum. Since three 
independent one-loop subgraphs of graph (a) are boxes, the numerator 
$N^{\rm (a)}$ 
cannot depend on any loop momenta.  Indeed, the iterated two-particle cuts,
or equivalently the rung insertion rule~\cite{BRY}, immediately fix
this contribution to be
\begin{equation}
N^{(\rm a)} = s^2 \,.
\label{NaValue}
\end{equation}
Solving the duality relations
(\ref{BCJjacobi}) to express $N^{\rm (a)}$ in terms of $N^{\rm (e)}$ we find,
\begin{eqnarray}
N^{(\rm a)} &=&
  N^{\rm (e)}(k_1, k_2, k_4, -k_3 + l_5 - l_6 + l_7, l_5 - l_6, -l_5) \nn \\
&& \null 
+ N^{\rm (e)}(k_2, k_1, k_4, -k_3 - l_5 + l_7, -l_5, l_5 - l_6) \nn \\
&& \null 
- N^{\rm (e)}(k_4, k_1, k_2, l_6 - l_7, l_6, l_5 - l_6)
- N^{\rm (e)}(k_4, k_2, k_1, l_6 - l_7, l_6, -l_5)\nn \\
&& \null  
- N^{\rm (e)}(k_3, k_1, k_2, l_7, l_6, l_5 - l_6)
-  N^{\rm (e)}(k_3, k_2, k_1, l_7, l_6, -l_5) \,.
\label{DualityEquationsForA}
\end{eqnarray}
Plugging in the value of the numerator factor $N^{\rm (e)}$ in
\eqn{Simplified_e}, we obtain
\begin{equation}
N^{\rm (a)} = s^2 + 
 (1 + 3 \alpha) \Bigl( (\tau_{16} - \tau_{46}) s
 - 2 (\tau_{17} + \tau_{37}) s 
 + (\tau_{16} - 2 \tau_{17} - \tau_{26} + 2\tau_{27}) t + 4 u t \Bigr)\,.
\end{equation}
Demanding that this expression matches the numerator factor $N^{\rm (a)}$
given in \eqn{NaValue}, or alternatively that it is independent of loop
momenta, fixes the final parameter to be $\alpha = -1/3$.
This constraint completely determines the numerator of graph (e) to be 
\be
N^{\rm (e)}= s (\tau_{45}  +\tau_{15}) + 
 \frac{1}{3}(t-s) (s+\tau_{15} - \tau_{25})\,,
\label{NumeratorE}
\ee
matching the result of ref.~\cite{BCJLoop}.

Remarkably, numerator (e) in \eqn{NumeratorE} generates all other
numerators $N^{(x)}$, via \eqn{BCJjacobi}, giving us the entire integrand at
three loops.  For all graphs, the resulting numerators reproduce the
expressions quoted in ref.~\cite{BCJLoop}, and the resulting amplitude
matches previous expressions~\cite{GravityThree,CompactThree} on
all $D$-dimensional unitarity cuts.  As already noted, it is highly
nontrivial to have a consistent solution where {\it all} duality
relations hold, all numerators have the graph symmetries and all
unitarity cuts are correct.  Squaring these numerators $N^{(x)}$,
using \eqn{DoubleCopy}, immediately yields the numerators for the
three-loop four-point $\NeqEight$ supergravity amplitude.  This form has 
also been confirmed against previous
expressions~\cite{GravityThree,CompactThree} on a spanning set
of $D$-dimensional unitarity cuts~\cite{BCJLoop}.

We shall use the same streamlined strategy to construct the four-loop
four-point amplitude in section~\ref{FourLoopSection}.  Before
carrying out this construction, however, we need to address an
important subtlety that appears in the construction of the three-loop
amplitude and affects the four-loop construction as well.

\subsection{Comment on one-particle-reducible graphs and snails}
\label{SnailSubtletySubsection}

Beyond tree-level, the on-shell three-point amplitudes of $\NeqFour$
sYM theory vanish.  The appearance of one-particle reducible (1PR)
graphs in the three-loop four-point amplitude may therefore seem
surprising. Indeed, graphs (i), (j) and (k)
of~\fig{ThreeLoopDiagramsFigure} do not appear in the original
representations of the same amplitude~\cite{GravityThree,CompactThree}.
The existence of 1PR
graphs may seem to imply that the three-point amplitude is
non-vanishing.  However, these graphs' numerators are proportional
to the Mandelstam invariant $s$, which is also the inverse propagator for
the sole line on which the graph is 1PR. Thus, the superficially 1PR
graphs are in fact just one-particle-irreducible (1PI) contact graphs.
Even though they are kinematically equivalent to
1PI graphs, the non-contact form of graphs (i), (j) and (k)
in~\fig{ThreeLoopDiagramsFigure}
is needed to describe easily their color structure, and to allow
the amplitude to obey the duality (\ref{BCJDuality}) between
color and kinematics.  
As we shall see, this feature continues at four loops,
where we encounter, not only graphs with three-point subgraphs, but also
nontrivial two-point subgraphs.  Some of these graphs
contain four-loop two-point bubble subgraphs on external legs,
and must be treated with particular care.

At first sight, it may appear surprising that two- and three-point
subgraphs show up; indeed in $\NeqFour$ sYM theory we expect the
vanishing of on-shell two- and three-point loop amplitudes.  This
property has been known in string theory for some
time~\cite{Martinec}.  By taking the low-energy limit, it should hold
in field theory as well.  A direct field theory argument for the
vanishing of the on-shell two-point function can be made as follows:
Quite generally, a (diagonal) on-shell two-point loop contribution
represents a correction to the mass of the corresponding field.  Gauge
invariance forbids such a term from being generated in the gluon
two-point function.  The chirality of $\NeqFour$ sYM gluinos forbids
such mass terms from being generated by perturbative quantum effects
for fermions as well. Thus, gluon and gluino two-point functions vanish
on shell.  Manifest off-shell ${\cal N}=1$ supersymmetry,
which can be maintained, then implies that the scalar field two-point
function also vanishes on shell.

We can also argue that three-point amplitudes vanish on shell.  Because
$\NeqFour$ supersymmetry relates all such amplitudes to each other, it
suffices to focus on the scattering amplitude of two fermions and one
scalar.  Up to an $SU(4)$ $R$-symmetry
transformation, we may further assume that
neither of the fermions is the ${\cal N}=1$ superpartner of the gluon.
Thus we consider only the interaction between ${\cal N}=1$ matter
multiplets.  Conservation of the matter $R$-symmetry subgroup $SU(3)$
requires that the three-field interaction is controlled by $SU(3)$
invariance, and thus is either holomorphic or antiholomorphic.  Now,
in the effective action language, the three-point amplitude originates
either from terms in the superpotential or the K\"ahler potential.
Due to the perturbative nonrenormalization of the
superpotential~\cite{SeibergNonrenormalization}, only the tree
amplitude comes from the former.  A nonvanishing loop amplitude can
only originate from a correction to the K\"ahler potential.  For
this case, a nonvanishing full superspace integral and Lorentz
invariance require that the product of three chiral superfields
containing the relevant wave functions must be accompanied by at least
two additional superderivatives.  In turn, this implies that the
product of one scalar and two fermion wave functions is always
accompanied by an external momentum invariant, originating from the
superspace integration measure. For
massless fields, any such product vanishes on shell. Thus, all quantum
corrections to three-point amplitudes in $\NeqFour$ sYM theory vanish on
shell, completing the argument.

While these arguments confirm the vanishing of two- and
three-point amplitudes at one-loop and beyond, we emphasize
that this does not mean that we cannot have graphs with two- and
three-point subgraphs. However, when such graphs appear they should always
carry factors that make their contributions vanish whenever legs are
cut (placed on shell) to isolate two- and three- point subamplitudes.
Indeed, we shall find that at four points, through four loops, all
such graphs with three- or four-point subgraphs can
be absorbed as contact terms in other graphs.  This property
is consistent with the fact that previous
representations of the three- and four-loop
amplitudes~\cite{GravityThree,CompactThree,Neq44np} do
not use any 1PR graphs with two- or three-point loop subgraphs.

\begin{figure}[t]
\includegraphics[scale=0.65]{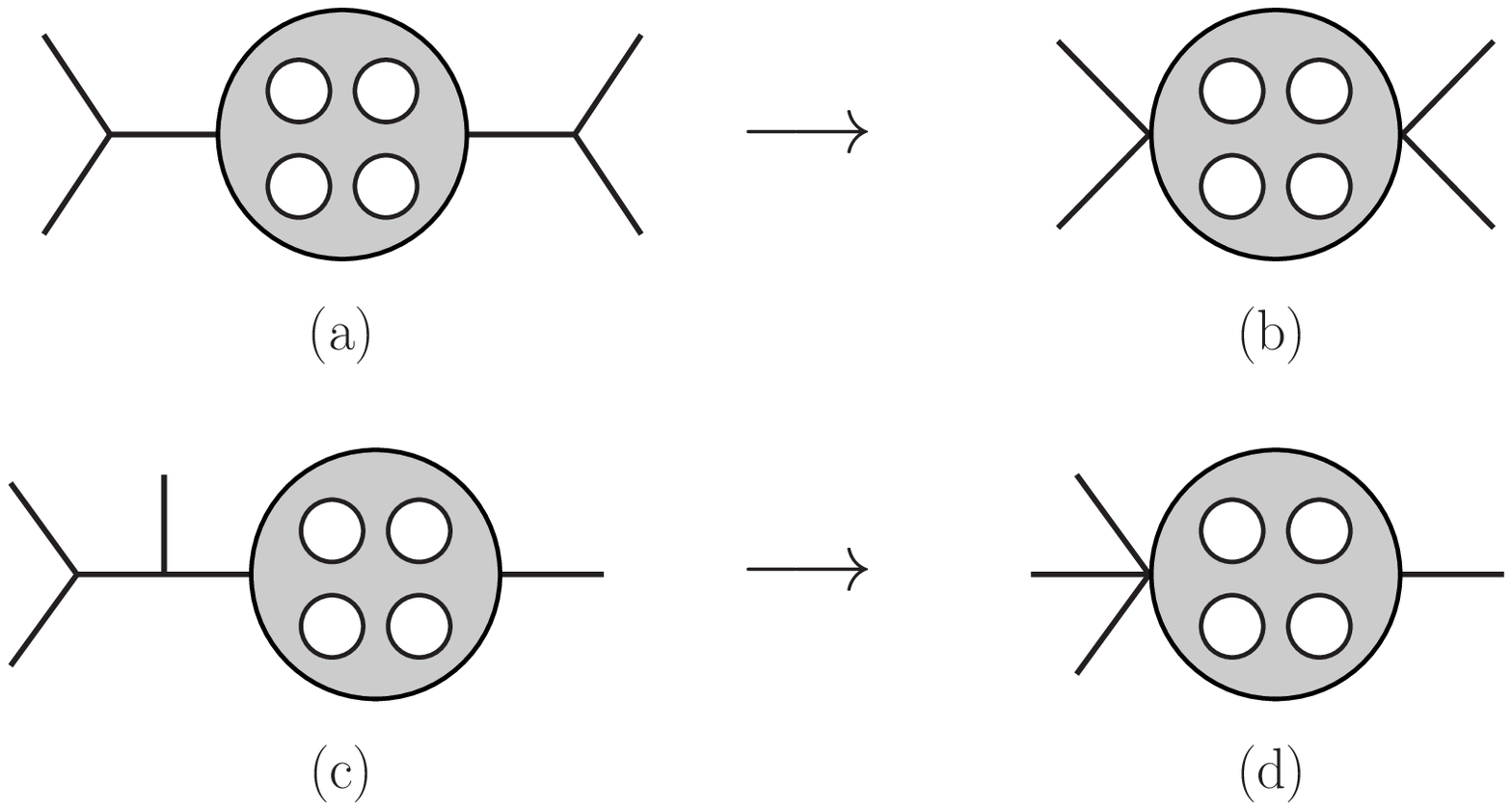}
\caption{Graphs (a) and (c) are generic propagator
  correction graphs that can appear at four loops and beyond if 
  we have a cubic organization of graphs.   Graphs (b) and (d) 
are rewritings of these graphs, which make explicit that in $\NeqFour$ 
sYM theory numerator factors always cancel the propagators that are
external to the loops in the four-point amplitude. }
\label{PropagatorFigure}
\end{figure}

At $L$ loops, the four-point amplitude in $\NeqFour$ sYM theory
is expected to have a representation with at most $2(L-2)$ powers of 
the loop momentum in the numerator of each 1PI cubic
graph~\cite{BDDPR,HoweStelleRevisited}.
At three loops, this power counting allows for cancellation of one internal
propagator, as in graphs (i), (j) and (k) of \fig{ThreeLoopDiagramsFigure}.
However, it precludes the existence of two-point graphs or
propagator corrections (and tadpole graphs), which would require
two inverse propagators or four powers of the loop momentum in the numerator.
On the other hand, at four loops and beyond, such graphs can and indeed
do appear.
Propagator corrections can be of two types: 
\begin{enumerate}
\item on internal legs, as shown in \fig{PropagatorFigure}(a), or
\item on external legs, as shown in \fig{PropagatorFigure}(c).
\end{enumerate}
In both cases the graph's numerators must contain momentum invariants
that cancel out the unwanted poles, so that they are kinematically
equivalent to the 1PI graphs shown in
\fig{PropagatorFigure}(b) and \fig{PropagatorFigure}(d),
respectively.

For case 1, this cancellation is straightforward because
the momentum invariant is nonvanishing for generic on-shell kinematics.
For case 2, the external leg corrections, the mechanism is more
subtle.  On the one hand, because the amplitudes have on-shell external
legs, a propagator in \fig{PropagatorFigure}(c) diverges:
$1/(k_1+ k_2 + k_3)^2=1/0$.  On the other hand, from the vanishing
of the on-shell two-point function we expect that the
numerator of \fig{PropagatorFigure}(c) is proportional to
$k_4^2=0$.  In order to resolve this 0/0 ambiguity, we need to regulate
the external leg by taking $k_4^2\neq0$, and cancel factors of $k_4^2$
between numerator and denominator.  This procedure yields the
``snail graph''\footnote{With suitable imagination, the graph
resembles a snail (as much as a penguin diagram resembles a penguin).}
in \fig{PropagatorFigure}(d), which is perfectly
well behaved at the level of the integrand, even with all external
momenta on-shell.

It is important to note that the snail graph in 
\fig{PropagatorFigure}(d) contains a scale-free integral, which
vanishes by the usual rules of dimensional regularization. We cannot,
however, simply ignore these contributions.  In dimensional
regularization, scale-free integrals evaluate to zero because of
cancellations between infrared and UV singularities.  Ignoring the
snail graph contributions in $\NeqFour$ sYM theory would lead to
incorrect values for the UV divergences.\footnote{In QCD, propagator
corrections on external legs can be ignored because the UV divergences
are known {\it a priori}. It is therefore quite simple to restore the
missing terms.  In $\NeqFour$ sYM theory, UV divergences in $D>4$ are unknown
{\it a priori}.}  Since we are interested in this paper in the coefficient
of the UV divergences, these snail graphs must be included.

While enforcing the duality constraints (\ref{BCJDuality}) brings the
phenomenon of snail graphs to the forefront, we emphasize that the
potential appearance of such contributions to amplitudes is
independent of the color-kinematic duality.  Snail contributions can
in principle occur within any representation; it is therefore
important to always check the unitarity cuts for such contributions.
Because snails are associated with external leg contributions,
ordinary unitarity cuts fail to detect them, and 
generalized cuts are required.  The momenta of the states crossing
the cut must either be complex, or else have an indefinite sign of their
energy.  (We have examined such cuts, and have confirmed thereby
that no snail contributions are
present in the representation found in ref.~\cite{Neq44np}.)

Although the snail contributions are important for $\NeqFour$ sYM
amplitudes, they will not infect the corresponding $\NeqEight$
supergravity amplitudes.  This may be understood heuristically as a
consequence of the double-copy formula (\ref{DoubleCopy}).  In
$\NeqFour$ sYM theory, graphs of the form in \fig{PropagatorFigure}(c)
carry a factor of 0 in their numerator to cancel the 1/0 from the
propagator.  In $\NeqEight$ supergravity we get a second factor of $0$
from the second copy, making the numerator vanish faster than the
denominator, and giving a vanishing snail contribution.
Below we confirm this heuristic argument directly from unitarity cuts.

Finally, we remark that very similar considerations appear in the analysis of
inverse derivative factors arising from the collision of vertex operators
in the discussion of nonrenormalization conditions for amplitudes in
superstring theory---see section 3.2 of
ref.~\cite{DoubleTraceNonrenormalization}.


\section{The four-loop four-point integrand}
\label{FourLoopSection}

We now turn to the construction of the four-loop four-point amplitude and 
follow the same strategy as described in the previous section
for the corresponding three-loop amplitude. 

\subsection{Overview of the result}

\begin{figure}[tb]
\includegraphics[scale=0.9]{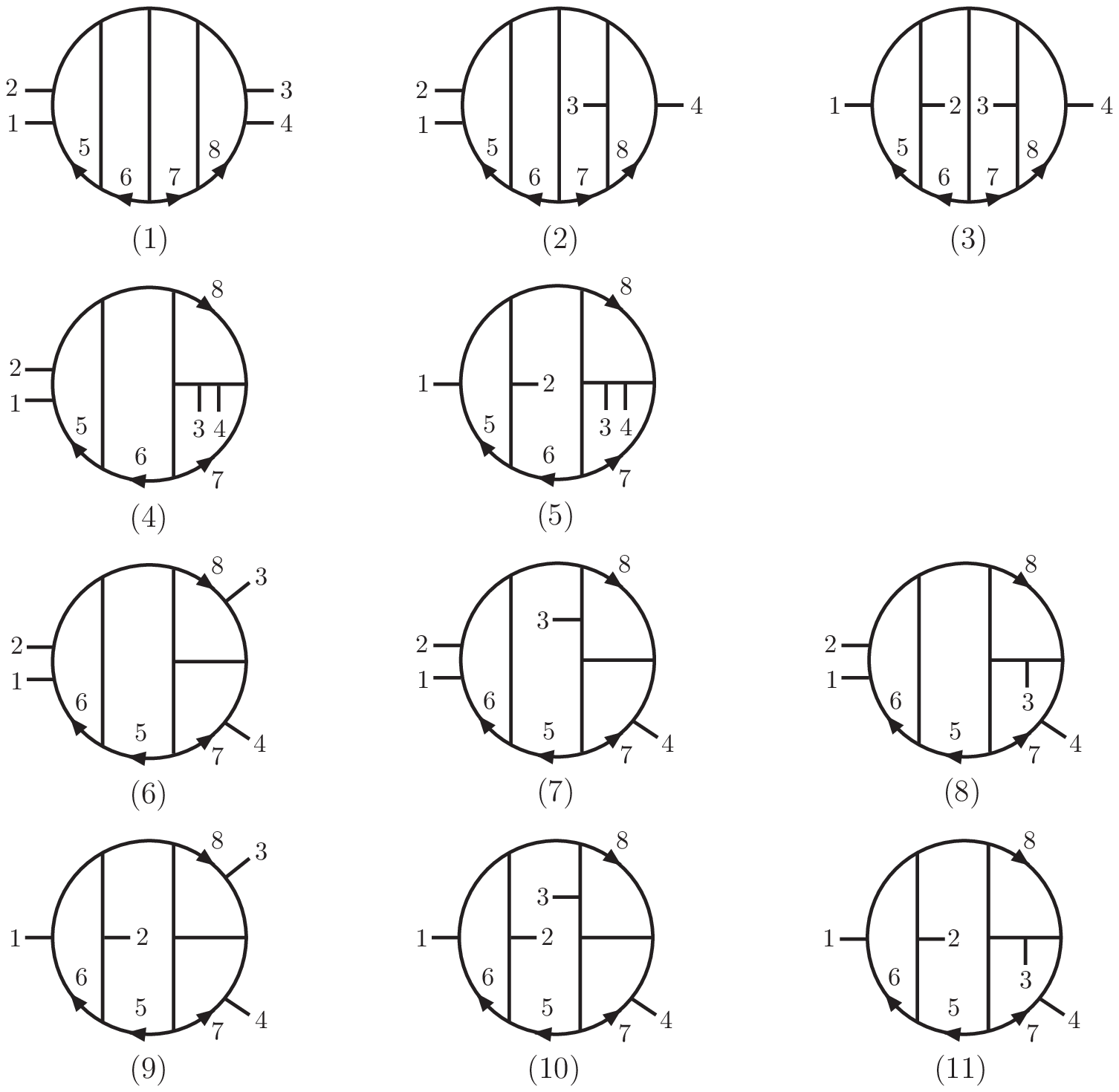}
\caption{Cubic graphs 1 to 11 that contribute to
the four-loop four-point amplitude of $\NeqFour$ sYM theory and
$\NeqEight$ supergravity.  The labels 1 to 4, indicate the legs
carrying external momenta $k_1$ to $k_4$. The labels 5 to 8 indicate
the propagators carrying the independent loop momenta $l_5$ to $l_8$.
The arrows indicate the direction of the momentum. The graphs also
specify the color factor of the graph, simply by dressing each cubic
vertex with an $\f^{abc}$, respecting the clockwise ordering of lines
at each vertex.}
\label{BC1Figure}
\end{figure}

\begin{figure}[tb]
\includegraphics[scale=0.875]{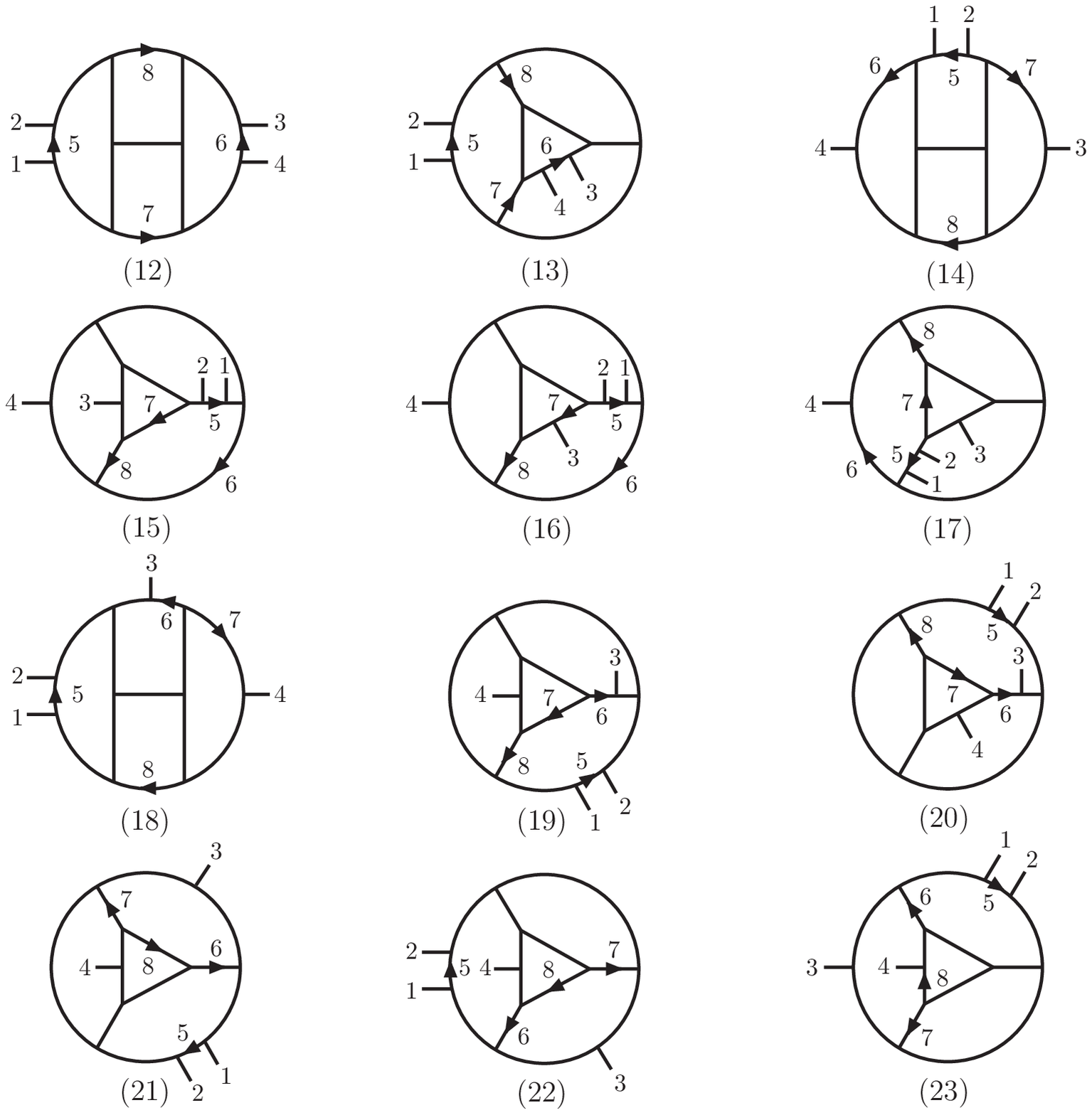}
\caption{Cubic graphs 12 to 23 that contribute to the
  four-loop four-point amplitude of $\NeqFour$ sYM theory and $\NeqEight$
  supergravity. The labeling is the same as for \fig{BC1Figure}.}
\label{D1Figure}
\end{figure}

\begin{figure}[tb]
\includegraphics[scale=0.875]{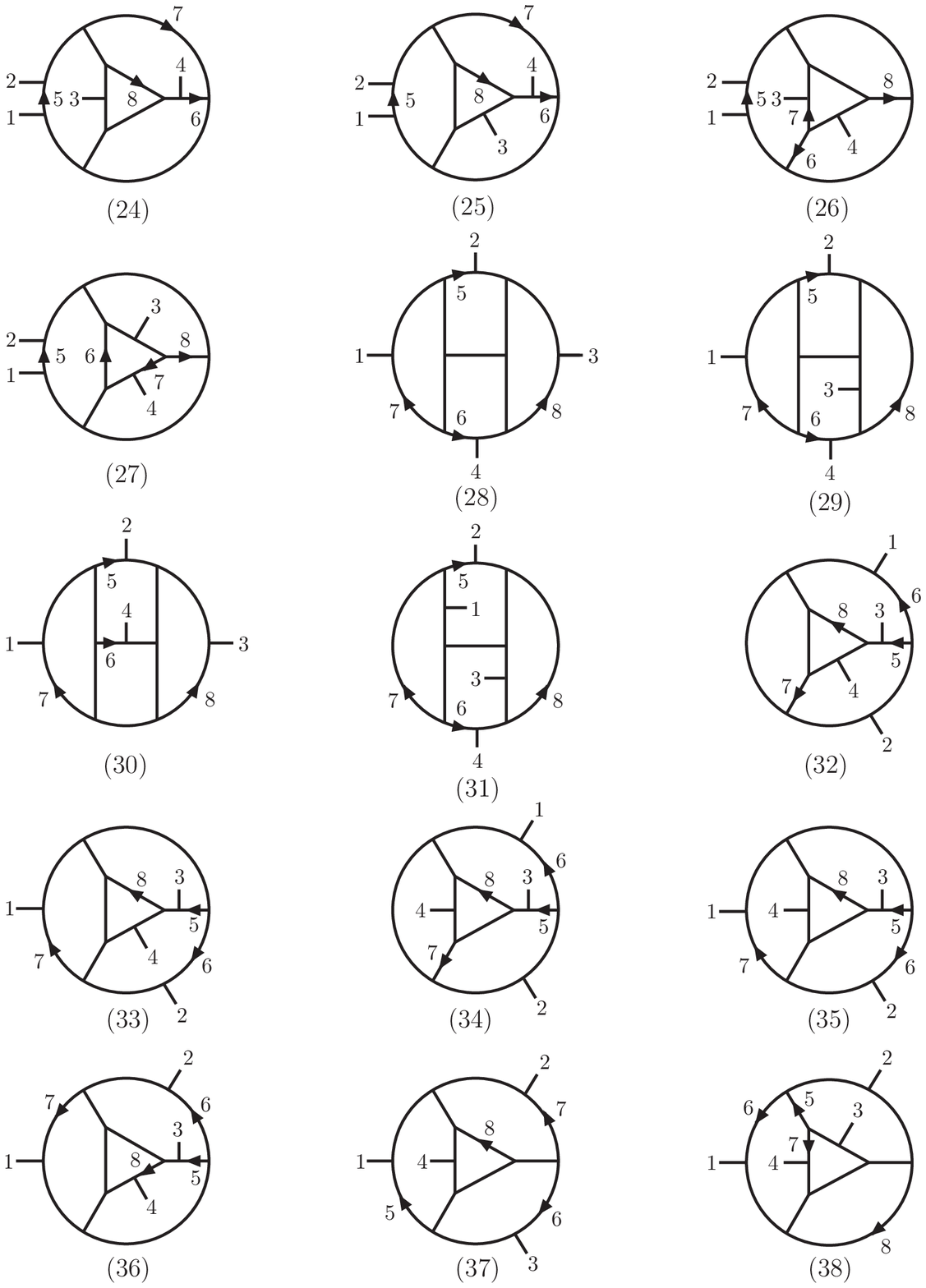}
\caption{Cubic graphs 24 to 38 that contribute to the
  four-loop four-point amplitude of $\NeqFour$ sYM theory and $\NeqEight$
  supergravity. The labeling is the same as for \fig{BC1Figure}. }
\label{D2Figure}
\end{figure}

\begin{figure}[tb]
\includegraphics[scale=0.86]{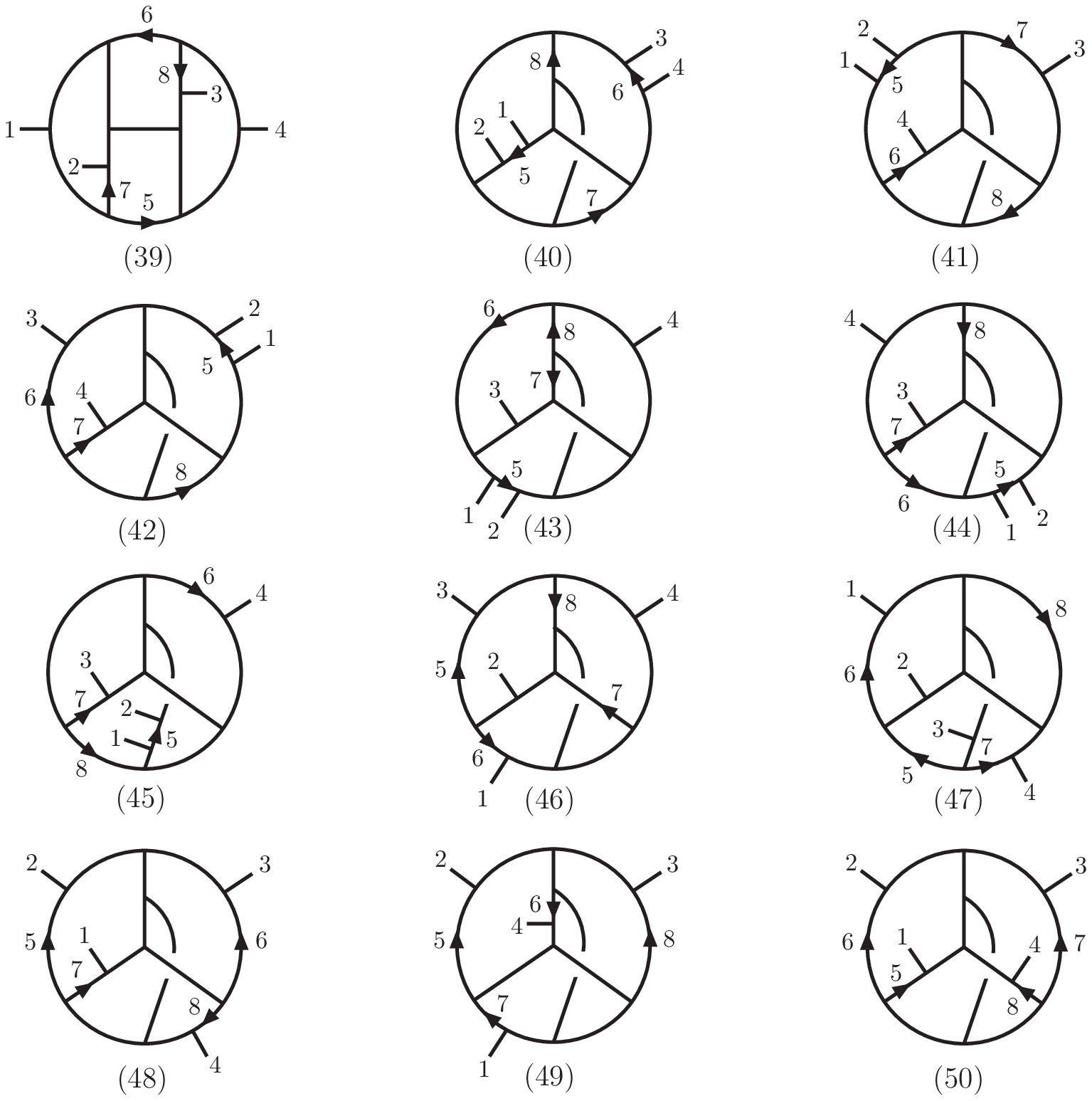}
\caption{Cubic graphs 39 to 50 that contribute to the
  four-loop four-point amplitude of $\NeqFour$ sYM theory and $\NeqEight$
  supergravity. The labeling is the same as for \fig{BC1Figure}. }
\label{E1Figure}
\end{figure}

\begin{figure}[t]
\includegraphics[scale=0.875]{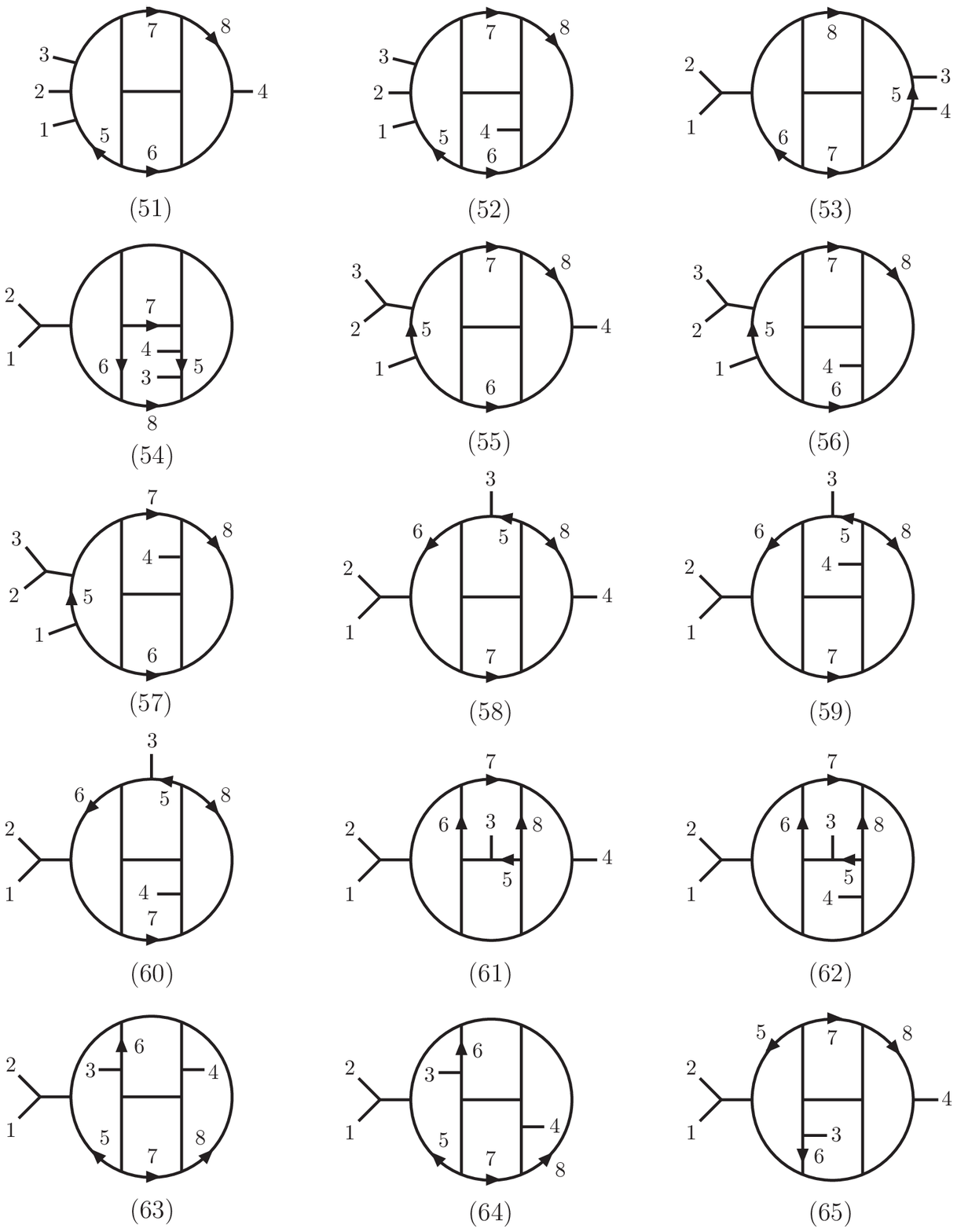}
\caption{Cubic graphs 51 to 65 that contribute to the
  four-loop four-point amplitude of $\NeqFour$ sYM theory and $\NeqEight$
  supergravity. The labeling is the same as for \fig{BC1Figure}. }
\label{D4Figure}
\end{figure}

\begin{figure}[t]
\includegraphics[scale=0.875]{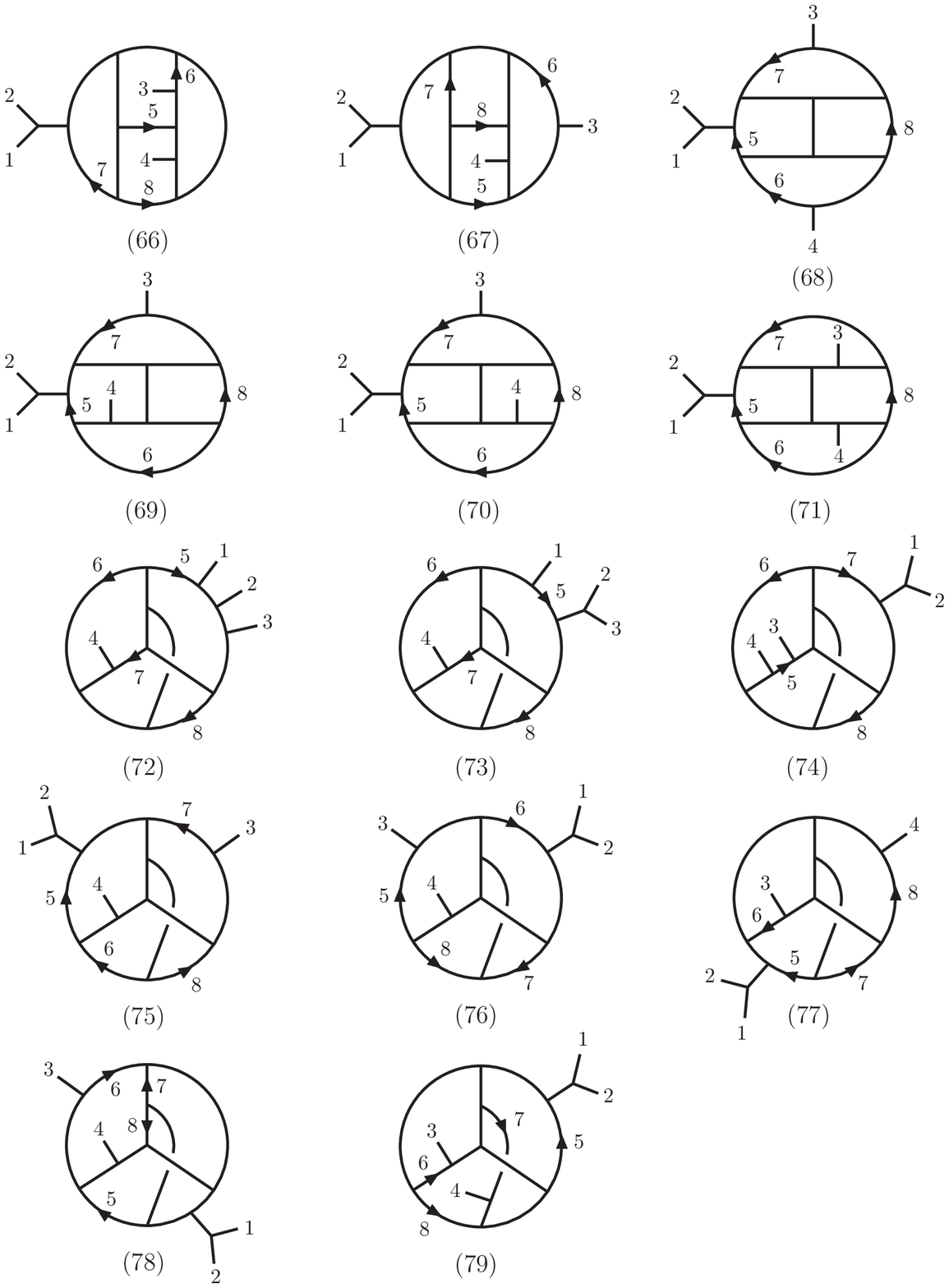}
\caption{Cubic graphs 66 to 79 that contribute to the
  four-loop four-point amplitude of $\NeqFour$ sYM theory and $\NeqEight$
  supergravity. The labeling is the same as for \fig{BC1Figure}.}
\label{D5Figure}
\end{figure}

\begin{figure}[t]
\includegraphics[scale=0.89]{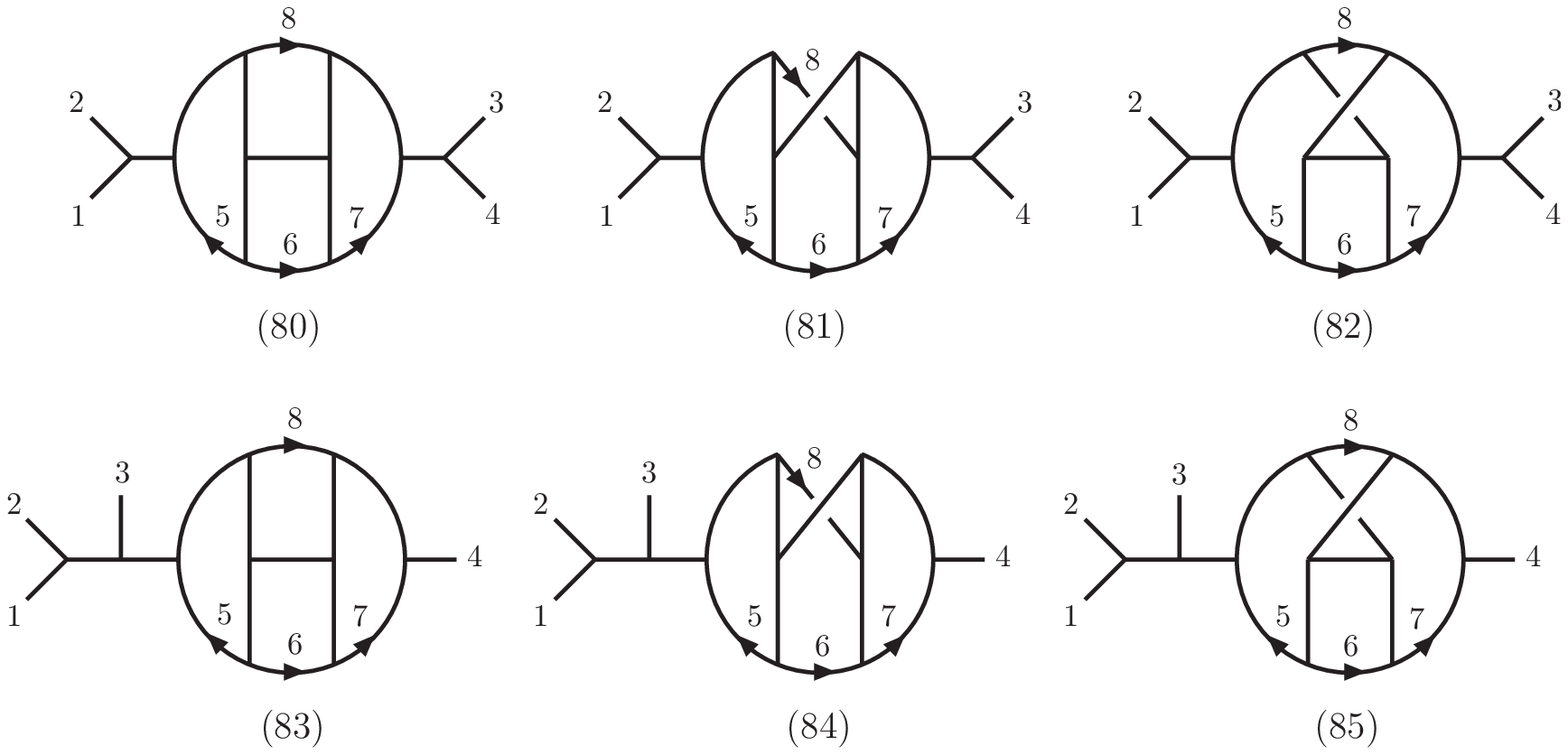}
\caption{Cubic graphs 80 to 85 that contribute to the
  four-loop four-point amplitude of $\NeqFour$ sYM theory. Graphs 
  83, 84, and 85 vanish identically for $\NeqEight$
  supergravity, but carry an UV singularity in the $D=11/2$
   $\NeqFour$ sYM case. The labeling is the same as for \fig{BC1Figure}.}
\label{D6Figure}
\end{figure}

We will find that, in terms of the 85 distinct graphs of
figs.~\ref{BC1Figure}--\ref{D6Figure}, the four-loop sYM amplitude is
given by
\begin{eqnarray}
{\cal A}_4^\fourloop &=& g^{10} s t A_4^\tree \sum_{{\cal S}_4}
\sum_{i=1}^{85} \int 
\biggl(\prod_{j=5}^8 \frac{d^{D} l_j}{(2\pi)^D}\biggr) \, 
\frac{1}{S_i}\frac{N_i(k_j,l_j) \, C_i }
{\prod_{\alpha_i =1}^{13} p_{\alpha_i}^2} \,, \hskip 1 cm 
\label{FourLoopYMAmplitude}
\end{eqnarray}
where $l_5,l_6,l_7,l_8$ are the four independent loop momenta and
$k_1,k_2,k_3$ are the three independent external momenta.  The
$p_{\alpha_i}$ are the momenta of the internal propagators
(corresponding to the internal lines of each graph $i$), and are
linear combinations of the independent loop momenta $l_j$ and the
external momenta $k_m$.  In the case of 1PR graphs, some
$p_{\alpha_i}$ will depend only on the external momenta.  As usual,
$d^{D} l_j/(2 \pi)^D$ is the $D$-dimensional integration measure for
the $j^{\rm th}$ loop momentum.  The numerator factors $N_i(k_j,l_j)$
are polynomial in both internal and external momenta, and are given in
\app{NumeratorAppendix}.  The color factors $C_i \equiv
C_i^{a_1a_2a_3a_4}$ are collected online~\cite{Online}, but they can
also be read directly off the figures.  The full amplitude is obtained
by summing over the group ${\cal S}_4$ of 24 permutations of the
external leg labels.  Overcounts are removed by the symmetry factors
$S_i$, which include both external symmetry factors (related to the
overcount from the sum over ${\cal S}_4$), as well as any internal
symmetry factors associated with automorphisms of the graphs holding
the external legs fixed. As at three loops, we extract the
crossing-symmetric, ${\cal S}_4$-invariant prefactor $st A_4^\tree$,
which contains all dependence on the external states.  (Notice that we
have used a slightly different notation for the independent loop momenta $l_j$
in \eqn{FourLoopYMAmplitude}, compared with $p_l$ in \eqn{LoopGauge}.)

Out of the 85 integrals in \eqn{FourLoopYMAmplitude},
graphs 50 and 79 are somewhat peculiar:  Their integrands are
nonvanishing, but they integrate to zero.  The vanishing of their
integrals can be seen from symmetry considerations alone.
For example, graph 50 has a symmetry exchanging legs 1 and 4, 
and legs 2 and 3, flipping the graph across a vertical midline.
It is easy to check that the color graph $C_{50}$ picks up a minus
sign under this operation; therefore the kinematic integrand must also
be antisymmetric, causing the integral to vanish.
In fact, as the duality between color and kinematics might
suggest, the color factors $C_{50}$ and $C_{79}$ vanish after the
internal color sum is carried out  (for any gauge group
$G$).
However, both graphs give nonvanishing contributions
to the $\NeqEight$ supergravity amplitude.  Therefore we retain them here.
(While the vanishing gauge-theory integrals are odd under the above relabeling
of the loop momenta, the double-copy property makes the gravity
integrals even under the same relabeling.)  

As we discussed in \sect{SnailSubtletySubsection}, graphs 83-85,
(displayed in \fig{D6Figure}), superficially appear as propagator
corrections on external legs.  These graphs give rise to the snail
contributions described there, after an external propagator is canceled
by a corresponding factor in the numerator.

Using the double-copy relation (\ref{DoubleCopy}), the four-loop
four-point $\NeqEight$ supergravity amplitude is obtained simply by
trading the color factor $C_i$ for $\tilde{n}_i = st \tilde{A}_4^\tree
N_i$ in \eqn{FourLoopYMAmplitude}.  Employing the relation $s^2t^2
A_4^\tree \tilde{A}_4^\tree = i stu \, M_4^\tree$ and changing the
gauge coupling to the gravitational coupling, we have
\begin{equation}
{\cal M}_4^\fourloop = - \Bigl(\frac{\kappa}{2}\Bigr)^{10} 
stu \, M_4^\tree 
\sum_{{\cal S}_4} \sum_{i=1}^{82} 
\int \biggl(\prod_{j=5}^8 \frac{d^{D} l_j}{(2\pi)^D}\biggr)
 \frac{1}{S_i}\frac{N_i^2(k_j,l_j)}
  { \prod_{\alpha_i=1}^{13} p_{\alpha_i}^2}  \, ,
\label{FourLoopGravityAmplitude}
\end{equation}
where $N_i(k_j,l_j)$ are the gauge-theory
numerator factors given in \app{NumeratorAppendix}.  In contrast to
the sYM amplitudes, potential snail contributions from graphs 83-85
vanish identically, as expected from our heuristic argument in
\sect{SnailSubtletySubsection}, and confirmed by an analysis of the
unitarity cuts. 

The symmetry factors appearing in
\eqns{FourLoopYMAmplitude}{FourLoopGravityAmplitude}
are given explicitly as:
\begin{align}
\sum_{i=1}^{85}\frac{1}{S_i}I_{i}&=
   \Frac{1}{4} I_{1} + \Frac{1}{4} I_{2} + \Frac{1}{16} I_{3}
 + \Frac{1}{4} I_{4} + \Frac{1}{8} I_{5} + \Frac{1}{2} I_{6}
 + \Frac{1}{2} I_{7} + I_{8} + \Frac{1}{4} I_{9}
 + \Frac{1}{4} I_{10} + \Frac{1}{2} I_{11} + \Frac{1}{4} I_{12}
 + \Frac{1}{2} I_{13} \nn \\&
\null+ \Frac{1}{2} I_{14} + \Frac{1}{4} I_{15} + I_{16} + \Frac{1}{2} I_{17}
     + I_{18} + I_{19} + I_{20} +  I_{21} + I_{22} + I_{23}
     + \Frac{1}{2} I_{24} + I_{25} + I_{26} \nn \\&
\null+ \Frac{1}{2} I_{27} + \Frac{1}{4} I_{28} + I_{29} + \Frac{1}{2} I_{30}
     + \Frac{1}{2} I_{31} + I_{32} + I_{33} + \Frac{1}{2} I_{34} + I_{35}
     + I_{36} + \Frac{1}{2} I_{37} +  \Frac{1}{4} I_{38} \nn \\&
\null+ \Frac{1}{2} I_{39} + \Frac{1}{4} I_{40} +  \Frac{1}{2} I_{41}
     + I_{42} + I_{43} + \Frac{1}{2} I_{44} +  \Frac{1}{4} I_{45}
     + \Frac{1}{2} I_{46} + \Frac{1}{8} I_{47} +  \Frac{1}{2} I_{48}
     + \Frac{1}{2} I_{49} + \Frac{1}{8} I_{50}  \nn \\&
\null+  \Frac{1}{2} I_{51} + I_{52} + \Frac{1}{4} I_{53} + \Frac{1}{2} I_{54}
     + \Frac{1}{2} I_{55} + \Frac{1}{2} I_{56} + \Frac{1}{2} I_{57}
     + \Frac{1}{2} I_{58} + \Frac{1}{2} I_{59} + \Frac{1}{2} I_{60}
     + \Frac{1}{4} I_{61} + \Frac{1}{2} I_{62}  \nn \\&
\null+ \Frac{1}{2} I_{63} + \Frac{1}{2} I_{64} + \Frac{1}{2} I_{65}
     + \Frac{1}{4} I_{66} + \Frac{1}{2} I_{67} + \Frac{1}{4} I_{68}
     + \Frac{1}{4} I_{69} + \Frac{1}{2} I_{70} + \Frac{1}{8} I_{71}
     + \Frac{1}{2} I_{72} + \Frac{1}{2} I_{73} + \Frac{1}{4} I_{74}  \nn \\&
\null+ \Frac{1}{4} I_{75} + \Frac{1}{2} I_{76} + \Frac{1}{2} I_{77}
     + \Frac{1}{4} I_{78} + \Frac{1}{8} I_{79} + \Frac{1}{16} I_{80}
     + \Frac{1}{8} I_{81} + \Frac{1}{16} I_{82} + \Frac{1}{4} I_{83}
     +  \Frac{1}{2} I_{84} + \Frac{1}{4} I_{85}
\, ,  \nn \\ 
\label{FourloopCombinatoricSum}
\end{align}
where $I_i$ should be interpreted only as placeholders for the graphs,
including both the numerator or color factors, in either theory. 

\begin{figure*}[tb] 
\begin{center}
\includegraphics[scale=0.95]{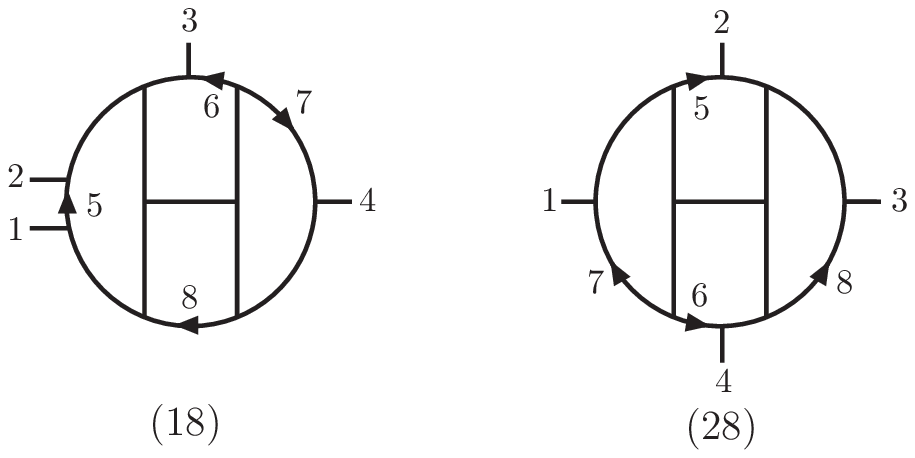}
\end{center}
\vskip -.9 cm 
\caption[a]{\small The planar master graphs, 18 and 28.  The numerators 
   and color factors of all other graphs are generated from the numerators
   and color factors of these two graphs through kinematic Jacobi relations.}
\label{Master4Figure}
\end{figure*}

\subsection{The calculation}
\label{FourLoopCalcSection}

As at three loops, the construction of the amplitude begins by writing
down a sufficient number of duality constraints so that a set of
master numerators can be identified.  We have constructed a set of
duality equations similar to the three-loop ones of \eqn{BCJjacobi}.
In \app{JacobiAppendix} we collect a set of simplified equations
derived from these duality constraints by imposing the four auxiliary
constraints presented in \sect{StrategySubsection}.  Because of these
additional simplifications, the duality equations in the appendix are
valid only for $\NeqFour$ sYM theory.  The duality equations
allow us to express all non-snail numerators directly as linear combinations
of the numerators $N_{18}$ and $N_{28}$.  The corresponding graphs are
shown in \fig{Master4Figure}; we will choose them as the master graphs.  

It is interesting to note that as an alternative we can use
a single nonplanar master graph that does the same job, such as
graph 33 (or equivalently 35 or 36, which have identical numerators up
to a sign).  However, we prefer to use planar graphs as master graphs
because their numerators have a somewhat simpler structure.
If we choose planar graphs as master graphs then the minimal number
is two.  In our treatment the snail contributions are only given partially
in terms of the master numerators, because the latter are
specified using on-shell external kinematics, whereas the numerators
of the former require an off-shell regularization to be nonvanishing.

Our next task is to determine the master numerators.  To this end we
begin by constructing an Ansatz for the numerator factors $N_{18}$ and
$N_{28}$ that satisfies the auxiliary constraints discussed in
\sect{StrategySubsection} and a restricted set of duality
relations. We then constrain the Ansatz by demanding that 
other duality relations are satisfied, and that the numerator
factors of the other integrals obey the auxiliary constraints.  For
both graphs, the numerator must be independent of loop momenta $l_7$ and
$l_8$ because they are assigned to one-loop box subgraphs, whose
momenta should not appear in their numerators.  Similarly, momenta
$l_5$ and $l_6$ are assigned to one-loop pentagon subgraphs, so
$N_{18}$ and $N_{28}$ should be no more than linear in these momenta,
according to our auxiliary constraints.  Thus, each of the two master
numerators should be a polynomial built from the monomials,
\be
M = \{ s^3, st^2, s^2t, t^3, \tau_{i5} s^2,  \tau_{i5} t^2,
    \tau_{i5} st, \tau_{i6} s^2,  \tau_{i6} t^2, \tau_{i6} st,
    \tau_{i5} \tau_{j6} s,  \tau_{i5} \tau_{j6} t, \tau_{56} s^2,
    \tau_{56} t^2, \tau_{56} st \} \,, \hskip .4 cm 
\ee
where $i = 1,2,3$ labels the three independent external momenta,
$k_1,k_2,k_3$.  In total this gives us a polynomial with 43 terms for
each master graph.  Labeling the monomials consecutively as $M_j$,
and including arbitrary coefficients, we have as our starting Ansatz,
\begin{equation}
N_{18} = \sum_{j=1}^{43} a_j M_j \,, \hskip 1 cm 
N_{28} = \sum_{j=1}^{43} b_j M_j \,.
\label{FourLoopAnsatz}
\end{equation}
The 86 free coefficients $a_j$ and $b_j$ are to be determined from various
consistency conditions obtained from the color-kinematic duality, graph 
symmetries and
unitarity cuts. The number of free parameters that need to be
determined in this construction is remarkably small, considering the
expected analytic complexity of amplitudes at four loops.

Using the 86-parameter Ansatz and the solution to the restricted set of
duality constraints listed in \app{JacobiAppendix} gives us
expressions for the numerator factors of any of the 82 non-snail 
graphs appearing in the amplitude. 
(The snail graphs will be determined below
in terms of the non-snail graphs using generalized unitarity cuts.)
These expressions do not yet satisfy all duality constraints; thus far
we have imposed only the relatively few relations in \app{JacobiAppendix}
sufficient to determine all numerators in terms of the master numerators, but
we have not yet accounted for the complete set of duality relations.
To further constrain the master Ansatz we could require that all other
dual Jacobi relations are satisfied; there are on the order of
$13\times85$ such functional relations (not all independent).  An
alternate strategy, which we follow here, is to first impose the
consistency constraints on the numerators of the graphs derived from
$N_{18}$ and $N_{28}$ through 
\eqns{smallJrel4loop}{trivialNrelations}.  After obtaining a complete
solution for all 86 parameters appearing in the ansatz, we then verify
that they indeed satisfy all remaining duality relations and unitarity cuts.
An advantage of this strategy is that it allows us to illustrate the
remarkably small number of unitarity cuts needed to find the complete
amplitude, including nonplanar contributions. 

As we shall see, to construct the complete amplitude we need only
information about the unitarity cuts of the four-loop planar amplitude, 
obtained previously in refs.~\cite{BCDKS,Neq44np}.  The list of
constraints needed to fix all 86 parameters in the Ansatz, thus
determining the amplitude, is remarkably short.  It is sufficient to enforce:
\begin{enumerate}
\item the graph automorphism symmetries on numerators
$N_{12}$, $N_{14}$ and $N_{28}$;
\item the maximal cut of graph 12;
\item the next-to-maximal cut of graph 14, where $l_5$ is the off-shell leg.
Graph 68 also contributes to this cut.
\end{enumerate}
Strikingly, only two rather simple planar cuts are needed to fully
determine the amplitude.  Let us now discuss some details of fixing
the parameters.

We start by analyzing the consequences of the symmetries of the
master graph 28: This graph is invariant under two independent
transformations:
\begin{eqnarray}
\{k_1 \leftrightarrow k_3,\,
  l_5 \rightarrow k_2 - l_5,\,
  l_6 \rightarrow k_4 - l_6,\,
  l_7 \leftrightarrow l_8 \}\,,
\label{N28sym1}
\end{eqnarray}
and
\begin{eqnarray}
\{k_2 \leftrightarrow k_4,\,
  l_5 \leftrightarrow l_6,\,
  l_7 \rightarrow k_1 - l_7,\,
  l_8 \rightarrow k_3 - l_8 \}\,.
\label{N28sym2}
\end{eqnarray}
Imposing the invariance of numerator $N_{28}$ under
\eqns{N28sym1}{N28sym2} reduces the number of its unknown coefficients
from 43 to 14. The other master graph, graph 18, does not have
any such automorphism relations; we are therefore left to determine a
total of 57 parameters.

We then impose similar symmetry conditions on $N_{12}$, which 
may be written in terms of $N_{18}$ and $N_{28}$ as
\bea
N_{12}&=&
-N_{18}(k_4, k_3, k_1, l_6, -l_5, -l_6, l_8) 
+ N_{18}(k_4, k_3, k_2, l_6, k_2 + l_8, l_5, l_7) \nn\\
 && \null
 +  N_{18}(k_4, k_3, k_2, k_3 + l_8, l_5 - l_8, -l_6, l_8) - 
 N_{28}(k_1, k_2, k_3, l_5 - l_8, k_3 - l_6 + l_8, -l_6, l_8) \nn\\
 && \null
 + N_{28}(k_2, k_1, k_3, -l_5, 0, -l_6, l_8) - 
 N_{28}(k_4, k_3, k_1, l_6 - l_8, k_2 - l_5 + l_8, l_8, l_8) \nn\\
 && \null
 + N_{28}(k_4, k_3, k_2, k_3, k_1 + l_5, -k_3 + l_6, l_8)\,,
\label{N12explicit}
\eea 
by combining the 2nd, 6th, 14th and 21st relations in
eq.~(\ref{smallJrel4loop}) in \app{JacobiAppendix}.  Invariance under
the automorphisms of graph 12 fixes 37 parameters, leaving
undetermined 20 parameters.  Similarly, imposing the graph symmetry
condition on $N_{14}$ reduces the total number of unknown parameters
to 17.  (These parameter counts are for the specific set of
duality relations given in \eqn{smallJrel4loop}. Using another set of
relations would result in somewhat different parameter counts;
however, the final solution would be the same.)

We could continue imposing more symmetry constraints on other
numerators, but we already have a very small set of undetermined
parameters. Ultimately, dynamical information provided by unitarity
cuts should become necessary.  Therefore we will now inspect some
cuts.  A good starting point is the maximal cut of graph 12.  Its
explicit value is easily obtained using the simple rung-rule numerator
of that graph~\cite{BRY,BCDKS},
\begin{equation}
N_{12}^{\rm rr} = s^2 (l_5 + l_6 + k_1+k_4)^2 \,.
\label{RungRuleNumerator12}
\end{equation}
The rung rule was originally designed to reproduce iterated
two-particle cuts. Since maximal cuts can be obtained from iterated
two-particle cuts by imposing additional cut conditions, the rung rule
reproduces the maximal cuts as well.  Our task is to match $N_{12}$,
as obtained from the duality relations, and $N_{12}^{\rm rr}$ in
\eqn{RungRuleNumerator12} on the maximal cut kinematics that uniquely
single out this graph, {\it i.e.}~we impose $p_i^2=0$ on all 13
propagators of graph 12.  Solving these conditions, we obtain 
\def\hs{\hskip .5 cm}
\begin{eqnarray}
&& 
l_i^2 =0\,, \hs  \tau_{15} = \tau_{25} = \tau_{36} = 0\,,\hs 
 \tau_{16} = -\tau_{26} \,, \hs
\tau_{17} = - \tau_{57},  \\
&& \tau_{38} = \tau_{68} \,,\hs \tau_{28} =  \tau_{58} \,,\hs 
\tau_{27} = - \tau_{37} + \tau_{57} +  \tau_{67} \,, \hs
\tau_{18} =  -s + \tau_{37} - \tau_{58} - \tau_{67} - \tau_{78}\,. \nn
\end{eqnarray} 
Thus, on the maximal cut, $N_{12}^{\rm rr}$ becomes
\begin{equation}
N_{12}^{\rm max.\, cut} = s^2 (t - \tau_{26} - \tau_{35} + \tau_{56})\,.
\end{equation} 
Requiring that the numerator obtained from the
Ansatz~(\ref{FourLoopAnsatz}) via \eqn{N12explicit}
matches $N_{12}^{\rm max.\, cut}$ on the maximal cut, the number of
undetermined parameters is reduced from 17 to 8.

Finally, all remaining parameters can be determined by requiring
the next-to-maximal cut of graph 14, where all propagators except for
$1/l_5^2$ are placed on shell, to be satisfied. 
Graph 68 also contributes
to this cut since it contains the same set of cut
propagators. Relabeling graph 68 so it matches graph 14, and
appropriately weighting the numerators by the remaining off-shell
propagators, we find under the cut kinematics, 
\begin{equation}
N_{14} + \frac{l_5^2}{s} 
\biggl(N_{68}\Bigr|_{\genfrac{}{}{0cm}{}{l_5 \rightarrow k_1-l_5,\, 
   l_6 \rightarrow -l_6}{l_7 \rightarrow -l_7,\, l_8 \rightarrow -l_8}} \biggr)
\ =\ s (l_5+k_2+k_3)^4
\,.
\end{equation}
The right-hand side of this equation is the numerator of graph 14,
as constructed using the rung rule; in the rung-rule representation
of the planar four-loop amplitude, graph 68 vanishes~\cite{BCDKS}.
This requirement fixes all remaining eight coefficients, giving us a
unique expression for the master numerators,
\begin{eqnarray}
N_{18} &=&
 \Frac{1}{4} (6 \uu^2 \t_{25} + \uu{} (2 \ss{} (5 \t_{25} + 2 \t_{26})
 - \t_{15} (7 \t_{16} + 6 \tt)) \nn\\
& & \null
 + \tt{} (\t_{15} \t_{26} - \t_{25} (\t_{16} + 7 \t_{26})) 
+ \ss{} (4 \t_{15} (\tt{} - \t_{26}) + 6 \t_{36} (\t_{35} - \t_{45}) \nn\\
 && \null
 - \t_{16} (4 \tt{} + 5 \t_{25}) - \t_{46} (5 \t_{35} + \t_{45})) 
 + 2 \ss^2 (\tt{} + \t_{26} - \t_{35} + \t_{36} + \t_{56}))
\,, \\
N_{28} &=& 
\Frac{1}{4} (\ss{} (2 \t_{15} \tt 
 + \t_{16} (2 \tt - 5 \t_{25} + \t_{35}) 
 + 5 \t_{35} (\t_{26} + \t_{36}) + 2 \tt{} (2 \t_{46} - \t_{56})
  - 10 \uu{} \t_{25})  \nn\\ 
&& \null
 - 4 \ss^2 \t_{25} - 6 \uu{} (\t_{46} (\tt - \t_{25} + \t_{45}) 
  + \t_{25} \t_{26}) - \tt{} (\t_{15} (4 \t_{36} + 5 \t_{46}) 
  + 5 \t_{25} \t_{36})) 
\nn \,.
\label{FourLoopMasters}
\end{eqnarray}
It is quite striking that the solution we obtain is unique and
relatively simple.  There are presumably other solutions to the
duality relations for this amplitude, but finding them would
require relaxing some of the auxiliary constraints described in
\sect{StrategySubsection}.

\begin{figure}[t]
\includegraphics[scale=0.84]{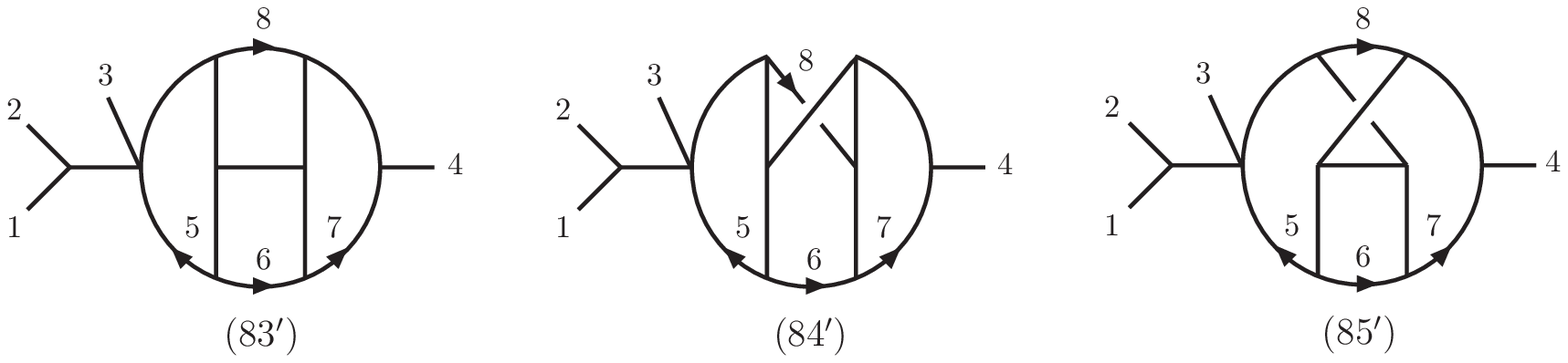}
\caption{The snail contributions in a form resolving the $0/0$
  ambiguity of graphs 83, 84 and 85.}
\label{SnailGraphsFigure}
\end{figure}

This construction completely fixes the values of all numerators from
$N_1$ to $N_{82}$ (subject to on-shell external kinematics) using the
duality relations in \app{JacobiAppendix}.

\subsection{Resolving the snails}

In the previous subsection we obtained the numerators for the master
graphs, which determine all numerator factors through duality
relations.  However, it does not resolve the 0/0 ambiguity appearing
in graphs 83-85 in \fig{D6Figure}, which needs at least one external
momentum off shell to be properly defined, if we insist on
representing the result in terms of graphs with only cubic vertices.
As explained in \sect{SnailSubtletySubsection},
these contributions are indeed finite, as can easily be confirmed
explicitly using generalized unitarity cuts,
as we do below.

\begin{figure}[t]
\includegraphics[scale=0.53]{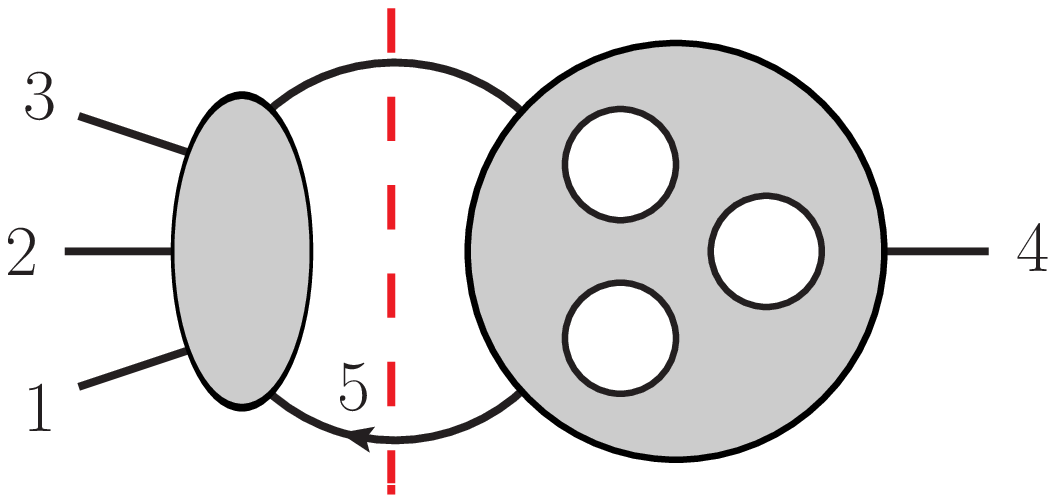}
\caption{A cut that determines the snail contribution 
in \fig{PropagatorFigure}(d). In $\NeqFour$ sYM theory
this cut vanishes at the level of the integrand.}
\label{SnailsCutFigure}
\end{figure}

\begin{figure}[t]
\includegraphics[scale=0.83]{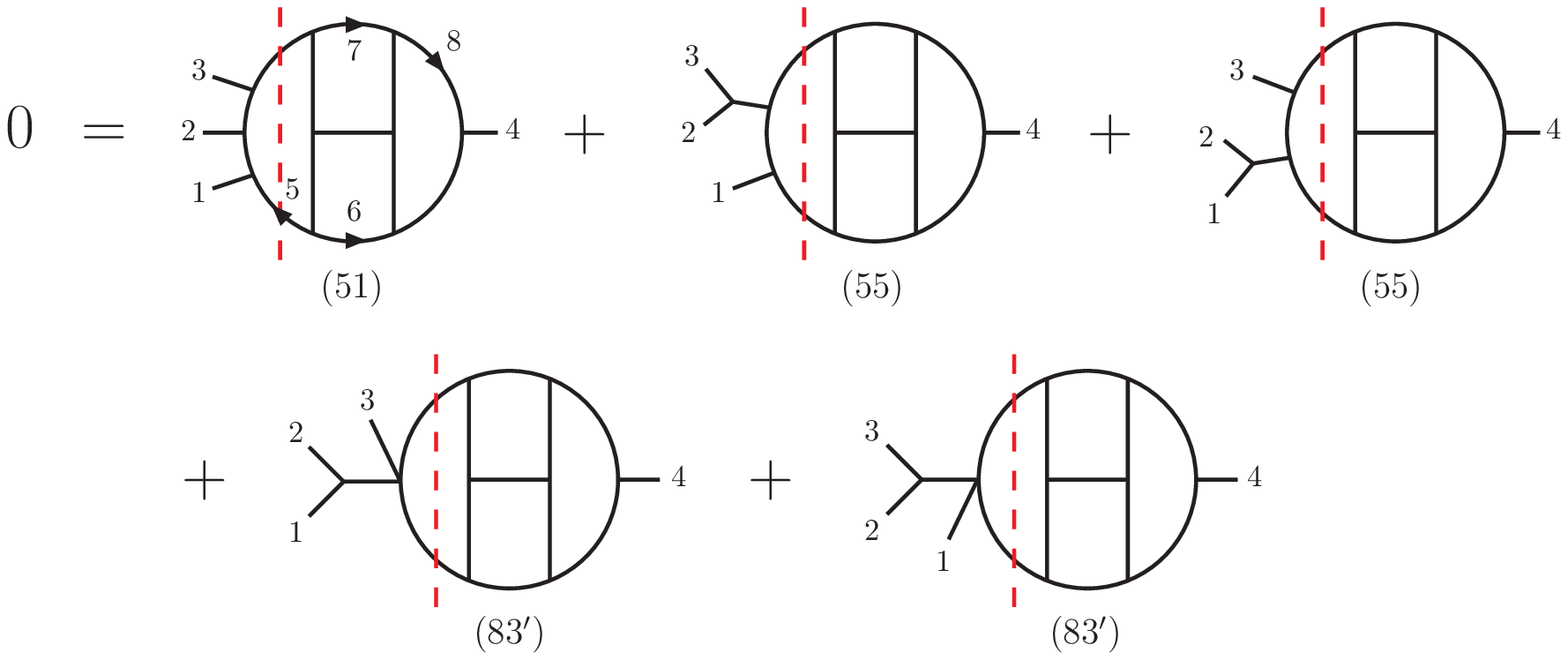}
\caption{The graphs contributing to the cut in
 \fig{SnailsCutFigure}.  The sum of these contributions to the cut
  vanishes in $\NeqFour$ sYM theory.}
\label{Snail4CutFigure}
\end{figure}

First we consider the planar snail contributions to the four-loop
four-point amplitude.  To determine them, we evaluate the unitarity
cut shown in \fig{SnailsCutFigure}.  In the planar limit, the
integrals contributing to
this cut are shown in \fig{Snail4CutFigure}.  Because the cut contains
an on-shell three-loop three-point subamplitude it must vanish,
following the arguments in \sect{SnailSubtletySubsection}. Therefore we can
determine the contribution of the snail graph 83 in terms of
that of graphs 51 and 55.  Dressing the numerators of the first three
graphs in \fig{Snail4CutFigure} with the appropriate ratios of
propagators, we find that
\bea
&&N_{51}+  
\frac{(l_5-k_1-k_2)^2}{t} \Bigl(N_{55}\Bigr|_{l_5 \rightarrow l_5 -k_1}\Bigr)
 + \frac{(l_5-k_1)^2}{s}  \Bigl(N_{55}\Bigr|_{k_1 \leftrightarrow k_3,
     l_5 \rightarrow k_1+k_2-l_5} \Bigr) \nn \\
&&~~~~~~~= \frac{27}{2}(l_5-k_1)^2(l_5-k_1-k_2)^2 \uu \,,
\label{snailcutline1}
\eea
where we used the relations $l_5^2\rightarrow 0$ and
$\tau_{45}\rightarrow 0$, valid on the cut. Here we relabeled the
momenta of the two contributions of graph 55 (compare to
\fig{D4Figure}) to match the labels of graph 51. The two factors on
the right-hand side of \eqn{snailcutline1} that depend on $l_5$ are simply
inverse propagators belonging to graph 51.  In order to 
compare to graph 83 we should remove these factors, because it
does not have these propagators.  Thus, in order for the sum of
contributions in \fig{Snail4CutFigure} to vanish, we must have
\begin{equation}
\frac{1}{s} N_{83'} +  \frac{1}{t}
  \Bigl(N_{83'}\Bigr|_{k_1 \leftrightarrow k_3} \Bigr) 
= - \frac{27}{2} \uu \,,
\label{SumSnails}
\end{equation}
where we start with the labeling in \fig{SnailGraphsFigure} and relabel
accordingly.  (Because the snail numerators turn out to have no loop-momentum
dependence, we need only specify the external momentum relabelings.)
Now numerator $N_{83'}$ should respect the $1 \leftrightarrow 3$ antisymmetry
of the graph.  This constraint, together with \eqn{SumSnails}, implies that
\begin{equation}
N_{83'} =  - \frac{9}{2}  \ss (\uu - \tt) \,.
\end{equation}
The original cubic graph in \fig{D6Figure} then has the
regulated numerator factor,
\begin{equation}
N_{83}=-\frac{9}{2} k_{4}^2 \ss (\uu - \tt) \,,
\end{equation} 
for the snail graph 83.  Here $k_4^2$
should not be set to zero until after this factor has canceled the
$1/k_4^2$ propagator of the graph.  The remaining two graphs, 84 and
85, can be determined in an analogous fashion using a nonplanar cut
which also isolates a three-point subamplitude. 
Alternatively, the duality relations in \app{JacobiAppendix} fix them to
be $N_{85}=N_{84}=N_{83}$.

We have carried out a very similar analysis of the corresponding cuts
of $\NeqEight$ supergravity and find that there are no snail
contributions.  This is in line with our heuristic expectations,
described in \sect{SnailSubtletySubsection}, that the gravity case
should have an extra vanishing factor of $k_i^2$ in the numerator,
setting all snail contributions to zero.

Now that we have a complete Ansatz for the amplitude, the remaining
task is to confirm that it has all desired properties and that it is a
correct representation of the amplitude.  Indeed, we verified that all
numerators respect the graph symmetries and that {\it all} duality
constraints hold on the sYM numerators.

To prove that our construction is correct, we verified that a spanning
set of $D$-dimensional generalized unitarity cuts are properly
reproduced.  It has been shown~\cite{Neq44np} that there are no
contributions to the four-loop four-point scattering amplitude in
$\NeqFour$ sYM theory beyond those exposed on next-to-next-to-maximal
cuts.  This result was later confirmed using six-dimensional
cuts~\cite{SixD}.  As long as a candidate representation manifestly
contains no worse loop-momentum behavior for individual terms than the
previous representation, as is true of our current form, the set of
next-to-next-to-maximal cuts are spanning (complete).  We employ those
cuts to verify our new expression, by comparing it against the previous
one.

We do not give the details here as the procedure is the same as given in
refs.~\cite{FiveLoop,Neq44np}, except that we generate the reference
$D$-dimensional analytic cuts using the previously-obtained forms of
the amplitude~\cite{Neq44np}.  For the gravity amplitude, obtained from
the double-copy formula (\ref{DoubleCopy}), we confirmed that
its cuts through (next-to)$^4$-maximal cuts match the corresponding
cuts of the result in ref.~\cite{GravityFour}.\footnote{The need for
checking this high an order arises because of the large numbers of
loop momenta occurring in individual terms in the original
representation of the $\NeqEight$ supergravity
amplitude~\cite{GravityFour}.}  The agreement with the
cuts of the earlier representation directly proves the duality and
double-copy properties for the four-loop four-point amplitudes of
$\NeqFour$ sYM theory and $\NeqEight$ supergravity.


\section{UV behavior of $\NeqEight$ supergravity
and $\NeqFour$ sYM theory}
\label{UVSection}

In this section, we examine the UV properties of the four-loop
four-particle $\NeqFour$ sYM theory and $\NeqEight$ supergravity
amplitudes derived in the previous section, after reviewing
lower-loop examples.  Unlike the original form obtained for the $\NeqEight$
supergravity four-loop amplitude~\cite{GravityFour}, the
representation derived in this paper through eq.~(\ref{DoubleCopy}) is
manifestly finite for $D<11/2$. This property makes it much easier to determine
its UV behavior.  We will see that, in complete harmony with the
corresponding $\NeqFour$ sYM amplitude, a divergence is indeed present
in the expected critical dimension $D_c=11/2$.  In $\NeqFour$ sYM theory
this divergence in the amplitude corresponds to a counterterm of the
schematic form $\Tr( {\cal D}^2 F^4 )$, where ${\cal D}$ represents a covariant
derivative and $F$ is the gauge field strength.
Similarly, in $\NeqEight$ supergravity the amplitude divergence
corresponds to a counterterm of the form ${\cal D}^8 R^4$,
with the indices of the
four Riemann tensors arranged in the supersymmetric combination
corresponding to the square of the Bel-Robinson
tensor~\cite{BelRobinson}. (For recent discussions of 
${\cal D}^{2k} R^4$ invariants in $\NeqEight$ supergravity see
ref.~\cite{FreedmanTonni}.)  We shall find a close connection between
coefficients of the UV counterterms in these theories.

In most of this section we will work in space-time dimension
$D = D_c-2\e = 11/2 - 2\e$.  In the last subsection we will discuss
the UV behavior of the color double-trace terms in the $\NeqFour$
sYM amplitude at four loops, for which the divergence in
$D = 11/2 - 2\e$ is not present.  Here we move to the next dimension
in which there can be a potential divergence, $D = 6 - 2\e$,
in order to assess whether there is one or not.

In $D = 11/2 - 2\e$, there are no UV divergences for either $\NeqFour$
sYM or $\NeqEight$ supergravity below four loops; in both theories,
the first divergences for $L=1,2,3$ are for dimensions $D_c = 8,7,6$,
respectively.  Therefore there can be no UV subdivergences in $D =
11/2 - 2\e$ at four loops.  Near this dimension, the overall
logarithmic UV divergences arise from integration regions in which the
loop momenta are much larger than the external momenta.  For the
representations of the $\NeqFour$ sYM and $\NeqEight$ supergravity
amplitudes in this paper, each integral is either finite or develops
at most a logarithmic divergence.  We can capture the leading UV
behavior by expanding in small external
momenta~\cite{Vladimirov,MarcusSagnotti}.  The logarithmic behavior of
each integral means that we need only the leading term in this
expansion.  The amplitudes in question reduce to a collection of
vacuum integrals that are relatively easy to evaluate.

We begin by reviewing the UV properties of $\NeqFour$ sYM theory, in
the light of the new representation of the four-loop four-point
amplitude. Then we will turn to $\NeqEight$ supergravity.

\subsection{Review of UV behavior of $\NeqFour$ sYM theory}

The UV properties of $\NeqFour$ sYM theory at one through four loops were
discussed in detail in ref.~\cite{Neq44np}. The first divergence appears
in the critical dimension
\begin{equation}
D_c=4+\frac{6}{L}~~~~~(L=2,3,4)\,,
\label{UsualBound}
\end{equation}
and $D_c=8$ at $L=1$.  In each case, the potential divergence is known
to appear with a nonzero coefficient~\cite{Neq44np}.  Thus the
bound~(\ref{UsualBound}), proposed in
refs.~\cite{BDDPR,HoweStelleRevisited}, is in fact saturated, at least
through four loops.

For gauge group $SU(N_c)$, gluon scattering amplitudes can be expressed
in the trace basis, {\it i.e.} in terms of traces of products of
generators in the fundamental representation.  For four external gluons,
only single-trace structures, of the form
$\Tr(T^{a_i}T^{a_j}T^{a_k}T^{a_l})$, and double-trace structures,
of the form $\Tr(T^{a_i}T^{a_j}) \Tr(T^{a_k}T^{a_l})$, can appear.  
At one and two loops, the color double-trace terms have the same UV behavior
as the overall amplitudes, obeying \eqn{UsualBound}.
However, starting at three loops they are less divergent~\cite{Neq44np}.
(See also the discussions in
refs.~\cite{DoubleTraceNonrenormalization,BHSUV}.)
In particular, the critical dimensions for finiteness for the 
double-trace terms satisfy
\begin{equation}
D_c^{\text{2-trace}}  = 4+\frac{8}{L}~~~~~(L=3,4)\,.
\label{DoubleTraceBound}
\end{equation}
Using the representations of the three- and four-loop amplitudes
described in previous sections, we will see that these bounds follow
more transparently than with the older representations in ref.~\cite{Neq44np}.
That is because the color factors associated with the most divergent
integrals now have a much simpler structure.

Ref.~\cite{Neq44np} showed that the bound~(\ref{DoubleTraceBound})
is saturated at three loops.  This computation did not involve
any UV subdivergences because $4+8/3 = 20/3 < 7$, and the first two-loop
divergences are at $D_c=7$.  In \sect{ColorDoubleTraceSubsection}
we will show that the double-trace bound is saturated in the four-loop
amplitude as well.  In this case there are subdivergences, because
$4+8/4=6$, and the first three-loop divergences are at $D_c=6$.

\begin{figure}
\centerline{\epsfxsize 2.6 truein \epsfbox{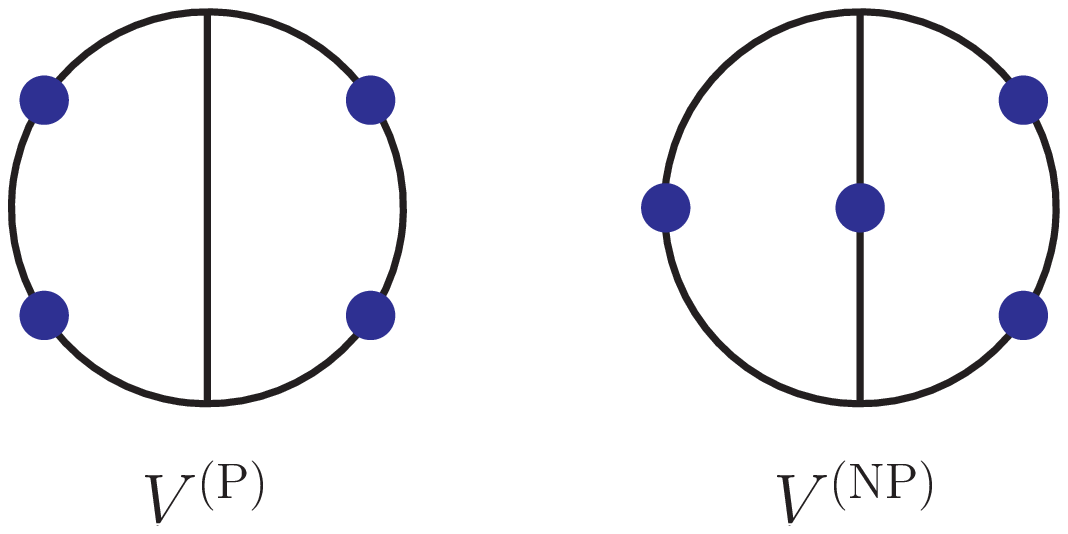}}
\caption[a]{\small
The two-loop vacuum integrals $V^{\P}$ and $V^{\NP}$.  Each (blue)
dot on a propagator indicates an additional power of the propagator.}
\label{Vacuum2loopsFigure}
\end{figure}

Before proceeding to three and four loops, we review the two-loop case.
The two-loop four-gluon amplitude as given in ref.~\cite{BDDPR}
already obeys the color-kinematic duality.  The full amplitude was
originally presented in the trace basis~\cite{BRY}; in the critical
dimension $D_c=7$, its divergence in terms of vacuum integrals
is~\cite{Neq44np}
\begin{eqnarray} 
\label{TwoLoopPoleVForm}
{\cal A}_4^{(2)} \Bigr|^{SU(N_c)}_{\rm pole} 
&=&  - \, g^6 \, {\cal K} \,
\biggl[ \Bigl( N_c^2 \, V^{\P} + 12 ( V^{\P} + V^{\NP} ) \Bigr)
\\ && \hskip1.5cm\null
 \times \Bigl( s \, ( \Tr_{1324} + \Tr_{1423} ) 
       + t \, ( \Tr_{1243} + \Tr_{1342} )
       + u \, ( \Tr_{1234} + \Tr_{1432} ) \Bigr)
\nonumber\\ &&\hskip1cm \null
- 12 \, N_c \, ( V^{\P} + V^{\NP} )
  \Bigl( s \Tr_{12} \Tr_{34}
       + t \Tr_{14} \Tr_{23}
       + u \Tr_{13} \Tr_{24} \Bigr) \biggr] \,,
\nonumber
\end{eqnarray}
where $V^{\P}$ and $V^{\NP}$, shown in \fig{Vacuum2loopsFigure}, are
the $k_i^\mu \rightarrow 0$ limit of the planar and nonplanar double-box
integrals.\footnote{We normalize our integrals as in
ref.~\cite{Neq44np}, so that at two loops there is a relative minus
sign compared to the normalization in ref.~\cite{BDDPR}.} 
Each blue dot denotes an extra power of the propagator on which it lies.
Here each dot coincides with the location of an external leg attachment
in the original four-point integral; adjacent propagators separated by
an external momentum become equal as that momentum vanishes.
The factor
\begin{equation}
 {\cal K} \equiv st A^\tree(1,2,3,4)\,,
\label{Kdef}
\end{equation}
contains the dependence on the external states. We use a shorthand
notation for the color traces,
\begin{equation}
\Tr_{ijkl}\ \equiv\ \Tr(T^{a_i}T^{a_j}T^{a_k}T^{a_l}), \qquad
\Tr_{ij}\ \equiv\ \Tr(T^{a_i}T^{a_j}) = \delta^{a_ia_j} \,.
\label{Trdef}
\end{equation}

The poles of the vacuum integrals $V^{\P}$ and $V^{\NP}$
in $D=7-2\eps$ are~\cite{BDDPR, Neq44np},
\begin{eqnarray}
V^{\P}  & = &  -\frac{\pi}{20 \, (4\pi)^7 \, \e}
\,, \nn \\
V^{\NP} & = &  -\frac{\pi}{30 \, (4\pi)^7 \, \e}
\,.
\label{Vacuum2Loop}
\end{eqnarray}
The UV divergence in the critical dimension, $D_c=7$, of 
the corresponding supergravity amplitude
is~\cite{BDDPR}
\be
{\cal M}_4^{(2)} \Bigl|_{\rm pole}= -2 \left(\frac{\kappa}{2} \right)^6
  s t u (s^2+t^2+ u^2)
 \, M_4^\tree \, (V^{\P} + V^{\NP})\,.
\ee
We note that the UV divergence of the supergravity amplitude and
that of the $1/N_c^2$-suppressed single-trace sYM amplitude are given
by the same linear combination of vacuum integrals, namely
$V^{\P} + V^{\NP}$.  We shall see that
this pattern repeats itself through four loops.  (This observation
holds for the one-loop four-point amplitude as well.  However, there
the relation is rather trivial because both amplitudes are expressed
in terms of the same scalar box integral; thus their UV divergences
must be expressed in terms of a unique vacuum integral.)

We now turn to three loops.  The integrals that appear in the three-loop
four-point $\NeqFour$ sYM amplitude in the
duality-satisfying form~\cite{BCJLoop}
all have 10 propagators.  However, three of them,
$I^{(\rm j)}$, $I^{(\rm k)}$ and $I^{(\rm l)}$ in
\fig{ThreeLoopDiagramsFigure}, have one propagator depending solely
on external momenta; we will refer to these integrals as nine-propagator
integrals.  From \eqns{Simplified_j}{BCJjacobi}, 
their numerator factors are independent of the loop momenta:
$N^{\rm (j)} = N^{\rm (k)} = N^{\rm (l)} = (t-u)s/3$.

If the other integrals had two powers of the loop momentum
in the numerator, then they would have the same generic large loop-momentum
behavior as the nine-propagator integrals.  However, the other integrals
all have numerators that are at most linear in the loop momenta.
\Eqn{NumeratorE} shows that the numerator $N^{\rm (e)}$ for the master
graph (e) is linear in the $\tau_{ij}$, which are in turn linear in the
loop momenta. The Jacobi relations~(\ref{BCJjacobi}) preserve this
linearity for all other numerators.  (In some cases the linear dependence
cancels down to a constant behavior.)  Therefore the leading UV divergence
of the three-loop $\NeqFour$ sYM amplitude comes from the three
nine-propagator integrals.

This result is consistent with a
rearrangement of the leading UV terms of the earlier
representation, as discussed in ref.~\cite{Neq44np}.
Because graphs (j), (k) and (l) contain an external three-point tree,
their color factors must be proportional to the product of two
structure constants, multiplied by a color Casimir operator
which has no free color indices~\cite{Neq44np}.  The product of the two
structure constants, $\colorf{a_1a_2b}\colorf{ba_3a_4}$ or a
permutation thereof, takes the same form as a tree amplitude,
namely a single color trace.
Hence the leading UV divergence at three loops, in the critical
dimension $D_c = 4 + 6/3 = 6$, contains no double color-trace
terms~\cite{Neq44np}, while they are present in the leading divergence
at two loops, \eqn{Vacuum2Loop}.
The duality-satisfying representation of the amplitude automatically
has the no-leading-double-trace feature. Very similar behavior is observed 
within a string theory analysis~\cite{DoubleTraceNonrenormalization}.
There the leading behavior at three (and four) loops is dominated
by the collision of pairs of vertex operators, producing inverse momentum
factors, reminiscent of the form of graphs (j), (k) and (l).
The finiteness of the three-loop double-trace terms in $D=6$ 
remains puzzling from the point of view of field-theoretic algebraic
nonrenormalization considerations~\cite{BHSUV}.

\begin{figure}
\centerline{\epsfxsize 2.7 truein \epsfbox{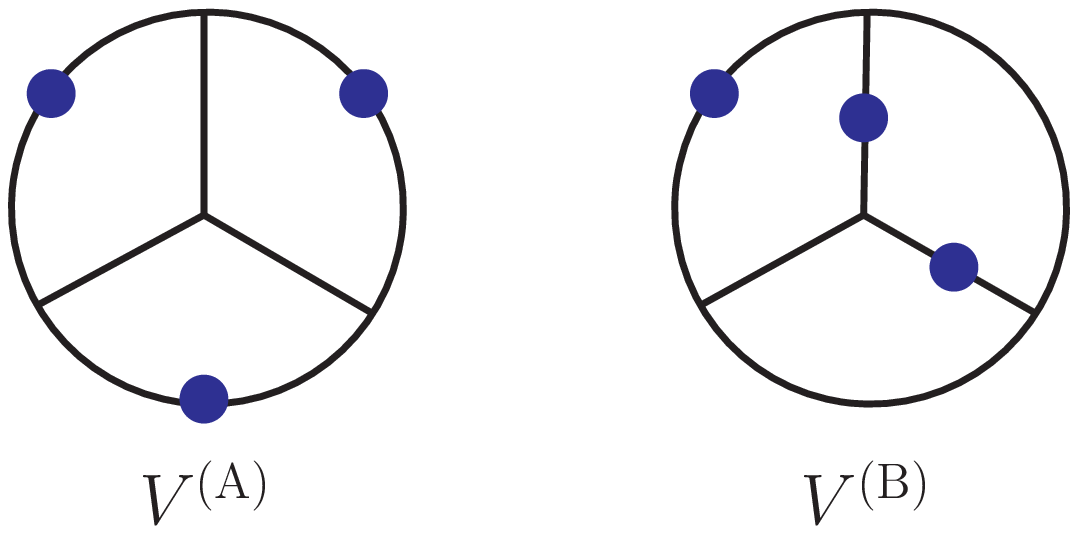}}
\caption[a]{\small
The three-loop vacuum integrals $V^{\A}$ and $V^{\B}$.}
\label{Vacuum3loopsFigure}
\end{figure}

To extract the UV divergence, we carry out the small momentum
expansion~\cite{Vladimirov,MarcusSagnotti}.  Integral (j) reduces to
the vacuum integral $V^{\A}$~\cite{CompactThree,Neq44np}
displayed in \fig{Vacuum3loopsFigure}, while integrals (k) and (l)
reduce to $V^{\B}$.  These integrals diverge first in $D=6$.  Their
color factors are closely related:  We can use a color Jacobi identity
involving the boxes in the upper right-hand corner of graphs (j) and
(k) of \fig{ThreeLoopDiagramsFigure} to show that the difference
between $C_{(\rm j)}$ and $C_{(\rm k)}$ contains a triangle subgraph.
A further color Jacobi identity allows us to replace the triangle by
a three-vertex, multiplied by the quadratic Casimir factor $C_A/2 =
N_c$.  Iterating this procedure, and also applying it to the
difference of $C_{(\rm j)}$ and $C_{(\rm l)}$, we find that
\be
C_{(\rm j)} - N_c^3 \colorf{a_1a_2b}\colorf{ba_3a_4}
 = C_{(\rm k)} = C_{(\rm l)} \,.
\label{cjklRelation}
\ee
Note that $C_{(\rm k)}$ and $C_{(\rm l)}$ are associated with
nonplanar graphs, and hence have only subleading-color terms.
\Eqn{cjklRelation} states that the subleading-color parts of
$C_{(\rm j)}$, $C_{(\rm k)}$ and $C_{(\rm l)}$ are all equal.

Taking into account that the combinatorial factor of $I^{(\rm l)}$
is twice as large as those for $I^{(\rm j)}$ and $I^{(\rm k)}$
in \fig{ThreeLoopDiagramsFigure}, and expressing the color factors
for (j), (k) and (l) in the trace basis, it is straightforward to see
that the UV divergence in $D_c=6$ is~\cite{Neq44np},
\begin{eqnarray} 
{\cal A}_4^{(3)}(1,2,3,4) \Bigr|^{SU(N_c)}_{\rm pole} 
&=&  2 \, g^8 \, {\cal K} \, 
   \Bigl( N_c^3 \, V^{\A}
       + 12 \, N_c \, ( V^{\A} + 3 \, V^{\B} ) \Bigr)
\label{ThreeLoopPoleVForm}\\ 
&&\hskip-1.2cm \null
\times \Bigl( s \, ( \Tr_{1324} + \Tr_{1423} )
           + t \, ( \Tr_{1243} + \Tr_{1342} )
           + u \, ( \Tr_{1234} + \Tr_{1432} ) \Bigr) \,.
\nonumber
\end{eqnarray}
The UV poles of the vacuum graphs are~\cite{CompactThree}
\begin{eqnarray}
V^{\A}\Bigr|_{\rm pole} &=& -\frac{1}{6 \, (4\pi)^9 \, \e} \,,
\label{VA} \\
V^{\B}\Bigr|_{\rm pole} &=& -\frac{1}{6 \, (4\pi)^9 \, \e}
\biggl( \zeta_3 - \frac{1}{3} \biggr) \,.
\label{VB}
\end{eqnarray}

We may compare this result to the
UV divergence in $D_c=6$ of the corresponding $\NeqEight$ supergravity
amplitude~\cite{CompactThree},
\be
{\cal M}_4^{(3)}\Bigr|_{\rm pole} =
-\left(\frac{\kappa}{2} \right)^8(s t u)^2 \, M_4^\tree \,
\left[10\,( V^{\A} + 3 \, V^{\B} )\right]\,.
\ee
In this case the leading UV divergence involves not only the 1PR
nine-propagator integrals, but also the 1PI ten-propagator integrals
whose sYM numerators are linear in the loop momenta, because their
(squared) supergravity numerators are quadratic.  Once again, as was
true for the two-loop amplitude, the UV divergence of the three-loop
supergravity amplitude and that of the $1/N_c^2$-suppressed
single-color-trace three-loop sYM amplitude are given by the same
combination of vacuum integrals, namely $V^{\A} + 3 \, V^{\B}$.  Below
we shall see that the same phenomenon persists through four loops.

\begin{figure}
\centerline{\epsfxsize 4.1 truein \epsfbox{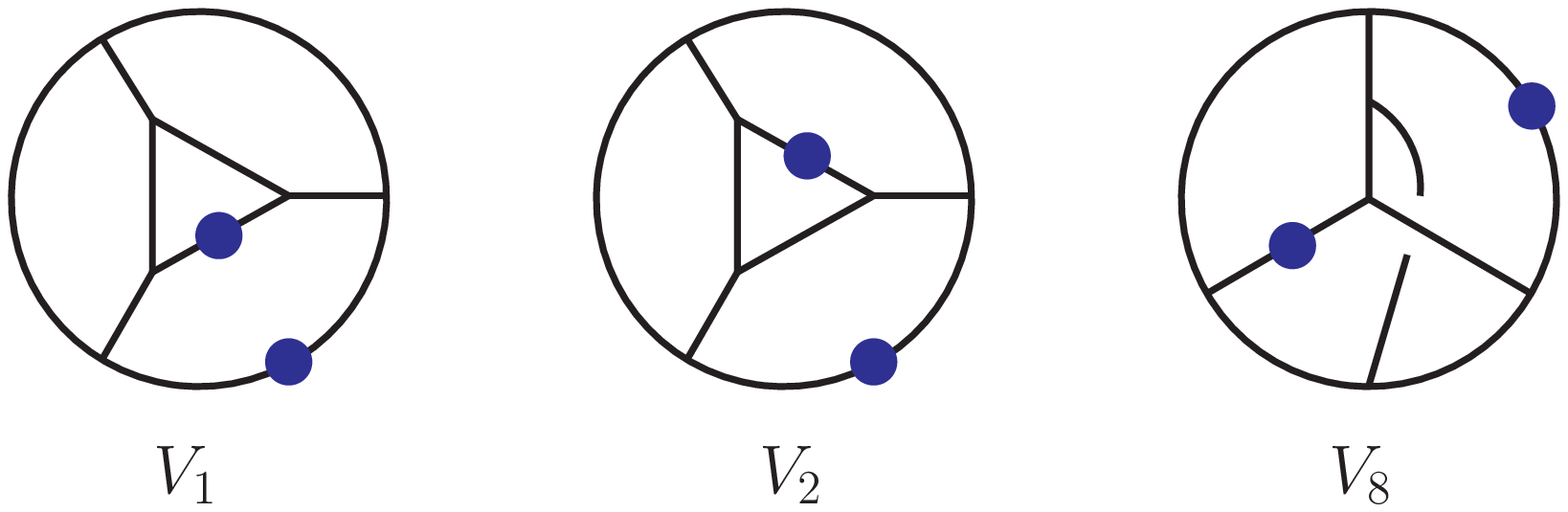}}
\caption[a]{\small
The basic four-loop vacuum integrals $V_1$, $V_2$ and $V_8$, 
to which all others can be reduced.
}
\label{vacuum_basis}
\end{figure}

Using the representation of the four-loop four-point $\NeqFour$ sYM
amplitude derived in earlier sections it is equally straightforward to
recover its UV divergence. Indeed, an inspection of the
integrals listed in figs.~\ref{BC1Figure}--\ref{SnailGraphsFigure} 
and of their numerator factors listed in \app{NumeratorAppendix} reveals that
the leading UV behavior comes solely from integrals $I_{80}$ through
$I_{85}$.  These six integrals have 11 internal propagators, and numerator
factors that are independent of the loop momentum.
Therefore they diverge first in $D_c=11/2$, which matches the expected
critical dimension, $D_c=4+6/L$ with $L=4$.

In contrast, the 1PI integrals $I_{1}$ through $I_{52}$ have 13 internal
propagators.  Their numerators would have to be quartic in the loop momenta
for them to diverge in $D=11/2$.  However, we note from \eqn{FourLoopMasters}
that the master numerators $N_{18}$ and $N_{28}$ are quadratic in the
$\tau_{ij}$, and hence merely quadratic in the loop momenta.
The Jacobi relations~(\ref{smallJrel4loop}) and (\ref{trivialNrelations})
preserve this quadratic behavior for all numerators.  Therefore integrals
$I_{1}$ through $I_{52}$ are finite in $D_c=11/2$.
The 1PR integrals $I_{53}$ through $I_{79}$ have 12 internal propagators.
If their numerators were quadratic in the loop momenta, then they would
diverge in $D=11/2$.  However, it is easy to see from
\eqns{trivialNrelations}{ListOfAllNumerators} that their numerators
are all linear in the loop momenta.

Integrals $I_{80}$ through $I_{85}$ reduce easily to vacuum integrals 
in the limit that the external momenta vanish.
The planar integrals $I_{80}$ and $I_{83}$ reduce to the vacuum
integral $V_1$ depicted in \fig{vacuum_basis}.  While integrals
$I_{81}$ and $I_{84}$ are nonplanar as four-point graphs, in the
vacuum limit they reduce to the planar vacuum integral $V_2$.
Finally, integrals $I_{82}$ and $I_{85}$ reduce to the nonplanar
vacuum integral $V_8$.

As was the case at three loops, the color factors for the leading
UV graphs are related by color Jacobi identities.  In this case
we can subtract, for example, $C_{81}$ from $C_{80}$, and use a Jacobi
identity operating on the box at the top center of the graphs 
in \fig{D6Figure}.  Then we reduce the resulting triangle subgraphs
iteratively, to find
\be
C_{80,83} - 2 \, N_c^4 \, \colorf{a_1a_2b}\colorf{ba_3a_4}
 = C_{81} = C_{82} = C_{84} = C_{85} \,.
\label{c8085Relation}
\ee
Again, the subleading-color parts of all contributing color factors
are equal.

From \eqn{trivialNrelations}, the numerator factors obey
$N_{80}=N_{81}=N_{82}$ and $N_{83}=N_{84}=N_{85}$.
On the other hand, the combinatorial factor of $I_{81}$
in \eqn{FourloopCombinatoricSum}
is twice as large as those for $I_{80}$ and $I_{82}$,
and similarly for $I_{84}$ with respect to $I_{83}$ and $I_{85}$.
Taking into account both combinatorial and numerator factors, the
contribution of $I_{83}$ is $-\frac{9}{8}$ times that of $I_{80}$,
and similarly for the other two pairs of graphs.
Combining all terms and switching to the color-trace basis,
we find that the UV divergence in the critical dimension $D=11/2$
is given by,
\begin{eqnarray} 
{\cal A}_4^{(4)}(1,2,3,4) \Bigr|^{SU(N_c)}_{\rm pole} 
&=& 
 - 6 \, g^{10} \, {\cal K} \, N_c^2
   \Bigl( N_c^2 \, V_1
       + 12 \, ( V_1 + 2 \, V_2 + V_8 ) \Bigr)
\label{FourLoopPoleVForm}\\ &&\hskip-1.2cm \null
\times \Bigl( s \, ( \Tr_{1324} + \Tr_{1423} )
           + t \, ( \Tr_{1243} + \Tr_{1342} )
           + u \, ( \Tr_{1234} + \Tr_{1432} ) \Bigr)
 \,,
\nonumber
\end{eqnarray}
in agreement with the results of ref.~\cite{Neq44np}. 
It is interesting to note from \eqn{FourLoopPoleVForm}
that the single-trace UV divergence in $D=11/2$ has
$N_c^4$ and $N_c^2$ components, but the $N_c^0$ component vanishes.
There is currently no general explanation for this fact,
apart from the explicit values of the color factors $C_{80}$ through
$C_{85}$.

The values of the three master integrals appearing in 
\eqn{FourLoopPoleVForm} are~\cite{Neq44np}
\bea
V_1 &=& \frac{1}{(4\pi)^{11} \, \e}
 \Biggl[ \frac{512}{5} \, \Gamma^4({\textstyle{\frac{3}{4}}})
      - \frac{2048}{105} \, \Gamma^3({\textstyle{\frac{3}{4}}})
 \Gamma({\textstyle{\frac{1}{2}}}) \Gamma({\textstyle{\frac{1}{4}}})
 \Biggr]\ +\ \Ord(1) \,,
 \cr
 V_2 &=& \frac{1}{(4\pi)^{11} \, \e}
 \Biggl[ - \frac{4352}{105} \, \Gamma^4({\textstyle{\frac{3}{4}}})
      + \frac{832}{105} \, \Gamma^3({\textstyle{\frac{3}{4}}})
 \Gamma({\textstyle{\frac{1}{2}}}) \Gamma({\textstyle{\frac{1}{4}}})
 \Biggr]\ +\ \Ord(1) \,,
 \label{valV1V2V8}
 \\
 V_8 &=& \frac{1}{(4\pi)^{11} \, \e}
\Biggl[
- \frac{20992}{2625} \, \Gamma^4({\textstyle{\frac{3}{4}}})
+ \frac{128}{75} \, \Gamma^3({\textstyle{\frac{3}{4}}})
   \Gamma({\textstyle{\frac{1}{2}}}) 
   \Gamma({\textstyle{\frac{1}{4}}}) 
+ \frac{8}{21 \,\Gamma({\textstyle{\frac{3}{4}}})} \, {\rm NO}_m
\Biggl]\ +\ \Ord(1) \, ,
\nonumber
 \eea
where NO$_m$ denotes a certain three-loop two-point nonplanar
integral. While its analytic expression in $D=11/2$ is not known, it
may be evaluated numerically~\cite{Neq44np} using the Gegenbauer
polynomial $x$-space technique (GPXT)~\cite{CKTGPXT},
with the result ${\rm NO}_m = -6.1983992267\ldots$.
We remark that an evaluation to much higher accuracy
is also possible~\cite{LSSPrivate}
using the DRA method~\cite{LeeDRA}, which involves combining
dimensional recurrence relations~\cite{TarasovDimRec}
with analyticity in the space-time dimension.
(This latter method has been applied to similar, and even more
complex, integrals in refs.~\cite{LSS}.)

In the following two subsections we will obtain the UV divergence of the
corresponding $\NeqEight$ supergravity amplitude.  We will find that,
as for the two- and three-loop case, it is given by the {\it same}
combination of vacuum integrals as the $1/N_c^2$-suppressed
single-trace UV divergence of the corresponding four-point $\NeqFour$
sYM amplitude, namely $V_1 + 2 \, V_2 + V_8$.

\subsection{$\NeqEight$ supergravity vacuum graphs at four loops}

With the duality-satisfying form of the corresponding $\NeqFour$ sYM
amplitude as the starting point, the double-copy formula~(\ref{DoubleCopy})
immediately gives us an expression for the four-loop $\NeqEight$
supergravity integrand.  We confirmed this integrand
by comparing its cuts with the cuts of the known four-loop $\NeqEight$
amplitude \cite{GravityFour}.  
The double-copy formula is equivalent to the following squaring
relation for the $\NeqEight$ supergravity numerators:
\be
N_i^\NeqEight\ =\ (N_i^\NeqFour)^2 \,.
\label{NSquaring}
\ee
By inspecting the squares of the numerator factors listed in
\app{NumeratorAppendix}, and counting the number of loop momenta
in the numerator of each integral, it is easy to see that,
in all cases, the integrals composing the resulting
$\NeqEight$ supergravity amplitude are manifestly finite for $D<11/2$.

Indeed, as remarked in the previous subsection,
the maximum degree in loop momenta of the numerator factors of
the sYM amplitude is 2 for the 13-propagator integrals, 1 for the
12-propagator integrals and 0 for the 11-propagator integrals, where
we count only those propagators carrying loop momentum.  Consequently,
the maximum degree in the loop momenta of each supergravity numerator
polynomial is 4, 2 and 0 for the 13-, 12- and
11-propagator integrals, respectively.  Such integrals all generically
diverge logarithmically in $D=11/2$.  Thus, the worst UV behavior
of any $\NeqEight$ supergravity integral matches that of the worst-behaved
integrals for the $\NeqFour$ sYM amplitude, {\it i.e.}~the 11-propagator
graphs (80)--(85) of \fig{D6Figure}, which have the form of
propagator corrections.  The main difference is that now graphs
other than propagator corrections carry this leading behavior.
The representation of the four-loop amplitude described here
reproduces the finiteness bounds of ref.~\cite{GravityFour}, but the
UV behavior is now manifest, allowing us to avoid performing any
loop integration to expose this feature.

The new representation also makes it far simpler to determine whether the
finiteness bound $D<11/2$ is saturated, by extracting the
precise UV divergence of $\NeqEight$ supergravity in $D_c=11/2$.
In our earlier representation~\cite{GravityFour}, the UV divergence
required the sixth-order terms in the expansion in small external
momenta, making it rather cumbersome to extract.
Now that the UV behavior is manifest, only the leading term in 
the expansion is required.  This feature means that for each integral,
we need only retain the terms in each sYM numerator with the highest 
powers of the loop momenta, and then square them.  

The result of the expansion in the external momenta is a collection of
tensor integrals, in which the numerator factors have a homogeneous degree
in the loop momenta and are polynomials in the scalar products of loop
and external momenta.  Such integrals may be further reduced by making
use of Lorentz invariance in order to extract the dependence on the
external momenta from the tensor integrals.  More precisely,
under integration we can replace a generic two-tensor by
\begin{eqnarray}
\label{2tensor}
l_i^{\mu_i} l_j^{\mu_j} \, \mapsto \,
\frac{1}{D} \, \eta^{\mu_i\mu_j} \, l_i\cdot l_j\,,
\end{eqnarray}
and a four-tensor by
\be
\label{4tensor}
l_i^{\mu_i} l_j^{\mu_j}l_k^{\mu_k} l_l^{\mu_l} \, \mapsto \,
\frac{1}{(D-1){} D {} (D+2)} \left(
A \, \eta^{\mu_i\mu_j} \, \eta^{\mu_k\mu_l}
+ B \, \eta^{\mu_i\mu_k} \, \eta^{\mu_j\mu_l}
+ C \, \eta^{\mu_i\mu_l} \, \eta^{\mu_j\mu_k}
\right)\,,
\ee
where
\bea
A &=& (D+1)l_i\cdot l_j\, l_k\cdot l_l
 - l_i\cdot l_k \, l_j\cdot l_l-l_i\cdot l_l \, l_j\cdot l_k\,,
\nn\\
B &=& -l_i\cdot l_j \, l_k\cdot l_l
 +(D+1) l_i\cdot l_k \, l_j\cdot l_l-l_i\cdot l_l \, l_j\cdot l_k\,,
\nn\\
C &=& -l_i\cdot l_j \, l_k\cdot l_l
 - l_i\cdot l_k \, l_j\cdot l_l+(D+1)l_i\cdot l_l \, l_j\cdot l_k\,.
\label{4tensorcoeffs}
\eea

\begin{figure}
\centerline{\epsfxsize 5.6 truein \epsfbox{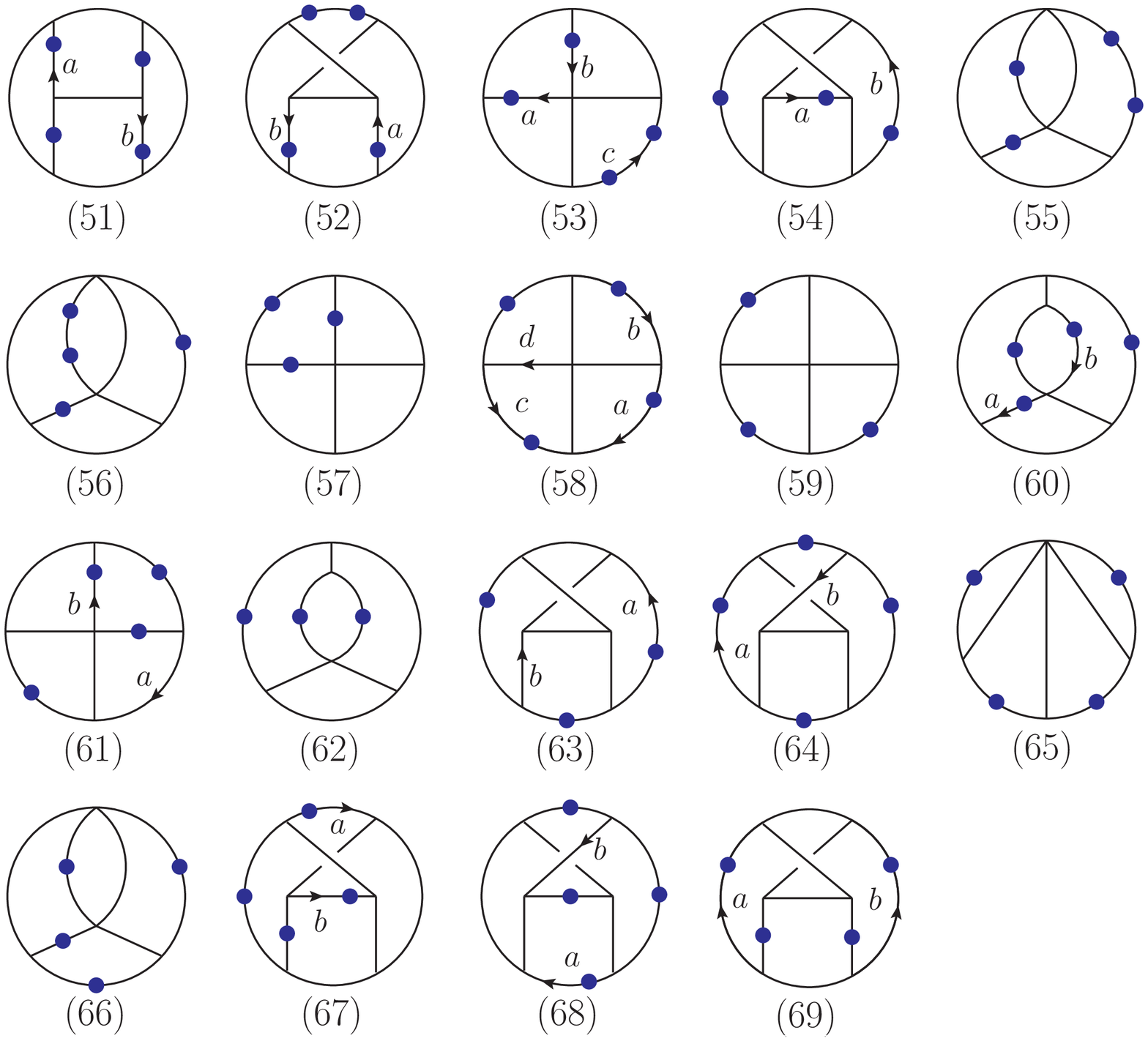}}
\caption[a]{\small
Vacuum graphs $I^v_{51}$ through $I^v_{69}$. }
\label{p3}
\end{figure}

Upon using these identities and summing over all permutations we find
that the UV pole has the form,
\be
{\cal M}^{(4)}_4 \Bigl|_{\rm pole} = - \frac{1}{4}
\Bigl(\frac{\kappa}{2}\Bigr)^{10} 
s t u\ (s^2 + t^2 + u^2)^2 \ M_4^\tree \ {\widetilde {\cal V}}^{(4)}\,,
\label{TableNormalization}
\ee
where $M_4^\tree$ is the four-point tree-level supergravity amplitude,
and
\be
{\widetilde {\cal V}}^{(4)} = \sum_{i=1}^{69} I_i^v
\label{V4def}
\ee
is the sum of integrals $I_i^v$ shown in figs.~\ref{p1}--\ref{p3}.
All the kinematic dependence has been extracted in
\eqn{TableNormalization}; each integral $I_i^v$ is a pure number
multiplied by a $1/\e$ pole.  The numerator factors for these
integrals are given in the first column of \tab{NumeratorTable}
(not counting the column labeling the integrals $I_i^v$) in
\app{VacuumNumeratorsAppendix}. The overall kinematic factor
of $(s^2 + t^2 + u^2)^2$ is guaranteed by complete permutation symmetry
of the four-point amplitude.

The precise combination of vacuum integrals $I_i^v$, in particular their
numerical coefficients and numerator tensor structures, is sensitive
to the choice of independent loop momenta in each of the integrals
appearing in the complete supergravity amplitude.  This choice
is inherited in turn from the parametrization of the integrals of
the $\NeqFour$ sYM amplitude, through the squaring
relation~(\ref{NSquaring}).  The ambiguity in choosing the loop
momenta may also be used to generate identities between different integrals.

\begin{figure}
\centerline{\epsfxsize 5.6 truein \epsfbox{vacuums_1_25.eps}}
\caption[a]{\small Vacuum graphs $I^v_{1}$ through $I^v_{25}$. The
momentum labels refer to the internal lines carrying an arrow. Their
numerator factors are listed in \tab{NumeratorTable} in
\app{VacuumNumeratorsAppendix}. }
\label{p1}
\end{figure}

\begin{figure}
\centerline{\epsfxsize 5.5 truein \epsfbox{vacuums_26_50.eps}}
\caption[a]{\small Vacuum graphs $I^v_{26}$ through $I^v_{50}$.}
\label{p2}
\end{figure}

We now discuss in some more detail how we arrived at the results
in the first column of \tab{NumeratorTable} in
\app{VacuumNumeratorsAppendix}.
After applying \eqns{2tensor}{4tensor}, the numerator factors of the
vacuum integrals appearing in ${\widetilde {\cal V}}^{(4)}$,
which no longer carry any external momenta, can be simplified
further using loop-momentum conservation.  Often,
inverse propagators can be identified, using for example
$l_i\cdot l_j = \frac{1}{2}(l_k^2-l_i^2-l_j^2)$
when $l_i+l_j=l_k$ corresponds to a propagator for the graph.
Such factors will cancel existing propagators
and lead to simpler integrals with fewer propagators and lower-degree
numerator factors.  In carrying out this procedure, it is useful to
ensure that propagators are not `over-canceled'.  For example,
no squared inverse propagator should appear in a vacuum
integral that does not have the corresponding propagator raised
to at least the second power. 
For certain integrals the complete numerator factor may be written as
a combination of different inverse propagators. In this case, the
integral reduces to a combination of scalar integrals of different
topologies, each of which is obtained from the initial integral by
collapsing some of its propagators.
For other integrals, the rank of the original numerator tensor is reduced
by only two units. For 13 integrals the numerator factor remains of
fourth order in loop momenta with no canceled propagators; in all
such cases the numerator factors carrying loop momenta
may be expressed as a perfect square, $(l_i\cdot l_j)^2$
for different $i$ and $j$.

To illustrate this procedure in more detail, we consider two examples
--- the reduction of integrals $I_{66}$ and $I_{12}$.
As we will see, the former
integral reduces to only scalar integrals, while the latter integral
leaves behind a four-tensor integral.
The numerator factor of the integral $I_{66}$ in the supergravity
amplitude is given by the squaring relation~(\ref{NSquaring}),
and formula~(\ref{ListOfAllNumerators}) for $N_{66}^\NeqFour$,
to be
\begin{eqnarray}
N_{66}^\NeqEight &=&
\ss{}^2 \Bigl[ 4 \tt{} (\t_{35} - 2 \t_{36})
 + 2 \uu{} (\t_{35} + 3 \t_{45} - 4 \t_{46})
 - \ss{} (6 \uu + \t_{15} - 6 \tt + 5 \t_{25} - 8 \t_{26}) \Bigr]^2
\,. \nn\\
\label{n66}    
\end{eqnarray}
For this 12-propagator integral, the leading UV behavior comes
from terms that are quadratic in the loop momenta in $N_{66}^\NeqEight$,
so the two $\tau$-independent terms inside the brackets in 
\eqn{n66} may be dropped.  For the remaining
terms, \eqn{2tensor} implies that $N_{66}$ is equivalent under
integration to 
\begin{equation}
\frac{1}{s} N_{66}^\NeqEight \, \mapsto \,
 \frac{8}{D} {} s^2 t u {} (11\t_{55}-48\t_{56}+48\t_{66})
- \frac{2}{D} {} s^4 (15\t_{55}-64\t_{56}+64\t_{66})
\,.
\label{N66i2}
\end{equation}
We have extracted a factor of $1/s$ on the left-hand side from 
the external propagator in the 1PR integral $I_{66}$.

Next we express all scalar products of internal momenta in terms
of inverse propagators, using $\tau_{55} = 2 l_5^2$
and $\tau_{66}-\tau_{56} = - l_5^2 + l_6^2 + (l_5-l_6)^2$.
Summing over the $4!$ permutations, and dividing by the symmetry factor
of 4 for this graph, we find that the contribution of $I_{66}$ is
\begin{equation}
\frac{1}{s} N_{66}^\NeqEight
\ \mapsto\ \frac{4}{D}\,(s^2+t^2+u^2)^2\,
 (17 l_5^2 - 32 (l_5-l_6)^2 - 32 l_6^2)\,.
\label{I66reductionA}
\end{equation}
In the permutation sum the first term in \eqn{N66i2}
cancels out completely due to $s+t+u=0$.

Each one of the factors $l_5^2$, $(l_5-l_6)^2$ and $l_6^2$
in \eqn{I66reductionA} cancels one propagator of $I_{66}$.
The topology resulting
from $l_5^2$ can be identified as graph (17) in \fig{p1}.
The $(l_5-l_6)^2$ and $l_6^2$ terms both lead to graph (1) in \fig{p1}.
Therefore, to leading order in the small external momentum expansion
in $D=11/2$, the contribution of integral $I_{66}$ to
${\widetilde {\cal V}}^{(4)}$ becomes
\begin{equation}
I_{66}\ \mapsto\ 
 \frac{32}{11} \, ( 17 \, {\widetilde I}_{17}^v - 64 \, {\widetilde I}_1^{v} )
\ =\ \frac{544}{11} \, {\widetilde I}_{17}^v 
  - \frac{2048}{11} \, {\widetilde I}_1^{v} \,.
\label{I66reduction}
\end{equation}
Here the ${\widetilde I_i^v}$ denote the vacuum integrals with
topologies shown in figures~\ref{p1}--\ref{p3}, and momentum-dependent
numerator factors shown in the first column of \tab{NumeratorTable},
but with the rational numerical coefficients set to unity.
For example, with this notation we have 
$I_1^{v}=  -\frac{117674}{1485}  {\widetilde I}_1^{v}$.
We have multiplied by $4/(s^2+t^2+u^2)^2$ in passing from
\eqn{I66reductionA} to \eqn{I66reduction}, to account for the 
relative prefactor of $(s^2+t^2+u^2)^2/4$ in \eqn{TableNormalization},
compared with \eqn{FourLoopGravityAmplitude}.
The difference between the numerical coefficients in
\eqn{I66reduction} and those in \tab{NumeratorTable}
is simply that the table collects the contributions to each 
vacuum graph from {\it all} the different integrals $I_i$
appearing in \eqn{FourLoopGravityAmplitude}.

As a second example, consider the reduction of $I_{12}$.
The numerator factor in the $\NeqEight$ supergravity amplitude is 
\begin{eqnarray}
N_{12}^\NeqEight&=&
\frac{s^2}{4} \Bigl[
s \, ( \t_{16} - \t_{26} - \t_{35} + \t_{45} + 2t )
       \nn\\ 
&&\null 
+ 2 \, s \, \t_{56}
- 2 \, (4 \t_{16} \t_{25} + 4 \t_{15} \t_{26} + 
  \t_{45} (\t_{36} - 3 \t_{46}) + \t_{35} (\t_{46} - 3 \t_{36}))\Bigr]^2
\,.
\end{eqnarray}
The leading UV divergence requires four powers of the loop momenta
from the numerator.  That can come only from the second and third terms
inside the square; the first term is linear in the loop momentum and
can be dropped.  Using \eqn{4tensor} and
the on-shell condition for the external legs we find, after some
straightforward calculation, that under the integral sign 
$N_{12}^\NeqEight$ is equivalent to
\begin{eqnarray}
N_{12}^\NeqEight&\mapsto&
- \frac{128}{(D-1){} D {} (D+2)} s^2 t u {}
\left[(D-2)\t_{56}^2 + D\t_{55}\t_{66} \right]\\ &&
+ \frac{1}{(D-1){} D{} (D+2)} s^4 \left[(D^3-19 D^2+146 D-96)
 \t_{56}^2+4(17D-25)\t_{55}\t_{66}\right]
\,.  \nn 
\end{eqnarray}
By inspecting the graph for $I_{12}$ it is straightforward to see that
$\tau_{56}$ cannot be completely expressed in terms of inverse
propagators. We therefore keep it in this form.  The sum over the $4!$
permutations leads, as in the case of $N_{66}^\NeqEight$, to the
complete cancellation of the term proportional to $s^2tu$.
Dividing by the symmetry factor of 4 for $I_{12}$, we find that 
\begin{equation}
N_{12}^\NeqEight \, \mapsto \, (s^2+t^2+u^2)^2 \,
\left[ \frac{D^3-19D^2+146 D-96}{(D-1)D(D+2)}\tau_{56}^2
  + 16 \frac{17D-25}{(D-1)D(D+2)}l_5^2l_6^2 \right] \,.
\end{equation}
Considering the graph associated with $I_{12}$,
we see that the $\tau_{56}^2$
term generates graph (3) in \fig{p1}, whereas the factor
of $l_5^2 l_6^2$ reduces the tripled propagators to doubled ones,
generating graph (4) in \fig{p1}.  Multiplying by an overall factor of
$4/(s^2+t^2+u^2)^2$ and setting $D=11/2$, we obtain for
the contribution of $I_{12}$ to ${\widetilde {\cal V}}^{(4)}$,
\begin{equation}
I_{12}\ \mapsto\ \frac{9556}{1485} \, {\widetilde I}_3^v
            + \frac{35072}{1485} \, {\widetilde  I}_4^v \,.
\label{I12final}
\end{equation}
It turns out that integral $I_{12}$ generates the only contribution to
$I_3^v$, so that $I_3^v = \frac{9556}{1485} \,{\widetilde I}_3^v$,
as given in \tab{NumeratorTable} in \app{VacuumNumeratorsAppendix}.
Several other integrals
$I_i$ contribute to ${\widetilde I}_4^v$, so the rational coefficient
given in \tab{NumeratorTable} for $I_4^v$ is different from
the one in \eqn{I12final}.

Carrying out the steps detailed above for all the integrals appearing
in the amplitude, we find that they reduce to 92 different vacuum
integral topologies. In the complete amplitude some of them cancel
out\footnote{Some of these cancellations are dimension-independent,
while others occur only in $D=11/2$.}, leaving only the 69 vacuum
topologies shown in figures \ref{p1}--\ref{p3}.  While these integrals
may be rearranged somewhat using momentum conservation, to find the full set
of relations between integrals we need more powerful methods.  We shall see
in the next section that integration-by-parts identities reduce them all
to linear combinations of the three scalar vacuum integrals shown in
\fig{vacuum_basis}.

\subsection{Integrating the vacuum graphs}

To evaluate vacuum integrals in their critical dimension $D_c=11/2$, we
use the same infrared rearrangement~\cite{Vladimirov} (related to the
$R^*$ operation \cite{Rstar}) that we used previously to evaluate the
three-loop vacuum integrals $V^{\A}$ and $V^{\B}$~\cite{CompactThree}
and the four-loop vacuum integrals $V_1$, $V_2$ and $V_8$~\cite{Neq44np}. 
This approach was already discussed in some detail in ref.~\cite{Neq44np},
so here we include only a brief summary.

The expansion in small external momentum introduces
unphysical infrared divergences.  To separate them from 
the UV singularities we inject and remove momentum $k_\mu$, with
$k^2\ne 0$, at two of the vertices of the vacuum integral, thus
transforming it into a four-loop two-point integral.  This two-point
integral possesses the same UV poles as the vacuum integral,
but no infrared divergences.  We always take the two vertices in question
to be connected by a single propagator in the four-loop vacuum integral.
In this case, the four-loop two-point integrals can be factorized into
products of three-loop two-point integrals and one-loop two-point integrals.
The one-loop two-point integrals are trivial, and contain the UV $1/\e$ pole.
The finite three-loop two-point propagator integrals are then evaluated
through the method of integration by parts (IBP)~\cite{CTIBP}, implemented 
in the MINCER algorithm.  For certain tensor integrals we also applied the
Laporta algorithm~\cite{Laporta} for solving the integration-by-parts
relations, as implemented in the computer code {\sc AIR}~\cite{AIR}.
In addition, we employed gluing relations~\cite{CTIBP},
which demand consistency of the various ways of factorizing the
four-loop UV-divergent integral into products of lower-loop
integrals.  These consistency conditions are nontrivial and
aid in the evaluation of some of the master integrals remaining after
IBP reduction in $D=11/2$.

Further consistency relations between vacuum integrals can be derived
as follows.  We start with a four-loop two-point integral with a numerator
factor of degree less than or equal to two in the loop momenta.
We expand it to next-to-next-to-leading order in the small external momenta,
using different parametrizations of the loop momenta.  We require that the
different parametrizations yield consistent results.
These relations may be used either to reduce the number of integrals that
need to be evaluated, or as consistency checks of the integral evaluation.
A similar strategy was used to show that the four-loop four-point 
$\NeqEight$ supergravity amplitude is UV finite in $D=5$~\cite{GravityFour}.
In that case, the complete amplitude was expanded for different
loop-momentum parametrizations.  This approach led to consistency relations
involving only integrals that naturally appear in the amplitude. 
The present situation is somewhat more involved because evaluation
of each of the 69 integrals implies that we need to expand integrals
which do not necessarily appear in the final result.  In this case, 
the consistency relations that are generated involve a much larger set
of integrals than just the 69 under consideration.

Following the strategies reviewed above, all the 69 integrals appearing in 
the small momentum expansion of the four-loop four-point $\NeqEight$ 
supergravity amplitude can be reduced to linear combinations of the 
integrals $V_1$, $V_2$ and $V_8$ shown in \fig{vacuum_basis}.
These integrals have already appeared in the UV divergence 
of the corresponding $\NeqFour$ sYM amplitude.  The coefficients
of $V_1$, $V_2$ and $V_8$ after this reduction are provided in the
second, third and fourth columns, respectively, of \tab{NumeratorTable}
in \app{VacuumNumeratorsAppendix}.  
The values of the UV poles of $V_1$, $V_2$ and $V_8$ were determined in 
ref.~\cite{Neq44np} and are given in \eqn{valV1V2V8}.

It is interesting to complete the evaluation of the two examples
discussed in the previous subsection, $I_{66}$ and $I_{12}$,
and thereby illustrate a general feature
regarding the positivity properties of the residues of the UV poles of
the integrals appearing in the $\NeqEight$ supergravity
amplitude in the double-copy representation.

The results of \app{VacuumNumeratorsAppendix} can be used
(after dividing by rational coefficients in the second column)
to express the following integrals ${\widetilde I}_i^v$ in terms
of the basis integrals $V_1$, $V_2$ and $V_8$:
\bea
{\widetilde I}_1^v &=& V_2 \,, \label{Itildev1Reduction}\\
{\widetilde I}_3^v &=& 
\frac{6671}{4800} \, V_1 - \frac{531}{800} \, V_2  \,,
\label{Itildev3Reduction}\\
{\widetilde I}_4^v &=& V_1 \,, \label{Itildev4Reduction}\\
{\widetilde I}_{17}^v &=& \frac{3}{4} \, V_1 + \frac{1}{2} \, V_2 \,,
\label{Itildev17Reduction}
\eea
Inserting these equations into \eqn{I66reduction}, 
for the contribution of $I_{66}$ to ${\widetilde {\cal V}}^{(4)}$
we obtain
\begin{equation}
I_{66}\ \mapsto\ \frac{544}{11} \, {\widetilde I}_{17}^v 
  - \frac{2048}{11} \, {\widetilde I}_1^{v}
  \ =\ \frac{408}{11} \, V_1 - \frac{1776}{11} \, V_2 \,.
\label{I66Finalreduction}
\end{equation}
Using the expressions~(\ref{valV1V2V8}) for $V_1$ and $V_2$  
we find that the sign of the UV divergence of $I_{66}$ is negative. 

Similarly, the contribution of the UV singularity of $I_{12}$
to ${\widetilde {\cal V}}^{(4)}$ is
\begin{equation}
 I_{12}\ \mapsto\ \frac{9556}{1485} \, {\widetilde I}_3^v
               + \frac{35072}{1485} \, {\widetilde I}_4^v 
 \ =\ \frac{58023419}{1782000} \, V_1
    - \frac{140951}{33000} \, V_2 \,.
\label{I12Finalreduction}
\end{equation}
Inserting the numerical values of $V_1$ and $V_2$ into
\eqn{I12Finalreduction} we find that, unlike $I_{66}$, the
UV pole of $I_{12}$ has a positive residue in $D=11/2$.

We see that the residue of the leading UV pole can be either positive
or negative, despite the fact that the initial numerator factors are
perfect squares.  This phenomenon occurs frequently in the reduction
of the various integrals appearing in the $\NeqEight$ supergravity
amplitudes.  It leads to strong cancellations among the corresponding
UV poles.  The origin of this phenomenon is simply that the
propagators occurring in the denominators of the integrand do not have
a fixed sign.  For example, the 1PR integral $I_{66}$ contains an
explicit factor of $1/s$ from a propagator external to the loops.  The
sum over its permutations contains factors of $1/t$ and $1/u$.  Not
all of these factors can be positive, given that $s+t+u=0$.  Even if
these factors are multiplied by positive numbers, the sum can be
negative.  Indeed, in any physical region (with one Mandelstam invariant
positive and two negative), if they were all multiplied
by the same number, the sum would be negative, because
$\frac{1}{s}+\frac{1}{t}+\frac{1}{u} < 0$ for $s+t+u=0$.

\subsection{The $\NeqEight$ supergravity UV behavior}

The expressions for the leading UV divergences of the 69 vacuum integrals
$I_i^v$ in $D=11/2 - 2\e$ are collected in \tab{NumeratorTable} in
\app{VacuumNumeratorsAppendix}.  Totaling up these contributions,
we determine the value of ${\widetilde {\cal V}}^{(4)}$, defined
through \eqns{TableNormalization}{V4def}, to be
\be
{\widetilde {\cal V}}^{(4)} = \frac{23}{2} \, ( V_1 + 2 V_2 + V_8 ) \,.
\label{V4tildefinalMainText}
\ee
Plugging this value for ${\widetilde {\cal V}}^{(4)}$ into
\eqn{TableNormalization}, the UV singularity of the
four-loop four-point $\NeqEight$ supergravity amplitude is given by
\begin{equation}
{\cal M}_4^{(4)}\Bigr|_{\text{pole}}= -\frac{23}{8} \,
\Bigl(\frac{\kappa}{2}\Bigr)^{10} 
stu \, ( s^2 + t^2 + u^2)^2 \, M_4^\tree \,( V_1 + 2 V_2 + V_8 )\, .
\label{FourLoopGpole}
\end{equation}
Because this result is nonvanishing, the four-loop four-point amplitude
of $\NeqEight$ supergravity diverges in exactly the same critical
dimension $D_c=11/2$ as that for $\NeqFour$ sYM theory. 

Comparing \eqns{FourLoopGpole}{FourLoopPoleVForm}, we see that the
leading UV divergence of the four-loop four-point $\NeqEight$
supergravity amplitude is given by the same combination of vacuum
integrals as the $1/N_c^2$-suppressed single-trace UV divergence of
the four-point $\NeqFour$ sYM amplitude, namely $V_1 + 2 V_2 + V_8$.
This pattern matches our observations above regarding the two- and
three-loop amplitudes.  While we do not understand the full
significance of this close relation between the UV divergences of the
gravity and subleading-color sYM amplitudes, it seems unlikely to be
accidental.  If it persists at higher loops as well, it could
potentially have interesting consequences for the higher-loop UV
behavior of $\NeqEight$ supergravity, especially given that
$\NeqFour$ sYM is UV finite in four dimensions.  The fact that this nontrivial
connection holds through at least four loops is a central observation
of this paper.

While the divergences of all integrals entering ${\cal M}_4^{(4)}$ are
expressed in terms of the vacuum master integrals $V_1$, $V_2$
and $V_8$, the expression~(\ref{FourLoopGpole}) arises as result of
extensive cancellations between the various integrals. For example,
integrals $I_{80}$ through $I_{85}$ are responsible for the
complete UV divergence of the four-loop four-point $\NeqFour$
sYM amplitude.  It is easy to see that they contribute to the UV
divergence of the $\NeqEight$ supergravity amplitude the following
amount:
\be 
{\cal M}_4^{(4)}\Bigr|_{\text{pole}}^{I_{80,81,82}}
= - 64 \, \Bigl(\frac{\kappa}{2}\Bigr)^{10}
s t u \, (s^2+t^2+u^2)^2 \, M_4^\tree \, ( V_1 + 2 V_2 + V_8 )\,.
\ee
Note that the snail integrals $I_{83}$, $I_{84}$ and $I_{85}$ yield
vanishing contributions in the supergravity case.
Because the rational coefficient in~\eqn{FourLoopGpole}, $23/8$, is
smaller than $64$, it follows that the integrals $I_1$ through
$I_{79}$ contribute to the UV pole with opposite sign, relative to
$I_{80}$, $I_{81}$ and $I_{82}$.
This fact is not surprising: although the numerators are squares,
the propagators do not have all the same signs.

To expose a more dramatic numerical cancellation, we have split the
above computation into two pieces: The 12-propagator contributions,
defined to be those from integrals $I_{53}$ through $I_{79}$, except
$I_{72}$, which all have explicit single inverse powers of the
Mandelstam variables, $s,t$ or $u$; and the remaining 11- and
13-propagator contributions, which have either no such explicit
factor, or else the factor is squared.  We find that
\bea
{\cal M}_4^{(4)}\Bigr|_{\text{pole}}^{\text{12-propagator}}
&=& + 142 \, \Bigl(\frac{\kappa}{2}\Bigr)^{10}
s t u \, (s^2+t^2+u^2)^2 \, M_4^\tree \, ( V_1 + 2 V_2 + V_8 )\,,
\label{NeqEight12prop}\\
{\cal M}_4^{(4)}\Bigr|_{\text{pole}}^{\text{11,13-propagator}}
&=& - \frac{1159}{8} \, \Bigl(\frac{\kappa}{2}\Bigr)^{10}
s t u \, (s^2+t^2+u^2)^2 \, M_4^\tree \, ( V_1 + 2 V_2 + V_8 )\,.
\label{NeqEight1113prop}
\eea
Remarkably, the numerical cancellation between these two sets of
contributions, in order to get the total~(\ref{FourLoopGpole}),
is quite significant, to within about 2\%.
That is, $(23/8)/142 = 0.0202\ldots$.

In summary, by carrying out the integration to extract the four-loop
four-point UV divergence of $\NeqEight$ supergravity in $D=11/2$, we
have found extensive cancellations, both analytical and numerical,
once again revealing surprising hidden structure.

\subsection{The $\NeqFour$ sYM color double-trace UV divergence}
\label{ColorDoubleTraceSubsection}
\begin{figure}
\centerline{\epsfxsize 4.1 truein \epsfbox{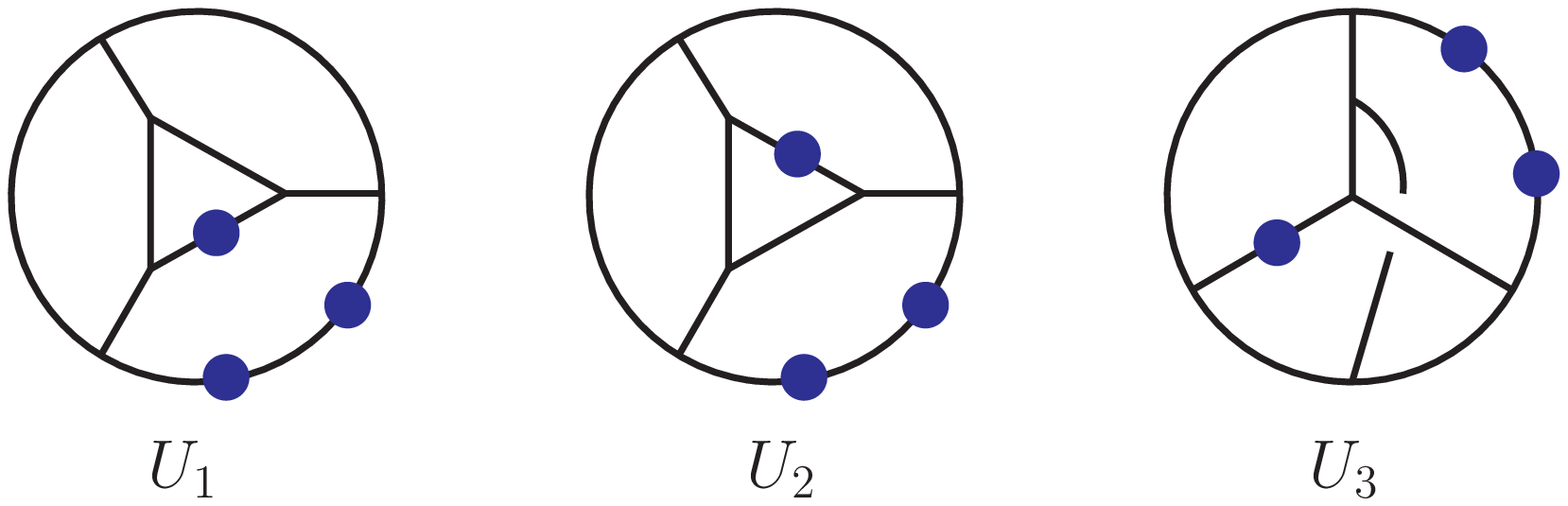}}
\caption[a]{\small The vacuum integrals $U_1$, $U_2$ and $U_3$
that would be generated by the graphs 51, 52 and 72 in the
double-trace part of the four-loop divergence of $\NeqFour$ sYM theory
in $D=6$.  Because these three graphs possess both an overall divergence
and a three-loop subdivergence, we only evaluate them after first
performing a subtraction of the subdivergence.}
\label{U_vacuum_basis}
\end{figure}

In ref.~\cite{Neq44np} we showed explicitly that the color double-trace
terms in the four-point $\NeqFour$ sYM amplitudes are better-behaved in
the UV than the single-trace terms, starting at three loops. In particular,
they obey the finiteness bound $D \le 4+8/L$ at three and four loops.
This improved behavior has been discussed from the vantage points of
both string theory and field
theory~\cite{DoubleTraceNonrenormalization,BHSUV}.  The form of
the four-loop four-point sYM amplitude constructed here allows us to
evaluate the four-loop double-trace divergence in $D=6-2\epsilon$
dimensions and probe whether the bound is saturated.  We will find that
the double-trace term does indeed diverge, and so the bound is saturated.
A subtlety arises in this calculation because the single-trace terms
in the three-loop four-point amplitude diverge in six dimensions.
Therefore the extraction of the four-loop four-point double-trace
divergence requires a careful subtraction of the contribution of the
three-loop counterterm.

In light of the presence of subdivergences in $D=6-2\eps$, it is
useful to separate the integrals appearing in the four-loop amplitude
into those that cannot have subdivergences (graphs 1 through 50)
and those that might (graphs 51 through 85).  The extraction of UV
divergences uses precisely the same methods as discussed above in the
analysis of the UV behavior of the $\NeqEight$ supergravity
amplitude.  For the first set of graphs, we simply quote the result without
providing further details.  The UV divergence from graphs 1 through 50 has the
following $\Tr_{12} \Tr_{34}$ component:
\bea
{\cal A}_4^{(4)}\Bigr|^{\Tr_{12} \Tr_{34}}_{\text{pole 1-50}} \hskip-0.1cm&=&
\frac{g^{10} {\cal K}}{3 \, (4\pi)^{12} \, \eps}
  \, \Tr_{12} \Tr_{34}\ N_c \nn \\
 && \hskip-0.1cm
\null \times \Big[ (s^2 + t^2 + u^2 )
     \bigl( N_c^2 ( 1 - 4 \zeta_3 + 10 \zeta_5 ) + 180 \zeta_5 \bigr)
   - 9 s^2 \bigl( N_c^2 \, \zeta_3 + 25 \zeta_5 \bigr) \Big]
\,, \hskip 1.1 cm 
\label{diags1to50}
\eea
where ${\cal K }$ is defined in \eqn{Kdef}.  The other double-trace
components can be obtained by permuting the external momenta in this
expression.  The absence of a $1/\epsilon^2$ pole signals that this
class of graphs indeed has no subdivergences.

Next we evaluate the remaining graphs, 51 through 85.  Because the
color factors of 1PR graphs do not contain double traces, it follows
that, in fact, only the integrals $I_{51}$, $I_{52}$ and $I_{72}$
contribute to the double-trace terms.  If they did not have subdivergences,
we could evaluate them directly by setting the external momenta to zero,
leading to the vacuum integrals $U_1$, $U_2$ and $U_3$ shown in 
\fig{U_vacuum_basis}.  Instead, we will perform a subtraction of the
subdivergence and evaluate the inner three-loop integral first,
before evaluating the integral over the outer loop momentum.

\begin{figure*}[tb]
\begin{center}
\vskip .7 cm 
\includegraphics[scale=.52]{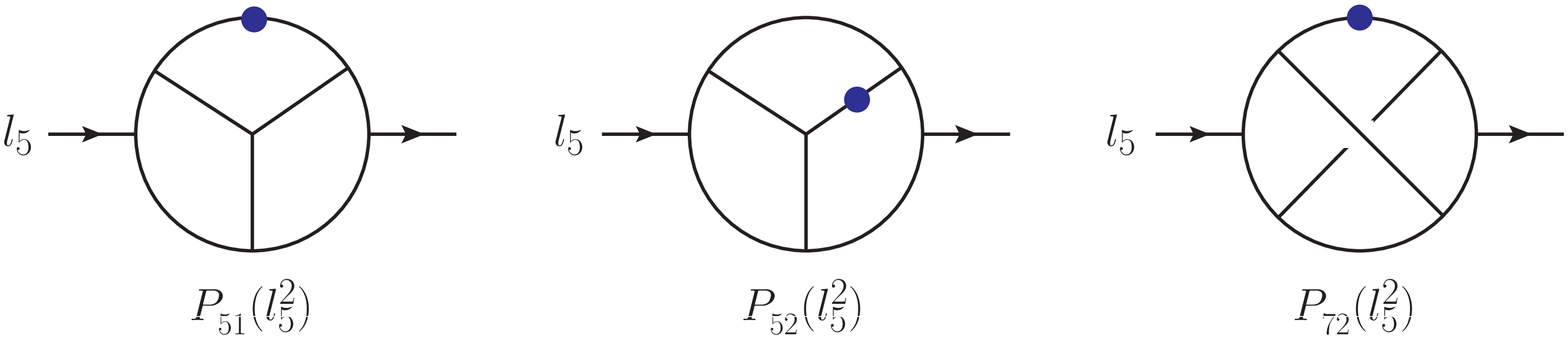}
\end{center}
\vskip -.7 cm 
\caption[a]{\small Three-loop propagator subgraphs $P_{51}(l_5^2)$, $P_{52}(l_5^2)$ and $P_{72}(l_5^2)$, 
generated by setting $k_4\rightarrow0$ in graphs 51, 52 and 72,
respectively.
\label{prop3loopsFigure}
}
\end{figure*}

We first note from \eqn{trivialNrelations}
that the numerator factors of these three integrals
are all equal, $N_{51} = N_{52} = N_{72}$.  All three integrals
contain an essentially identical subdivergence, from a three-point
three-loop subgraph whose external legs carry momentum $k_4$, $l_5$
and $l_5+k_4$.  In the evaluation of this inner graph, and its subtraction
term, if we are only interested in the final $1/\e^2$ and $1/\e$
contributions, we can neglect the dependence on $k_4$: 
the $1/\e^2$ divergence is independent of the details of the 
external momenta, and the $1/\e$ contribution, arising from the finite 
part of the three-loop subgraph, comes from the integration region 
$l_5\gg k_4$ where the outer loop diverges.
Then the three-point subgraph reduces to a propagator (two-point) 
subgraph, as shown in \fig{prop3loopsFigure}
for the respective cases of graphs 51, 52 and 72.  The blue dot indicates
the location of the doubled propagator that is generated in the limit
$k_4 \to 0$.

These integrals can be evaluated in $D=6-2\e$ using
IBP identities and gluing relations through the necessary order, 
$\Ord(\e^0)$.  The results are:
\bea
P_{51}(l_5^2) &=& -(-l_5^2)^{-3\e} \, 
\frac{e^{-3\gamma\e}}{(4\pi)^{9-3\e}} \biggl[
\frac{1}{6\,\e}
 + \frac{25}{9} + \zeta_3 - \frac{10}{3} \zeta_5 + \Ord(\e) \biggr]
\,, \label{P51answer}\\
P_{52}(l_5^2) &=& -(-l_5^2)^{-3\e} \, 
\frac{e^{-3\gamma\e}}{(4\pi)^{9-3\e}} \biggl[
\frac{1}{6\,\e}\biggl( \zeta_3 - \frac{1}{3} \biggr)
 - \frac{25}{27} + \frac{17}{18} \zeta_3 + \frac{1}{4}\zeta_4
 + \Ord(\e) \biggr]
 \,, \label{P52answer}\\
P_{72}(l_5^2) &=& -(-l_5^2)^{-3\e} \, 
\frac{e^{-3\gamma\e}}{(4\pi)^{9-3\e}} \biggl[
\frac{1}{6\,\e}\biggl( \zeta_3 - \frac{1}{3} \biggr)
 - \frac{25}{27} + \frac{17}{18} \zeta_3 + \frac{1}{4}\zeta_4
 + \Ord(\e) \biggr]
 \,. \label{P72answer}
\eea
We note that the propagator subgraphs for graphs 52 and 72 happen
to have identical values through $\Ord(\e)$.  Also,
inspecting the form of the integrals, and comparing with \eqns{VA}{VB},
we see that the leading $1/\e$ singularities correspond to the 
three-loop vacuum integrals $V^{\A}$ (for the planar graph 51)
and $V^{\B}$ (for graphs 52 and 72).

In the trace basis, the color factors for graphs 51, 52 and 72 have
the following form:
\bea
C_{51} &=& N_c^2 (N_c^2 + 12) \bigl( \Tr_{1234} + \Tr_{1432} \bigr)\nn\\
&&\hskip2cm \null
+ 2 N_c (N_c^2+12)
 \bigl( \Tr_{12} \Tr_{34}+\Tr_{13} \Tr_{24}+\Tr_{14} \Tr_{23}) \,,
\label{C51tracebasis}\\
C_{52} &=& C_{72} \, = \,
12 N_c^2 \bigl( \Tr_{1234} + \Tr_{1432} \bigr)
+ 24 N_c 
\bigl( \Tr_{12} \Tr_{34}+\Tr_{13} \Tr_{24}+\Tr_{14} \Tr_{23}) \,.~~~~
\label{C5272tracebasis}
\eea
The double-trace parts of these color factors contain
a piece proportional to $N_c^3$ and one proportional to $N_c^1$.
The $N_c^3$ part comes only from graph 51.  The $N_c^1$ parts of
these color factors are equal for the three graphs.  Taking into
account the relative symmetry factors in \eqn{FourloopCombinatoricSum},
we see that the relevant linear combination of propagator integrals
for the $N_c^1$ part is $P_{N_c^1} \equiv P_{51} + 2P_{52} + P_{72}$,
which is given by,
\be
P_{N_c^1}(l_5^2)\ =\ -(-l_5^2)^{-3\e} \, 
\frac{e^{-3\gamma\e}}{(4\pi)^{9-3\e}} \biggl[
\frac{\zeta_3}{2\,\e}
 + \frac{23}{6} \zeta_3 + \frac{3}{4}\zeta_4
 - \frac{10}{3} \zeta_5 + \Ord(\e) \biggr] \,.
\label{P515272answer}
\ee

Next we need to identify a subtraction that accounts for the
three-loop counterterm needed to cancel the pole given in 
\eqn{ThreeLoopPoleVForm}.  We could compute the four-loop
counter-amplitude by sewing a tree amplitude onto the matrix element
of the counterterm (essentially the negative of \eqn{ThreeLoopPoleVForm}),
in close analogy to how we evaluated the UV divergence of the
three-loop four-point supergravity amplitude in odd dimensions
above $D=6$~\cite{CompactThree}.  However, it is safer to perform
the subtraction within the integrals for the individual graphs 51,
52 and 72, in order to ensure that there are no spurious contributions
arising from potentially different ways of regularizing the infrared
behavior.

In discussing higher-loop divergences of a theory, there is always
a freedom associated with additional, finite renormalizations of
the theory at lower loops, in this case three loops.  Here we will choose,
for definiteness and simplicity, an $\overline{\rm MS}$ scheme for the 
three-loop renormalization.  In this scheme, the necessary counterterms
are 
\bea
P_{51}^{{\rm c.t.}} &=&  \frac{e^{-3\gamma\e}}{(4\pi)^{9-3\e}}
 \, \frac{1}{6\,\e}
\,, \label{P51ct}\\
P_{N_c^1}^{{\rm c.t.}} &=&  \frac{e^{-3\gamma\e}}{(4\pi)^{9-3\e}}
 \, \frac{\zeta_3}{2\,\e} \,. \label{P515272ct}
\eea
Notice that the factor of $(-l_5^2)^{-3\e}$ in 
\eqns{P51answer}{P515272answer} is absent in the
counterterm contributions~(\ref{P51ct}) and (\ref{P515272ct}).

The evaluations of the remaining one-loop integrals and of their subtraction 
are essentially identical for the graphs 51, 52 and 72.  We will discuss in 
detail only the graph 51 for which we consider the following subtracted
integral,
\be
I_{51}^{{\rm sub}} \equiv -i \int \frac{d^{6-2\e} l_5}{(2\pi)^{6-2\e}}
\frac{N_{51}(l_5)}{l_5^2 (l_5-k_1)^2(l_5-k_{12})^2(l_5-k_{123})^2}
\, \biggl[ P_{51}(l_5^2) + P_{51}^{{\rm c.t.}} \biggr] \,.
\label{I51subA}
\ee
If a term in the outer integral over $l_5$ is UV convergent, one can
set $\e\to0$ and the $1/\e$ poles will cancel in the brackets,
leaving no UV contribution at all at four loops.  On the other hand,
if a term in the outer integral generates a logarithmic divergence,
as in the terms in $N_{51}$ that are quadratic in $l_5$, then
the leading $1/\e^2$ pole from $P_{51}(l_5^2)$ is not cancelled by
the counterterm contribution $P_{51}^{{\rm c.t.}}$.
Accounting for a factor of $(-l_5^2)^{-\e}$ from the one-loop
integration measure, we see that there is a mismatch by a factor
of four. That is, the leading $1/\e^2$ contribution
from $P_{51}^{{\rm c.t.}}$ is of opposite
sign to that from $P_{51}(l_5^2)$ and is four times larger in magnitude,
due to a factor of $(-l_5^2)^{-\e}$ instead of $(-l_5^2)^{-4\e}$.

The relative factor of four originates physically from the locality
of the UV poles in the subtracted four-loop amplitude.
The four-loop amplitude carries fractional dimension $\propto s^{-4\e}$,
while the one-loop counterterm amplitude carries fractional dimension
$\propto s^{-\e}$.  Expanding these $s$-dependent factors in $\e$,
we see that a ratio of $(-4)$ between the $1/\eps^2$ poles
from the amplitude and its counter-amplitude is required, in order
for non-local terms of the form $1/\epsilon \times \ln(-s_{ij})$ to cancel
in the subtracted amplitude.

From \eqn{ListOfAllNumerators}, the quadratic terms in the numerator
factor $N_{51}$ are given by,
\bea
N_{51}^{\rm quad} &=&
 \Frac{1}{2} \Bigl[
  - 6 \bigl( \tt \t_{15}^2 + \uu \t_{25}^2 + \ss \t_{35}^2 \bigr)
  + 5 \bigl( \ss \t_{15} \t_{25} + \tt \t_{25} \t_{35}
           + \uu \t_{15} \t_{35} \bigr) \nn\\
&&\hskip0.5cm\null
  - (s^2+t^2+u^2) l_{5}^2 \Bigr] \,.
\label{N51quad}
\eea
We insert \eqn{N51quad} into \eqn{I51subA} and Feynman parametrize
the resulting one-loop integral.  As usual with a one-loop integral with
numerator quadratic in the loop momentum, there are two types of terms:
\begin{enumerate}
\item terms that depend in a fairly complicated way on the Feynman
parameters and external momentum invariants, originating from 
the shift $l_5^\mu = q^\mu + \Delta^\mu$ in the loop momentum $l_5$
needed to complete the square in the denominator, and
\item terms from integrating over the shifted loop momentum $q^\mu$,
in which one can use the identity
$q^\mu q^\nu = q^2/D \times \eta^{\mu\nu}$.
\end{enumerate}
The former terms do not contain an ultraviolet divergence from the outer
integral; therefore they can be dropped as discussed above.
In the latter terms, the first set of terms in $N_{51}^{\rm quad}$ in
\eqn{N51quad}, containing the $-6$ prefactor, drop out because
$k_1^2 = k_2^2 = k_3^2 = 0$.  In the second set of terms in
$N_{51}^{\rm quad}$, one can effectively make the
replacement~(\ref{2tensor}), because any extra terms due to external
momentum dependence are finite and unrelated to the UV divergence and 
therefore drop out.  (We checked this statement
by performing a full Feynman parametrization.)

Therefore, the $1/\e^2$ and $1/\e$ terms in $I_{51}^{\rm sub}$
in \eqn{I51subA} are correctly captured by
\be
I_{51}^{\rm sub}\ =\
-i \biggl( \frac{5}{D} - \frac{1}{2} \biggr) (s^2+t^2+u^2)
 \int \frac{d^D l_5}{(2\pi)^D}
\frac{P_{51}(l_5^2) + P_{51}^{{\rm c.t.}}}
{(l_5-k_1)^2(l_5-k_{12})^2(l_5-k_{123})^2} +\Ord(\e^0)\,.
\label{I51subB}
\ee
We do not need to evaluate this full integral; to extract the UV pole it
is sufficient to simplify it to the form of a massive bubble
integral by rearranging the external momenta,
\be
I_{51}^{\rm sub}\ =\
-i \biggl( \frac{5}{D} - \frac{1}{2} \biggr) (s^2+t^2+u^2)
 \int \frac{d^D l_5}{(2\pi)^D}
\frac{P_{51}(l_5^2) + P_{51}^{{\rm c.t.}}}{(l_5^2)^2 (l_5-k_{12})^2} +\Ord(\e^0) \,.
\label{I51subC}
\ee
In the term coming from $P_{51}(l_5^2)$ in \eqn{I51subC}, the powers
to which the two propagators in the $D=6-2\e$ bubble (after analytic
continuation) are raised are $2+3\e$ and $1$; while in the term coming
from $P_{51}^{{\rm c.t.}}$ they are $2$ and $1$.

The momentum-independent parts of these bubble integrals are given by
\bea
P^{(1)}(2+3\e,1;6-2\e) &=& -\frac{e^{-\gamma\e}}{(4\pi)^{3-\e}}
\biggl[ \frac{1}{8\,\e} + \frac{7}{16} + \Ord(\e) \biggr] \,, 
\label{Gamplitude}\\
P^{(1)}(2,1;6-2\e) &=& -\frac{e^{-\gamma\e}}{(4\pi)^{3-\e}}
\biggl[ \frac{1}{2\,\e} + 1 + \Ord(\e) \biggl] \,.
\label{Gsubtraction}
\eea
(Other ways of regulating the infrared behavior, such as keeping
the complete dependence on external momenta, will give different
$\Ord(\e^0)$ terms in \eqns{Gamplitude}{Gsubtraction}.  However,
the difference must always be the same; otherwise a $1/\eps$
UV pole would be generated from a UV convergent integral.)
Including the overall factors, we get,
\bea
I_{51}^{\rm sub} &=&
\biggl( \frac{5}{6-2\e} - \frac{1}{2} \biggr) 
\frac{e^{-4\gamma\e}}{(4\pi)^{12-4\e}}
\, (s^2+t^2+u^2)
 \nonumber\\
&&\hskip2cm\null\times \biggl\{
\biggl[ \frac{1}{6\,\e}
        + \frac{25}{9} + \zeta_3 - \frac{10}{3} \zeta_5 \biggr]
\biggl( \frac{1}{8\,\e} + \frac{7}{16} \biggr)
-  \frac{1}{6\,\e} 
 \biggl( \frac{1}{2\,\e} + 1 \biggr)+\Ord(\e^0)
\biggr\}
\nonumber\\
&=&  \frac{e^{-4\gamma\e}}{(4\pi)^{12-4\e}}
 \, \frac{s^2+t^2+u^2}{24}
 \biggl[ -\frac{1}{2\,\e^2}
 + \frac{1}{\e} \biggl( \frac{29}{18} + \zeta_3
 - \frac{10}{3} \zeta_5 \biggr) \biggr] + \Ord(\e^0) \,.
\label{I51subFinal}
\eea
Similarly, the $N_c^1$ double-trace contribution is
obtained using the same formula~(\ref{I51subC}) with $P_{51}$
replaced by $P_{N_c^1}$, taken from \eqns{P515272answer}{P515272ct},
\bea
I_{N_c^1}^{\rm sub} &=&
\biggl( \frac{5}{6-2\e} - \frac{1}{2} \biggr)
\frac{e^{-4\gamma\e}}{(4\pi)^{12-4\e}}
\, (s^2+t^2+u^2) \nonumber\\
&&\hskip2cm\null\times \biggl\{ \biggl[
\frac{\zeta_3}{2\,\e}
 + \frac{23}{6} \zeta_3 + \frac{3}{4}\zeta_4
 - \frac{10}{3} \zeta_5 \biggr] 
 \biggl( \frac{1}{8\,\e} + \frac{7}{16} \biggr)
 - \frac{\zeta_3}{2\,\e} 
\biggl( \frac{1}{2\,\e} + 1 \biggr) + \Ord(\e^0)\biggr\}
\nonumber\\
&=&   \frac{e^{-4\gamma\e}}{(4\pi)^{12-4\e}}
 \, \frac{s^2+t^2+u^2}{24}
 \biggl[ -\frac{3\, \zeta_3}{2\,\e^2}
 + \frac{1}{\e} \biggl( \frac{1}{3} \zeta_3 + \frac{3}{4} \zeta_4
 - \frac{10}{3} \zeta_5 \biggr) \biggr] + \Ord(\e^0) \,.
\label{I515272combsubFinal}
\eea
We notice that, similarly to the numerator factor $N_{51}^{\text{quad}}$
in \eqn{N51quad}, both \eqns{I51subFinal}{I515272combsubFinal}
have manifest permutation symmetry.

Plugging \eqns{I51subFinal}{I515272combsubFinal} into the full
amplitude, including the double-trace part of the color factors,
the sum over all 24 permutations, and the overall prefactor,
we obtain,
\bea
{\cal A}_4^{(4)}\Bigr|^{\rm sub}_{\text{pole 51-85}}
&=& \frac{g^{10} \, {\cal K} \, e^{-4\gamma\e} }{ (4\pi)^{12-4\e} } 
\, N_c \, 
(\Tr_{12} \Tr_{34}+\Tr_{13} \Tr_{24}+\Tr_{14} \Tr_{23}) \,
( s^2 + t^2 + u^2 )\nn\\
&&\hskip0.3cm\null\times \biggl\{
 - \frac{N_c^2 + 36 \zeta_3}{2 \, \e^2}
 + \frac{1}{\e} \biggl[
      N_c^2 \biggl( \frac{29}{18} + \zeta_3 - \frac{10}{3} \zeta_5 \biggr)
    + 4 \zeta_3 + 9 \zeta_4 - 40 \zeta_5 \biggr] \biggr\} \,. 
\nn\\
&~& \label{A515272sub}
\eea
We have also evaluated the difference between the one-loop outer integrals of $I_{51}, I_{52}$
and $I_{72}$ and their corresponding $\overline{\rm MS}$ subtraction
terms, $I_{51}^{\text{sub}}, I_{52}^{\text{sub}}$  and $I_{72}^{\text{sub}}$, prior to reduction 
to bubble integrals. That is, we directly integrated eq.~(\ref{I51subA}) for general external 
momenta and separately the outer integral of $I_{51}$ and its subtraction for two special kinematic points.  
In this way we verified the required 
cancellation of the nonlocal divergent terms $\ln(-s_{ij})/\eps$.

Finally, we add the contribution~(\ref{diags1to50}) 
from the graphs 1--50 that have no subdivergences, in order
to obtain the total four-loop divergence (after subtraction
of three-loop subdivergences):
\bea
{\cal A}_4^{(4)}\Bigr|^{\text{double trace}}_{\text{pole}}
&=& {\cal A}_4^{(4)}\Bigr|^{\text{double trace}}_{\text{pole 1-50}} 
    + {\cal A}_4^{(4)}\Bigr|^{\rm sub}_{\text{pole 51-85}}
\nn\\
 &=& 
\frac{g^{10} \, {\cal K} \, e^{-4\gamma\e} }{ (4\pi)^{12-4\e} }
 \, N_c \, \biggl\{
(\Tr_{12} \Tr_{34}+\Tr_{14} \Tr_{23}+\Tr_{13} \Tr_{24})
 \, ( s^2 + t^2 + u^2 )
\nn\\
&&\hskip1.2cm\null \times \biggl[
 - \frac{N_c^2 + 36 \zeta_3}{2 \, \e^2}
 + \frac{1}{\e} \biggl( 
      N_c^2 \biggl( \frac{35}{18} - \frac{\zeta_3}{3} \biggr)
    + 4 \zeta_3 + 9 \zeta_4 + 20 \zeta_5 \biggr) \biggr]
\nn\\
&&\hskip0.0cm\null
- \frac{3}{\e} ( N_c^2 \, \zeta_3 + 25 \zeta_5 ) \, 
( \Tr_{12} \Tr_{34} \, s^2 +\Tr_{14} \Tr_{23} \, t^2
 + \Tr_{13} \Tr_{24} \, u^2 ) \biggr\} \,. \nn\\
&~& \label{fourloopDoubleTracepole}
\eea

Of course, the double-trace part of the four-loop counterterm
must be chosen to cancel these poles,
\be
{\cal A}_{4;\text{c.t.}}^{(4)}\Bigr|^{\text{double trace}}
= - {\cal A}_4^{(4)}\Bigr|^{\text{double trace}}_{\text{pole}}
\ +\ {\cal O}(1)\,,
\label{countertermDefinitionfourloop}
\ee
corresponding to a nonvanishing divergent coefficient for a
counterterm of the schematic form,
$\Tr({\cal D}^{4-k} F^2) \Tr({\cal D}^{k} F^2)$.
(The covariant derivatives may be distributed among the two traces
in various ways.)  It is possible to perform finite shifts of the
coefficient of the single-trace operator $\Tr( {\cal D}^2 F^4 )$
which has a divergent coefficient at three loops.  One can trace through
the effect of such a shift by shifting $P_{51}^{{\rm c.t.}}$
and $P_{N_c^1}^{{\rm c.t.}}$ by $\Ord(\e^0)$ constants.
In principle, one could remove the $1/\e$ terms in
\eqn{fourloopDoubleTracepole} that are proportional to $(s^2+t^2+u^2)$.
However, one cannot remove the $1/\e^2$ pole.  Nor can one remove
the term proportional to $( N_c^2 \, \zeta_3 + 25 \zeta_5 )$,
because the dependence on color and kinematics is different from the one
induced by the three-loop counterterm.

In conclusion, the double-trace terms in the
four-point $\NeqFour$ sYM amplitude do diverge at four loops,
saturating the double-trace finiteness bound of $D_c = 4 + 8/L$.


\section{Conclusions}
\label{ConclusionSection}

In this paper, we recomputed the four-loop four-point amplitudes of
$\NeqFour$ sYM theory and $\NeqEight$ supergravity, first obtained in
refs.~\cite{GravityFour,Neq44np}. By exploiting the conjectured
duality between color and kinematics~\cite{BCJ,BCJLoop} we found
greatly simplified representations. It also allowed us to find the
form of the complete amplitude, including nonplanar contributions,
using only planar cut information as input.  We
confirmed the correctness of the construction by comparing the
unitarity cuts of the new expressions to the cuts of the earlier
forms~\cite{GravityFour,Neq44np}.  This provides new nontrivial evidence in
favor of the duality conjecture and the associated gravity double-copy
property. An important advantage of the current construction is that
once the sYM amplitude has been arranged into a duality-satisfying
form, the construction of the corresponding supergravity integrand is
trivial: one simply replaces the color factors with kinematic
numerator factors in each graph.

The new form of the four-loop four-point amplitude of $\NeqEight$ has
an important advantage over the previous one~\cite{GravityFour},
because no integral displays a worse UV power counting than the
complete amplitude.  This feature greatly simplifies the extraction of
the UV divergence of the four-loop $\NeqEight$ supergravity amplitude
in the critical dimension $D_c=11/2$, corresponding to the lowest
dimension where both $\NeqFour$ sYM theory and $\NeqEight$
supergravity first diverge at four loops.  To carry out the required
integration we used techniques similar to those described in
refs.~\cite{GravityThree,GravityFour,Neq44np}.  Our results prove that
the four-loop four-point amplitude of $\NeqEight$ supergravity does
indeed diverge in the same critical dimension as the corresponding
amplitude of $\NeqFour$ super-Yang-Mills theory.  Thus, the
$\NeqEight$ supergravity finiteness bound~\cite{GravityFour} is, in
fact, saturated at four loops.  The amplitude divergence in $D_c =
11/2$ means that the ${\cal D}^8 R^4$ supergravity counterterm has a
nonzero divergent coefficient, in much the same way as the $\Tr({\cal
  D}^2 F^4)$ counterterm of $\NeqFour$ sYM has a nonvanishing
divergent coefficient in this dimension.  Moreover, we found that the
four-loop finiteness
bound~\cite{Neq44np,DoubleTraceNonrenormalization}, $D<6$, for the
double-color-trace terms of $\NeqFour$ sYM theory is also saturated.
In other words, the corresponding $D=6$ double-trace counterterms
$\Tr({\cal D}^{4-k} F^2) \Tr({\cal D}^{k} F^2)$ are also present with
non-vanishing coefficients.

More generally, the duality between color and kinematics offers the promise of
carrying advances from the planar sector of gauge theory to the
nonplanar sector and then to gravity theories.  Its underlying origin
is, however, still poorly understood; recent progress suggests that,
at least in the self-dual case~\cite{Oconnell}, underlying it is an
infinite-dimensional Lie algebra of area-preserving diffeomorphisms.
Progress has also been made in finding explicit representations
of tree amplitudes that manifestly satisfy the
duality~\cite{ExplicitForms}.  It would be interesting and very useful
to devise effective rules that would generate directly
duality-satisfying representations for loop amplitudes, thus
eliminating the need to solve the system of duality constraints on a
case-by-case basis.  A step towards finding a Lagrangian with the
desired properties has been given in ref.~\cite{Square}.  It would
also be interesting to explore whether the color-kinematic duality
extends beyond weak-coupling perturbation theory as well as whether
the existence of such a duality has practical consequences after carrying
out the loop-momentum integrations.

Explicit calculations often lead to surprises.  The results described
here are no different.  In particular, in the critical dimension
$D=11/2$ we found that, after reducing the integrals containing UV
divergences to a basis of vacuum integrals encoding the numerical
factors in front of the divergent operator, the UV divergence is given
by exactly the same combination of basis integrals as found in the
single-trace $1/N_c^2$-suppressed terms of $\NeqFour$ sYM theory.  It
seems unlikely that this is accidental because similar behavior is
found at lower loops.  It would obviously be important to understand
the origin of this curious connection and implications it may have at
higher loops on UV divergences.  Another interesting property is the
existence of strong cancellations between the contributions of various
graphs to the UV divergence in the critical dimension.  This suggests
that different integral contributions may be related to each other by
a hidden symmetry.  

In summary, the duality between color and kinematics offers a powerful
means for streamlining the construction of multiloop amplitudes,
carrying advances in the planar sector to the nonplanar sector.  It
allowed us to express the numerators of the four-loop four-point
amplitudes of $\NeqFour$ sYM theory and $\NeqEight$ supergravity in
terms of the numerators of two planar graphs. Using this simplified
form, in the critical dimension $D_c =11/2$, we found a surprising
coincidence between the UV divergences of $\NeqEight$ supergravity and
those of subleading color single-trace terms of $\NeqFour$ sYM theory.
This hints at further new relations between gauge and gravity theories
to be unraveled and that further surprises await us at five and higher
loops.  We look forward to using the tools described in this paper to
further explore the multiloop structure of gauge and gravity
amplitudes and to unravel their UV properties.


\section*{Acknowledgments}
\vskip -.3 cm We thank Nima Arkani-Hamed, Scott Davies, Tristan
Dennen, Michael Green, Yu-tin Huang, Harald Ita, David Kosower, Kelly Stelle and
Pierre Vanhove for
many stimulating discussions.  We thank Academic Technology Services
at UCLA for computer support.  This research was supported by the US
Department of Energy under contracts DE--AC02--76SF00515,
DE--FG03--91ER40662 and DE--FG02--90ER40577 (OJI), and by the 
US National Science Foundation under grants PHY--0756174, PHY--0855356 and PHY05--51164.
R.~R. acknowledges support from the A.~P. Sloan Foundation.
H.~J.'s research is supported by the European Research Council under
Advanced Investigator Grant ERC-AdG-228301.

\vskip4cm

\appendix

\section{Useful numerator functional equations}
\label{JacobiAppendix}

In this appendix we list a set of numerator equations that determine
the four-loop four-point \NeqFoursYM{} amplitude (up to snail contributions),
starting from the two planar master graphs 18 and 28 in
\fig{Master4Figure}. These equations follow directly from the dual
Jacobi relations.  However, to make the equations more convenient for
generating numerator factors from the our two planar master
numerators, we performed various simplifications which follow from the
\NeqFoursYM{} auxiliary constraints described in
\sect{StrategySubsection}.  In particular, we use the two-term
relations (see \eqn{trivialNrelations}) which rely on the no one-loop
triangle subgraph constraint to eliminate numerators appearing in
other dual Jacobi relations.  We also simplified the functional
arguments of the numerators using the auxiliary constraint that
numerators are independent of the loop momenta of one-loop box subgraphs.  For
example, instead of the dual relation,
\begin{eqnarray}
N_{50}(k_1,k_2,k_3,l_5,l_6,l_7,l_8) &= &
N_{28}(k_2,k_1,k_4,l_5,k_3-l_7,k_2-l_6,l_8)\nn \\
&& \null \hskip 2 cm 
- N_{28}(k_1,k_2,k_3,l_6,k_4-l_8,k_1-l_5,l_7)\,, 
\end{eqnarray}
we simplify this to 
\begin{eqnarray}
N_{50}(k_1,k_2,k_3,l_5,l_6,l_7,l_8) &=&
N_{28}(k_2,k_1,k_4,l_5,k_3-l_7,l_7,l_8) \nn \\
&& \null \hskip 2 cm 
- N_{28}(k_1,k_2,k_3,l_6,k_4-l_8,l_7,l_8)\,,
\end{eqnarray}
using the fact that $N_{28}$ is, in fact, independent of the values of the last
two arguments since these momenta are those of one-loop box subgraphs
(see \fig{Master4Figure}).
In this sense, the last two arguments of $N_{28}$ are effectively placeholders,
and can be assigned any value without altering the numerators.
These simplifications, however, imply that the equations given below are
specific to \NeqFoursYM{} theory and will not hold for corresponding
numerators of amplitudes of theories with fewer supersymmetries.
They are also not in direct correspondence with the color-Jacobi
equations, because numerators of graphs with triangle subgraphs are set 
to zero, although corresponding color factors are nonvanishing. 

On the left-hand side of each duality equation, for
simplicity, we will suppress the canonical arguments, which are the
three external momenta and the four independent loop momenta following
the graph labels in figs.~\ref{BC1Figure}--\ref{D6Figure}, 
{\it i.e.}
\begin{equation}
N_i\equiv N_i(k_1, k_2, k_3, l_5, l_6, l_7, l_8)\,,
\end{equation}
and we take $k_4 \equiv -k_1 -k_2 -k_3$ throughout.  
We have ordered the equations so that the substitutions
that are required to express the given numerators in terms of the two
master numerators always come from previous equations in the list.
With the above notation, the required equations are 
\def\Pl{(}
\def\Pr{)}
{\setlength{\jot}{.09pt}  
\begin{align}
N_{58} & =
 N_{18}\Pl k_1, k_2, k_3, k_2 - l_6, l_5, l_7, l_8 \Pr
- N_{18}\Pl k_2, k_1, k_3, k_1 - l_6, l_5, l_7, l_8 \Pr
 \,, \nn \\ 
N_{33}& =
  N_{28}\Pl k_4, k_3, k_2, k_3 - l_5, k_2 - l_6 + l_7, l_7, l_8 \Pr
 -N_{18}\Pl k_1, k_2, k_3, k_2 - l_6, k_3 - l_5, l_7, l_8 \Pr
 \,, \nn \\
N_{50}& =
 N_{28}\Pl k_2, k_1, k_4, l_5, k_3 - l_7, l_7, l_8 \Pr
-N_{28}\Pl k_1, k_2, k_3, l_6, k_4 - l_8, l_7, l_8 \Pr
 \,, \nn \\
N_{6}& =
 -N_{33}\Pl k_1, k_2, k_4, l_7, l_5 - l_6, k_1 - l_6, l_8 \Pr
 - N_{33}\Pl k_2, k_1, k_4, l_7, l_6, k_2 - l_5 + l_6, l_8 \Pr
 \,, \nn \\
N_{14}& =
 -N_{33}\Pl k_3, k_2, k_1, l_5, -l_5 - l_7, k_3 - l_7 + l_8, l_6 \Pr
 - N_{33}\Pl k_3, k_2, k_1, l_5, k_2 + l_7, l_7 - l_8, l_8 \Pr
 \,, \nn \\
N_{24}& =
 -N_{28}\Pl k_1, k_2, k_3, l_5 - l_7, -l_6, l_7, l_8 \Pr
 - N_{33}\Pl k_1, k_2, k_4, -l_6, -l_7, -l_5, l_8 \Pr
 \,, \nn \\
N_{32}& =
 -N_{28}\Pl k_4, k_2, k_1, l_7, k_3 - l_5, l_7, l_8 \Pr
 - N_{33}\Pl k_2, k_1, k_3, l_5, l_6, k_2 + l_5 + l_6 - l_7, l_8 \Pr
 \,, \nn \\
N_{48}& =
 N_{28}\Pl k_3, k_4, k_1, l_8, k_2 - l_5, l_7, l_8 \Pr
 - N_{33}\Pl k_1, k_2, k_3, k_3 - l_6, k_2 - l_5, l_7, l_8 \Pr
 \,, \nn \\
N_{49}& =
 -N_{33}\Pl k_1, k_2, k_3, k_3 - l_8, k_2 - l_5, -l_7, l_8 \Pr
 - N_{33}\Pl k_4, k_1, k_2, l_5, -l_7, l_6, l_8 \Pr
 \,, \nn \\
N_{66}& =
 N_{58}\Pl k_1, k_2, k_4, l_5 - k_3  - l_6, l_6, l_7, l_8 \Pr
  -N_{58}\Pl k_1, k_2, k_3, k_3 + l_6, l_6, l_7, l_8 \Pr
 \,, \nn \\
N_{1}& =
 -N_{6}\Pl k_1, k_2, k_3, l_6, l_5, l_7, l_8 \Pr
 - N_{6}\Pl k_1, k_2, k_4, l_6, l_5, l_7, l_8 \Pr
 \,, \nn \\
N_{68}& =
 N_{14}\Pl k_1, k_2, k_3, k_1 - l_5, -l_6, -l_7, -l_8 \Pr
 - N_{14}\Pl k_1, k_2, k_4,  l_5 - k_2, -l_7, -l_6, l_8 \Pr
 \,, \nn \\
N_{21}& =
 -N_{14}\Pl k_2, k_1, k_3, l_5, l_6, l_7, l_8 \Pr
 - N_{18}\Pl k_2, k_1, k_3, -l_5, k_1 + k_3 + l_5 - l_6, l_7, l_8 \Pr
 \,, \nn \\
N_{26}& =
 N_{24}\Pl k_2, k_1, k_3, -l_5, -k_4 - l_6 - l_7, l_8, l_6 \Pr
 - N_{24}\Pl k_2, k_1, k_4, -l_5,  l_7 - k_3, l_6 -k_1 - l_5 - l_8, l_6 \Pr
 \,, \nn \\
N_{27}& =
 -N_{18}\Pl k_2, k_1, k_4, -l_5, l_7, l_7, l_8 \Pr
 - N_{24}\Pl k_1, k_2, k_4, l_5, -k_3 - l_7 - l_8,
            k_3 - l_6 + l_7 + l_8, l_8 \Pr
 \,, \nn \\
N_{37}& =
 -N_{28}\Pl k_2, k_1, k_3, k_1 - l_5, k_4 + l_8, l_7, l_6 \Pr
 - N_{49}\Pl k_2, k_1, k_3, k_1 - l_5, -l_8,  l_7 - k_2, l_6 \Pr
 \,, \nn \\
N_{39}& =
 N_{28}\Pl k_2, k_1, k_3, -l_5 - l_7, k_4 + l_6 + l_8, l_5, l_6 \Pr
 - N_{48}\Pl k_1, k_2, k_3, l_7, l_8, -l_5 - l_7, -l_6 - l_8 \Pr
 \,, \nn \\
N_{45}& =
 N_{49}\Pl k_1, k_2, k_3, l_5 - l_6 - l_7 - l_8, k_4 - l_6, l_5, l_7 \Pr 
\nn \\ & \null \hskip 1 cm 
+ N_{49}\Pl k_1, k_2, k_4, k_2 + l_6 + l_7 + l_8, l_7, l_5, k_4 - l_6 \Pr
 \,, \nn \\
N_{38}& =
 N_{49}\Pl k_2, k_1, k_4, l_6, k_3 + l_5 + l_7, -l_5 + l_6, k_4 - l_8 \Pr 
 \nn \\ & \null \hskip 1 cm 
 - N_{49}\Pl k_1, k_2, k_4, l_5 - l_6, k_3 + l_5 + l_7, -l_6, l_7 + l_8 \Pr
 \,, \nn \\
N_{53}& =
 N_{58}\Pl k_1, k_2, k_3, k_3 - l_8, l_6, l_7, l_8 \Pr
 + N_{66}\Pl k_1, k_2, k_4, l_8, -k_4 - l_5, l_7, l_8 \Pr
 \,, \nn \\
N_{12}& =
 N_{18}\Pl k_4, k_3, k_2, l_6, k_2 + l_8, l_5, l_7 \Pr
 + N_{26}\Pl k_3, k_4, k_1, -l_6, l_8, -l_5, l_8 \Pr
 \,, \nn \\
N_{51}& =
 N_{18}\Pl k_3, k_2, k_1, k_1 + k_2 - l_5, -l_6, l_7, l_8 \Pr
 - N_{21}\Pl k_2, k_3, k_1,  l_5 -k_1 - k_2, -l_6, l_7, l_8 \Pr
 \,, \nn \\
N_{63}& =
 N_{21}\Pl k_1, k_2, k_3, k_2 - l_5, k_1 + k_2 - l_5 - l_6, l_7, l_8 \Pr
 \nn \\ & \null \hskip 1 cm 
 - N_{21}\Pl k_2, k_1, k_3, k_1 - l_5, k_1 + k_2 - l_5 - l_6, l_7, l_8 \Pr
 \,, \nn \\
N_{79}& =
 N_{45}\Pl k_1, k_2, k_3, k_2 - l_5, k_4 - l_7, l_6, -l_6 - l_8 \Pr 
 \nn \\ & \null \hskip 1 cm 
- N_{45}\Pl k_1, k_2, k_3, l_5 - k_1, l_7,
            k_3 - l_6, k_4 + l_5 - l_7 - l_8 \Pr
 \,, \nn \\
N_{80}& =
 N_{53}\Pl k_1, k_2, k_3, k_3 - l_7, l_6, l_7, l_8 \Pr
 + N_{53}\Pl k_1, k_2, k_3,  l_7 - k_4, l_5, l_6, l_8 \Pr
 \,, \nn \\
N_{55}& =
 N_{51}\Pl k_1, k_2, k_3, k_1 + l_5, l_6, l_7, l_8 \Pr
 - N_{51}\Pl k_1, k_3, k_2, k_1 + l_5, l_6, l_7, l_8 \Pr
 \,, \nn \\
N_{83}& =
 -N_{55}\Pl k_3, k_1, k_2, k_1 + k_2 - l_5, l_8, l_6, l_7 \Pr
 - N_{55}\Pl k_3, k_1, k_2,  l_5 -k_3, l_6, l_7, l_8 \Pr
\,.
\label{smallJrel4loop}
\end{align}
}
There are also a set of simpler two-term relations whenever one of the
three numerators vanishes due to the appearance of a forbidden one-loop
triangle subgraph,
{\setlength{\jot}{.09pt} 
\begin{align}
N_{5} &= N_{4} = N_{3} = N_{2} = N_{1}
 \,, \nn\\
%
N_{11} &=  N_{10} = N_{9} = N_{8} = N_{7} = N_{6}  
\,, \nn\\
%
N_{40} &= N_{13} = -N_{12}
 \,, \nn\\
%
N_{41} &= -N_{17} = -N_{16} = -N_{15} = N_{14} 
 \,, \nn\\
N_{42} &= N_{20} = -N_{19} = N_{18} 
\,, \nn\\
N_{43} &= -N_{23} = -N_{22} = -N_{21} 
\,, \nn\\
N_{25} &= N_{24} 
\,, \nn\\
N_{44} & = -N_{26} 
\,, \nn\\
N_{31} &= -N_{30} = N_{29} = N_{28} 
\,, \nn\\
N_{46} &= N_{34} = N_{32} 
\,, \nn\\
N_{36} &= -N_{35} = -N_{33} 
\,, \nn\\
N_{47} &= N_{38}  
\,, \nn\\
N_{72} &= N_{52} = N_{51} 
\,, \nn\\
N_{74} &= -N_{54} = -N_{53} 
 \,, \nn\\
N_{73} &= N_{57} = N_{56} = N_{55} 
 \,, \nn\\
N_{76} &= -N_{62} = -N_{61} = -N_{60} = -N_{59} = -N_{58} 
\,, \nn\\
N_{77} &= -N_{65} = N_{64} = N_{63} 
 \,, \nn\\
N_{78} &= -N_{67} = N_{66}  
\,, \nn\\
N_{75} &= N_{71} = N_{70} = N_{69} = N_{68} 
\,, \nn\\
N_{82} &= N_{81} = N_{80}  
\,, \nn\\
N_{85} &= N_{84} = N_{83} 
\,. 
\label{trivialNrelations}
\end{align}
}
Plain-text, computer-readable versions of both the original duality
relations (using only the no one-loop triangle subgraph property)
and the simplified ones presented above may be found
online~\cite{Online}.  Many other functional equations can be obtained
from the dual Jacobi relations, which we will not list here.  Although
important, as they provide additional independent constraints to the
full system, they are not needed to specify the solution once the
system and master graph numerators have been solved. However, we have
confirmed that the numerators presented in \app{NumeratorAppendix}
automatically satisfy all these remaining dual Jacobi relations.

\section{Explicit numerators for graphs}
\label{NumeratorAppendix}

In this appendix we give the explicit values $N_i$ of the distinct
graph numerators in the $\NeqFour$ sYM amplitude. (The remaining ones
are given directly in terms of these via \eqn{trivialNrelations}.)
These values are obtained by taking the numerators of the master
graphs~(\ref{FourLoopMasters}) and substituting their values into
\eqn{smallJrel4loop}. We have performed some algebraic simplifications
to obtain the results collected here.  The $\NeqEight$ supergravity
numerators are squares of the $N_i$, as in
\eqns{FourLoopGravityAmplitude}{NSquaring}.

The explicit values of the distinct numerators are 
\def\Frac#1#2{{\textstyle \frac{#1}{#2}}}
\begin{align}
N_{1} &= 
\ss^3 
\,, \nn\\
N_{6} &= 
 \Frac{1}{2} \ss^2 (\t_{45} - \t_{35} - \ss)
\,, \nn\\
N_{12} & = 
\Frac{1}{2} \ss{} (\ss{} 
(\t_{16} - \t_{26} - \t_{35} + \t_{45} + 2 \t_{56} + 2 \tt) \nn\\ &
- 2 (4 \t_{16} \t_{25} + 4 \t_{15} \t_{26} + \t_{45} 
(\t_{36} - 3 \t_{46}) + \t_{35} (\t_{46} - 3 \t_{36})))
\,, \nn\\
N_{14} &= 
 \Frac{1}{4} (\ss{} 
(9 \t_{15}^2 + 9 \t_{25}^2 + 4 \tt^2 + 8 \tt \t_{35} + 2 \t_{35}^2
 + 2 \t_{45}^2) + 8 \t_{25} (\uu^2 - \ss^2) \nn\\ &
 - 5 \tt{} (\t_{15} \t_{35} + \t_{25} \t_{45}) 
+ \uu{} (4 \tt{} (2 \t_{15} + l_{5}^2) - 5 \t_{15} \t_{45}
 - 5 \t_{25} \t_{35}))
\,, \nn\\
N_{18} &=
 \Frac{1}{4} (6 \uu^2 \t_{25} + \uu{} (2 \ss{} (5 \t_{25} + 2 \t_{26})
 - \t_{15} (7 \t_{16} + 6 \tt)) \nn\\ &
 + \tt{} (\t_{15} \t_{26} - \t_{25} (\t_{16} + 7 \t_{26}))
 + \ss{} (4 \t_{15} (\tt - \t_{26}) + 6 \t_{36} (\t_{35} - \t_{45}) \nn\\ &
 - \t_{16} (4 \tt + 5 \t_{25}) - \t_{46} (5 \t_{35} + \t_{45}))
 + 2 \ss^2 (\tt + \t_{26} - \t_{35} + \t_{36} + \t_{56}))
\,, \nn\\
N_{21} &= 
 \Frac{1}{4} (\tt{} (12 \t_{15}^2 - 7 \t_{15} \t_{16}
 + \t_{25} (\t_{16} - 10 \t_{35}))
 + \uu{} (\t_{25} (12 \t_{25} - 8 \tt - 7 \t_{26})
+ \t_{15} (8 \tt + \t_{26} - 10 \t_{35}))  \nn\\ &
 + 4 l_{5}^2 (\uu^2 - \ss \tt)
 - \ss{} (2 \t_{15} (6 \t_{16} + 5 \t_{25})
 + 4 \uu{} (\t_{16} - \tt+ \t_{26} + 2 \t_{45})
 + \t_{35} (\t_{36} - 12 \t_{35} - 10 \t_{46})  \nn\\ &
 - \t_{45} (11 \t_{36} + 6 \t_{46}) + 12 \t_{25} \t_{26})
 + 2 \ss^2 (\uu - \t_{16} - \t_{35} + \t_{36} - \t_{56}))
\,, \nn\\
N_{24} &= 
 \Frac{1}{4} (\ss{} (2 \t_{36} (3 \t_{35} + \t_{45})
 + \t_{46} (\t_{35} + 13 \t_{45}) -4 \uu{} (2 \t_{15} + \t_{16}
 - 2 \t_{25} + \t_{26})- 11 \t_{15} \t_{26}) \nn\\ &
 - 2 \ss^2 (2 \t_{15} - 2 \t_{25} + \t_{26} - \t_{36} - \t_{37}
 + \t_{47} - \t_{56})  + \uu{} (11 \t_{16} \t_{25}
 + \t_{15} (7 \t_{16} + \t_{26}))  \nn\\ &
 + \tt \t_{25} (12 \t_{16} + 7 \t_{26}))
\,, \nn\\
N_{26} &= 
 \Frac{1}{4} (\uu \t_{15} (7 \t_{16} + 12 \t_{26})
 + \tt{} (11 \t_{15} \t_{26} + \t_{25} (\t_{16} + 7 \t_{26}))
 + \ss{} (16 \t_{15} \t_{17} \nn\\ &
 - 4 \uu{} (2 \t_{15} + \t_{16} - 2 \t_{25} + \t_{26})
 + \t_{25} (16 \t_{27} - 11 \t_{16})
 + \t_{35} (6 \t_{36} - 4 \t_{37} + \t_{46} - 20 \t_{47}) \nn\\ &
 + \t_{45} (2 \t_{36} - 20 \t_{37} + 13 \t_{46} - 4 \t_{47}))
 + 2 \ss^2 (\t_{17} - 2 \t_{15} + 2 \t_{25} - \t_{26} - \t_{27}
 + \t_{36} + \t_{56} + 2 \t_{57}))
\,, \nn\\
N_{27} &= 
\Frac{1}{4} (\uu \t_{18} (\t_{25} - 7 \t_{15})
 + \t_{28} (\tt{} (\t_{15} - 11 \t_{25}) - 4 \uu \t_{25})
 + \ss{} (4 \t_{15} (3 \t_{17} + \t_{18} - \t_{27})  \nn\\ &
 + 4 \uu{} (2 \t_{15} - 2 \t_{25} + \t_{28})
 + \t_{45} (5 \t_{18} + 5 \t_{28} - 16 \t_{37} - 5 \t_{38})
 + \t_{35} (2 \t_{38} - 16 \t_{47} - 9 \t_{48}) - 4 \t_{18} \tt  \nn\\ &
 - 4 \t_{25} (\t_{17} - 3 \t_{27}))
 + 2 \ss^2 (2 \t_{15} + 2 \t_{17} - 2 \t_{25} + \t_{28} - \t_{36}
 + 2 \t_{37} + \t_{38} + \t_{46} + 2 \t_{57} + \t_{58}))
\,, \nn\\
N_{28} &= 
\Frac{1}{4} (\ss{} (2 \t_{15} \tt
 + \t_{16} (2 \tt - 5 \t_{25} + \t_{35})
 + 5 \t_{35} (\t_{26} + \t_{36}) + 2 \tt{} (2 \t_{46} - \t_{56})
 - 10 \uu \t_{25})  \nn\\ &
 - 4 \ss^2 \t_{25} - 6 \uu {}(\t_{46} (\tt - \t_{25} + \t_{45})
 + \t_{25} \t_{26}) - \tt{} (\t_{15} (4 \t_{36} + 5 \t_{46})
 + 5 \t_{25} \t_{36}))
\,, \nn\\
N_{32} &= 
 \Frac{1}{4} (\tt{} (\t_{25} (\t_{16} - 11 \t_{25} - 12 \t_{26})
 + \t_{25} \t_{35} - 6 \t_{35}^2 - \t_{15} (\t_{26} - 4 \t_{45}))
 - \uu{} (5 \t_{25} \t_{26}  \nn\\ &
+ \t_{15} (7 \t_{16} - 5 \t_{25} + 5 \t_{35}))
 + \ss{} (\t_{15} (5 \t_{16} - 4 \tt)
 + \t_{16} (8 \tt + \t_{25}) + \t_{35} \t_{36}
 + 5 \t_{45} (\t_{25} - \t_{35} - 2 \t_{36})  \nn\\ &
- \t_{46} (11 \t_{35} + 6 \t_{45}) + 2 \uu{} (6 \t_{25}
 + 4 \t_{26} - 4 \t_{35} - 2 l_{5}^2))
 + 2 \ss^2 (3 \t_{25} - \t_{35} - \t_{36} - 3 \t_{46} + \t_{56}))
\,, \nn\\
N_{33} &= 
\Frac{1}{4} (\ss^2 (4 \t_{17} - 2 (4 \t_{26} + \t_{35} + 2 \t_{36}))
 - 6 \uu^2 \t_{35} + \uu{} ((4 \t_{16} + 5 \t_{26}) \t_{45}  \nn\\ &
 - \t_{17} (11 \t_{25} + 7 \t_{35} + 6 \t_{45}) + \t_{35} (11 \t_{37}
 - 5 \t_{46})) - \tt{} (5 \t_{17} \t_{25} + 6 \t_{15} \t_{26}
 + (6 \t_{26} - 5 \t_{17} - 4 \t_{27}) \t_{45}  \nn\\ &
 + \t_{35} (7 \t_{26} - 11 \t_{36} + 5 \t_{47}))
 + \ss{} (\t_{15} (5 \t_{26} + 4 \t_{46}) - 5 \t_{35} (\t_{16} + \t_{27})
 + 4 \t_{25} \t_{47} + 2 \uu{} (5 \t_{17} + 2 \t_{25} \nn\\ &
 - 5 (\t_{26} + \t_{35}) + \t_{56})
 + 2 \tt{} (\t_{16} - \t_{15} - \t_{27} + \t_{57})))
\,, \nn\\
N_{37} &= 
 \Frac{1}{4} (\uu^2 (4 \t_{15} - 2 \t_{27})
 - 2 \ss^2 (2 \t_{15} + 3 \t_{27} + 4 \t_{36})
 + \tt{} (6 \t_{26} \t_{27} + 5 \t_{27} \t_{35} - 6 \t_{27} \t_{36}  \nn\\ &
 + 6 \t_{36} \t_{37} - \t_{25} (5 \t_{36} + 4 \t_{46})
 - \t_{15} (6 \t_{26} + 5 \t_{46}) + 4 \t_{35} \t_{47})
 + \uu{} (5 \t_{27} \t_{46} - 6 \t_{35} \t_{36} + (5 \t_{27} \nn\\ &
 + 4 \t_{37}) \t_{45} + \t_{17} (5 \t_{36} + 4 \t_{46} - 2 \tt)
 + \t_{15} (12 \t_{36} + 5 \t_{37} + 6 \t_{46} - 5 \t_{47})
 + 2 \tt{} (\t_{57} - \t_{25})) \nn\\ &
 + \ss{} (6 \t_{25} \t_{27} - (4 \t_{26} + 5 \t_{36}) \t_{45}
 - 5 (\t_{26} + \t_{46}) \t_{47} - \t_{16} (4 \tt - 5 \t_{27} + \t_{47})
 - \t_{15} (11 \t_{26} + 12 \t_{27} \nn\\ &
 + 6 \t_{37} + 11 \t_{47}) + 2 \tt{} (\t_{35} - \t_{56})
 + 2 \uu{} (\tt + 2 \t_{26} - 5 \t_{27} - 8 \t_{36} + 2 \t_{37} + \t_{67})))
\,, \nn
\end{align}
\begin{align}
N_{38} &=
\Frac{1}{4} (\tt{} (\t_{16} \t_{25} + 6 \t_{25}^2 - 11 \t_{25} \t_{26}
 - \t_{25} \t_{35} + 11 \t_{35}^2 + 7 \t_{35} \t_{37} + \t_{37} \t_{45}
 + \t_{15} (\t_{45} - 4 \t_{16})  \nn\\ &
 - \t_{35} \t_{47}) - \ss{} (5 \t_{15}^2 - 6 \t_{17} \t_{25}
 + 20 \t_{17} \t_{26} + 6 \t_{25} \t_{27} + 4 \t_{26} \t_{27}
 + 4 \t_{16} (\t_{17} + 5 \t_{27}) - 9 \t_{25} \t_{36}  \nn\\ &
 - 11 \t_{35} \t_{36} - 16 \t_{36} \t_{37} - 5 \t_{37} \t_{45}
 + 10 \t_{35} \t_{46} + 5 \t_{45} \t_{46} - 4 (\t_{35}
 + 4 \t_{46}) \t_{47} + \t_{15} (\t_{37} + \t_{47} \nn\\ &
 - 4 \t_{27} + 10 \t_{35} - 9 \t_{36}))
 + \uu{} (\t_{45} (7 \t_{47} - 5 (2 \t_{35} + \t_{45}))
 - 4 \t_{25} \t_{26} + \t_{15} (\t_{26} - 11 \t_{16})) \nn\\ &
 + 2 \ss{} (4 \tt \t_{37} + \uu{} (4 (\t_{16} + \t_{25} - \t_{26} - \t_{35}
 + \t_{47}) + 2 l_{5}^2) + \ss{} (2 \t_{16} - 2 \t_{17} + 2 \t_{25}
 - 2 \t_{26} \nn\\ &
 - 2 \t_{27} - 2 \t_{35} + \t_{56} - \t_{57} + 2 \t_{67})))
\,, \nn\\
N_{39} &= 
 \Frac{1}{4} (\tt{} (\t_{15} (12 \t_{16} + 6 \t_{26} + \t_{36})
 - 10 \t_{38} \t_{45} + \t_{27} (5 \t_{26} + 12 (\t_{36} + \t_{46}))
 + \t_{35} (\t_{48} - 7 \t_{38})) \nn\\ &
 - \uu{} (\t_{17} (7 \t_{16} + \t_{26}) + 11 \t_{16} \t_{27}
 - \t_{25} (4 \t_{36} + 5 \t_{46}) + \t_{45} (9 \t_{38} + 6 \t_{46}
 + 7 \t_{48})) + \ss{} (4 \t_{17} \t_{18} \nn\\ &
 + 11 \t_{17} \t_{26} + 20 \t_{18} \t_{27} + 20 \t_{17} \t_{28}
 + 4 \t_{27} \t_{28} + \t_{15} (5 \t_{16} - 11 \t_{18} + 10 \t_{26}
 + 5 \t_{28}) - 4 \t_{18} \t_{35} 
\nn\\ &
+ 4 \t_{26} \t_{35} - 6 \t_{35} \t_{38} - 16 \t_{37} \t_{38}
 + 2 \tt{} (\t_{25} + \t_{36} + 4 \t_{38}) + 4 \t_{28} \t_{45}
 - 5 \t_{25} \t_{46} - \t_{37} \t_{46}
 \nn\\ &
 - 11 \t_{45} \t_{46} - 13 \t_{46} \t_{47}
 - 2 \t_{36} (3 \t_{37} + \t_{47}) + 2 (3 \t_{35} - \t_{45}
 - 8 \t_{47}) \t_{48} 
 \nn\\ &
+ 2 \uu{} (3 \t_{15} + 4 \t_{17} - 4 \t_{27} + 3 \t_{46} + 4 \t_{48}
 + \t_{56}))
 \nn\\ &
 - 2 \ss^2 (\uu - \t_{15} - 2 \t_{17} + 2 \t_{18} + \t_{26} + 2 \t_{27}
 + 2 \t_{28} + \t_{35} - \t_{46} + \t_{58} + \t_{67} + 2 \t_{78}))
\,, \nn\\
N_{45} &= 
\Frac{1}{4} (8 \uu^2 \t_{15} + \tt{} (\t_{15} (\t_{16} + 7 \t_{17}
 + \t_{26} + \t_{27} - 5 \t_{35} + 5 \t_{45}) - (\t_{16} + \t_{17}
 + 8 \tt) \t_{25}) \nn\\ &
 + \ss{} (\t_{15} (12 \t_{17} - 18 \t_{25} + 11 \t_{26} + \t_{27})
 + 2 (5 \t_{16} \t_{25} - 2 \t_{17} \tt - 2 \tt{} (6 \t_{25} + \t_{26})
 + 3 \t_{25} (\t_{26} + 2 \t_{27})) \nn\\ &
 + 11 \t_{35}^2 + \t_{45} (6 (\t_{45} - 2 \t_{46} - \t_{47}) - \t_{36}
 - 11 \t_{37}) + \t_{35} (\t_{37} - 5 \t_{36} + 13 \t_{45} - 10 \t_{47}))
  \nn\\ &
 + \uu{} (4 \ss{} (\t_{16} - \tt+ \t_{27} + 2 \t_{35})
 - 6 \t_{15} \t_{16} + \t_{25} (7 \t_{26} + 7 \t_{27} - 5 \t_{35}
 + 5 \t_{45}) + 4 \tt l_{5}^2)  \nn\\ &
 + 2 \ss^2 (\t_{16} + \t_{27} + 5 \t_{35} + \t_{36} + \t_{37}
 + \t_{38} + 3 \t_{45} - \t_{48} - \t_{56} + \t_{57}))
\,, \nn\\
N_{48} &= 
\Frac{1}{4} (\ss{} (\t_{18} (5 \t_{25} + 4 \t_{35}) + \t_{36}
 (\t_{15} - 5 \t_{27} - 4 \t_{45}) - (\t_{17} - 5 \t_{25}) \t_{46}
 + 5 \t_{35} \t_{48})- 2 \ss \uu \tt\nn\\ &
  - 5 \t_{15} \t_{18} \tt - \tt{} (5 \t_{17} \t_{26} + 7 \t_{35} \t_{36}
 + \t_{25} (11 \t_{26} - 5 \t_{38}) + \t_{15} (5 \t_{28} + \t_{38})
 + 4 \t_{37} \t_{46} \nn\\ &
 + \t_{36} \t_{47} - 4 \t_{16} (\t_{45} + \t_{47}))
 + \uu{} (4 \t_{15} \t_{26} + 6 \t_{25} \t_{28}
 + \t_{16} (7 \t_{17} + \t_{25} - 4 \t_{35}) + \t_{36} (11 \t_{37} \nn\\ &
 + \t_{45}) - 4 \t_{26} (\t_{17} + \t_{47}) + 6 (\t_{45} - \t_{25}) \t_{48})
 + 2 (\ss^2 (\t_{25} - \t_{35} + 2 \t_{36}) + 3 \uu \tt{} (\t_{17} + \t_{48})
 \nn\\ &
 + \ss{} (\uu{} (3 \t_{25} + \t_{26} + 4 \t_{36} - \t_{56})
 + \tt{} (\t_{15} - \t_{16} - 2 \t_{17} - \t_{18} - \t_{47} - 2 \t_{48}
 + \t_{58} + \t_{67}))))
\,, \nn\\
N_{49} &= 
\Frac{1}{4} (2 \ss^2 \tt - \tt{} (11 \t_{35} \t_{38}
 + 2 \t_{15} (2 \t_{36} + 5 \t_{46}) + \t_{25} (11 \t_{28} - 4 \t_{37}
 + 4 \t_{38} + \t_{47})  \nn\\ &
 + 4 \t_{27} \t_{48} + 5 \t_{17} (\t_{35} + \t_{48}))
 + \ss{} (\t_{15} (11 \t_{17} + 5 \t_{38}) - 5 \t_{17} \t_{28}
 - 2 \t_{16} (5 \t_{25} + 2 \t_{35}) + \t_{45} (5 \t_{46}  \nn\\ &
 - 4 (\t_{18} + \t_{37})) + 5 \t_{38} (\t_{27} - \t_{47})
 - 4 \t_{28} \t_{47} + \t_{25} (7 \t_{27} + 5 \t_{48} - 5 \t_{26}))
 + \uu{} (\t_{27} (11 \t_{38} + 4 \t_{45})  \nn\\ &
 - \t_{25} (5 \t_{16} - 5 \t_{18} + 11 \t_{26} + \t_{37})
 - \t_{17} (6 \t_{18} - 12 \t_{38} + \t_{45})
 + (7 \t_{25} + 6 \t_{35}) \t_{46} + \t_{38} (5 \t_{45} + 6 \t_{47})  \nn\\ &
 + \t_{15} (\t_{46} - 4 (\t_{47} + \t_{48}))) + 2 (\ss^2 (\t_{25}
 - 4 \t_{17} - \t_{35} + 2 \t_{38}) + \uu{} (4 \uu \t_{17}
 + \tt{} (2 \t_{27} - 2 \t_{15}  \nn\\ &
 - 2 \t_{25} - 3 \t_{46} + \t_{57})) + \ss{} (\t_{16} \tt
+ \uu{} (\t_{28} + 4 \t_{38} - \t_{58})
 - \tt{} (\t_{18} + 2 \t_{37} - 2 \t_{46} + \t_{56} + \t_{78}))))
\,, \nn\\
N_{50} &= 
 \Frac{1}{4} (\tt{} (\t_{25} (5 \t_{37} + 4 \t_{47})
 - \t_{38} (4 \t_{16} + 5 \t_{26}) - 5 \t_{15} \t_{27})
 + \ss{} (\t_{45} (4 \t_{27} + 5 \t_{37}) \nn\\ &
  - \t_{18} (5 \t_{26} + 4 \t_{36}) - 5 \t_{15} \t_{47}
 + 5 (\t_{16} - \t_{36}) \t_{48}) + \uu{} (6 \t_{35} \t_{37}
 - 6 \t_{26} \t_{28} + \t_{15} (11 \t_{17} - \t_{37})  \nn\\ &
  + (11 \t_{16} + 6 (2 \t_{26} + \t_{36})) \t_{48})
 + 2 \tt{} (3 \uu{} (\t_{15} - \t_{26} + \t_{37} - \t_{48})
 + \ss{} (\t_{18} - 2 \t_{15} + 2 \t_{26} - \t_{27}  \nn\\ &
  + \t_{36} - 2 \t_{37} - \t_{45} + 2 \t_{48} + \t_{57} - \t_{68})))
\,, \nn\\
N_{51} &= 
 \Frac{1}{2} (4 \tt^2 (\t_{15} + \t_{25}) - 6 \t_{15}^2 \tt
 - 4 \ss \tt{} (\ss - \uu) - 6 \uu \t_{25}^2
 + \t_{35} (5 \tt \t_{25} - 6 \ss \t_{35}) \nn\\ &
  + 5 \t_{15} (\ss \t_{25} + \uu \t_{35}) + 14 \ss^2 \t_{45}
 - \ss{} (6 \uu \t_{15} + \tt{} (\t_{15} + 6 \t_{25} + 13 \t_{35}
 - 2 l_{5}^2)) - 2 \uu^2 l_{5}^2)
\,, \nn\\
N_{53} &=
8 \ss{} (\tt \t_{35} + \uu \t_{45} - \ss \t_{25})
\,, \nn\\
N_{55} &= 
\Frac{1}{2} \tt{} (\tt{} (\t_{25} - 8 \t_{15} + 5 \t_{45})
 + \uu{} (9 \t_{45} - 17 \t_{15}))
\,, \nn\\
N_{58} &=  
\ss{} (2 \uu{} (\t_{45} - 3 \t_{35})
 - \ss{} (\uu - \tt + 4 \t_{25} + 5 \t_{35} + \t_{45}))
\,, \nn
\end{align}
\begin{align}
N_{63} &=  
-\Frac{1}{2} \ss{} (5 \tt \t_{35}
 + \uu{} (12 \t_{36} + 5 \t_{45} - 4 \t_{46})
 + \ss{} (2 \tt - 2 \uu + 2 \t_{25} + 2 \t_{46} + 8 \t_{26} + 10 \t_{36}
 - 7 \t_{15}))
\,, \nn\\
N_{66} &= 
\ss{} (4 \tt{} (\t_{35} - 2 \t_{36}) + 2 \uu{} (\t_{35} + 3 \t_{45}
 - 4 \t_{46}) - \ss{} (6 \uu + \t_{15} - 6 \tt + 5 \t_{25} - 8 \t_{26}))
\,, \nn\\
N_{68} &=  
\Frac{1}{2} \ss{} (\ss{} (2 (\t_{15} + \tt - \uu) - 7 \t_{25})
 + 5 \uu \t_{35} + 5 \tt \t_{45})
\,, \nn\\
N_{79} &= 
 \Frac{1}{2} \ss{} (5 \tt \t_{35} + \uu{} (5 \t_{45} + 4 \t_{46}
 - 4 \t_{37} + 12 \t_{47} - 12 \t_{36}) \nn\\ &
+ \ss{} (2 (\t_{16} + \t_{25} - \t_{27} + 3 \t_{17} - 3 \t_{26}
 + 4 \t_{47} - 4 \t_{36}) - 7 \t_{15}))
\,, \nn\\
N_{80} &= 
16 \ss^2 (\uu - \tt)
\,, \nn\\
N_{83} &= 
-\Frac{9}{2} k_{4}^2 \ss{} (\uu - \tt)
\label{ListOfAllNumerators} \,.
\end{align}
The other numerators are given directly in terms of these in
\eqn{trivialNrelations}.  For $N_{83}$ it should be understood that
$k_4^2 \rightarrow 0$ only after canceling the $1/k_4^2$ propagator.
Alternatively, we can rewrite the snail contributions in terms of the
numerators of the graphs in \fig{SnailGraphsFigure}, we have
\begin{equation}
N_{83'}= N_{84'} = N_{85'} = -\frac{9}{2} \ss{} (\uu - \tt) \,.
\end{equation}
Plain-text, computer-readable versions of these expressions may
be found online~\cite{Online}.

It is interesting to note that $N_{33}$ can be used as a non-planar
master graph numerator, as discussed in
\sect{FourLoopCalcSection}. This implies that the single numerator
$N_{33}$ contains the same amplitude-specific information as the two
planar master numerators $N_{18}$ and $N_{28}$ combined.

\begin{table}
\caption{Effective numerators for the vacuum integrals $I_i^v$
entering the UV pole of the four-loop $\NeqEight$ supergravity amplitude,
and the coefficients arising from writing them as a linear combination
of basis vacuum integrals $V_1$, $V_2$, and $V_8$. }
\label{NumeratorTable}
\bea
\begin{array}{c|c|c|c|c}
I^v & \text{Effective numerator}
& V_1 & V_2&  V_8   \cr 
\hline
 I^v_{1} &
   -\frac{117674}{1485}  
 %
 &
 0
 &
-\frac{117674}{1485}
&  
0   
\cr
 I^v_{2} &  
 \frac{19112  }{1485}~\tau_{{a}, {b}}^2 
 %
 &  
 \frac{8798687}{5346000}
 &  
 \frac{212621}{27000}
 &  
 0   
 \cr
 I^v_{3} &  
 \frac{9556  }{1485}~\tau_{{a}, {b}}^2 
%
 &  
 \frac{15937019}{1782000}
 &  
 -\frac{140951}{33000}
 &  
 0
    \cr
 I^v_{4} &  
 -\frac{16427}{495}    
%
 &  
 -\frac{16427}{495}
 &  
 0
 &  
 0   
 \cr
 I^v_{5}  &   
 \frac{19112  }{1485}~\tau_{{a}, {c}} - \frac{19112 }{1485} ~\tau_{{b}, {c}}
%
 &  
 -\frac{2389}{2970}
 &  
 -\frac{2389}{1485}
 &  
 0   
 \cr
 I^v_{6}  &   
 -\frac{4778 }{495} ~\tau_{{a}, {c}} + \frac{4778  }{1485}~\tau_{{b}, {c}} 
%
 &  
 \frac{16723}{2970}
 &  
 -\frac{4778}{1485}
 &  
 0   
 \cr
 I^v_{7}  &   
 -\frac{9556}{1485} 
%
 &  
 \frac{109894}{7425}
 &  
 \frac{90782}{2475}
 &  
 0
    \cr
 I^v_{8}  &   
 \frac{38224}{1485}  ~\tau_{{a}, {b}}   
%
 &  
 -\frac{2389}{675}
 &  
 -\frac{19112}{2475}
 &  
 0   
 \cr
 I^v_{9}  &   
 \frac{38224 }{1485} ~\tau_{{a}, {b}}^2 
%
 &  
 -\frac{1617353}{148500}
 &  
 \frac{2606399}{74250}
 &
   0   
   \cr
 I^v_{10}  &   
 -\frac{19112 }{1485} ~\tau_{{a}, {c}} - \frac{19112  }{1485}~\tau_{{b}, {c}}
%
 &  
 -\frac{2389}{2970}
 &  
 -\frac{2389}{1485}
 &  
 0   
 \cr
 I^v_{11}  &   
 -\frac{38224  }{1485}~\tau_{{a}, {b}} 
%
 &  
 \frac{90782}{22275}
 &  
 -\frac{9556}{825}
 &  
 0  
  \cr
 I^v_{12}  &   
 -\frac{19112}{1485}    
%
 &  
 \frac{31057}{990}
 &  
 \frac{38224}{495}
 &  
 0   
 \cr
 I^v_{13}  &   
 \frac{10048}{99}  
%
 &  
 \frac{2512}{99}
 &  
 \frac{10048}{99}
 &  
 0   
 \cr
 I^v_{14}  &   
 -\frac{19112}{1485}    
%
 &  
 -\frac{4778}{275}
 &  
 -\frac{324904}{7425}
 &  
 0   
 \cr
 I^v_{15}  &  
  \frac{19112}{1485} 
%
 &  
 \frac{66892}{4455}
 &  
 \frac{19112}{495}
 & 
  0   
  \cr
 I^v_{16}  &   
 \frac{19112 }{1485} ~\tau_{{a}, {b}}^2     
%
 &  
 \frac{977101}{267300}
 &  
 \frac{88393}{14850}
 &  
 0   
 \cr
 I^v_{17}  &  
  \frac{39676}{1485} 
%
 &  
 \frac{9919}{495}
 &  
 \frac{19838}{1485}
 &  
 0   
 \cr
 I^v_{18}  &   
 \frac{9556  }{1485}~\tau_{{a}, {b}}^2  
%
 &  
 -\frac{1478791}{297000}
 &  
 \frac{661753}{148500}
 &  
 \frac{2389}{396}  
  \cr
 I^v_{19}  &   
 -\frac{64441}{1485} 
%
 &  
 0
 &  
 0
 &  
 -\frac{64441}{1485}  
 \cr
 I^v_{20}  &   
 \frac{38224  }{1485}~\tau_{{a}, {b}}    
%
 &  
 -\frac{102727}{14850}
 &  
 -\frac{74059}{7425}
 &  
 0   
\cr
 I^v_{21}  &   
 \frac{5284}{1485}  
%
 &  
 \frac{18494}{7425}
 &  
 \frac{34346}{7425}
 &  
 0   
 \cr
 I^v_{22}  &   
 \frac{934}{165}    
%
 &  
 \frac{467}{165}
 &  
 \frac{1868}{165}
 &  
 -\frac{934}{165}  
  \cr
 I^v_{23}  &   
 \frac{526  }{135}~\tau_{{a}, {b}} - \frac{91  }{1485}~\tau_{{a}, {c}}     
%
 &  
 \frac{279199}{297000}
 &  
 \frac{72052}{37125}
 &  
 0   
 \cr
 I^v_{24}  &   
 \frac{3736 }{495} ~\tau_{{a}, {b}}    
%
 &  
 \frac{26152}{12375}
 &  
 -\frac{91532}{12375}
 &  
 0  
  \cr
 I^v_{25}  &   
 -\frac{9556  }{1485}~\tau_{{a}, {b}} 
%
 &  
 -\frac{2389}{2475}
 &  
 -\frac{16723}{7425}
 &  
 0 
\cr
 I^v_{26}  &   
 -\frac{11048}{1485}    
%
 &  
 -\frac{17953}{2475}
 &  
 -\frac{44192}{2475}
 &  
 0
    \cr
 I^v_{27}  &  
 -\frac{1228}{135}   
%
 &  
 -\frac{307}{50}
 &  
 -\frac{10438}{675}
 & 
  0   
 \cr
 I^v_{28}  &  
 -\frac{3736}{495}    
%
 &  
 \frac{934}{825}
 &  
 -\frac{14944}{825}
 &  
 0 
   \cr
 I^v_{29}  &   
 \frac{3736  }{495}~\tau_{{a}, {b}}  
%
 &  
 -\frac{48568}{4125}
 &
   \frac{76588}{4125}
   &  
   0
   \cr
 I^v_{30}  &   
 \frac{934}{495}    
%
 &  
 \frac{90131}{24750}
 &  
 \frac{119552}{12375}
 &  
 0   
 \cr
 I^v_{31}  &   
 \frac{4778 }{1485} ~\tau_{{a}, {b}} + \frac{9556 }{1485} ~\tau_{{a}, {c}} 
 %
 &  
 -\frac{45391}{19800}
 &  
 -\frac{112283}{14850}
 &  
 0   
 \cr
 I^v_{32}  &   
 \frac{19112  }{1485}~\tau_{{a}, {b}}^2    
 %
 &  
 \frac{2721071}{148500}
 &  
 -\frac{327293}{74250}
 &  
 -\frac{2389}{495}   
 \cr
 I^v_{33}  &   
 -\frac{3736 }{495} ~\tau_{{a}, {b}} 
 %
 &  
 -\frac{1868}{2475}
 &  
 -\frac{1868}{275}
 &  
 \frac{1868}{165}   
 \cr
 I^v_{34}  &   
 \frac{4778}{297}  
%
 & 
  -\frac{155285}{2376}
  &  
  -\frac{47780}{297}
  &  
  0  
   \cr
 I^v_{35}  &   
 -\frac{7904}{1485} 
%
 &  
 -\frac{3952}{495}
 &  
 -\frac{27664}{1485}
 &  
 0   
\end{array}
\nonumber 
\eea
\end{table}


\begin{table}
\bea
\begin{array}{c|c|c|c|c}
I^v & \text{Effective numerator}
& V_1 & V_2&  V_8   \cr
\hline
 I^v_{36}  &   
 -\frac{3736  }{495}~\tau_{{a}, {b}} 
 %
 &  
 \frac{467}{4950}
 &  
 \frac{7939}{2475}
 &  
 0   
 \cr
I^v_{37}  &   
\frac{9556  }{1485}~\tau_{{a}, {b}} 
%
 &  
 -\frac{45391}{59400}
 &  
 -\frac{54947}{29700}
 &  
 0   
 \cr
 I^v_{38}  &   
 \frac{3736}{495}    
%
 &  
 \frac{21482}{2475}
 &  
 \frac{76588}{2475}
 &  
 0   
 \cr
 I^v_{39}  &   
 \frac{4778}{495} 
%
 &  
 \frac{45391}{6600}
 &  
 \frac{54947}{3300}
 &
   0   
   \cr
 I^v_{40}  &  
 -\frac{1228}{135}    
%
 &  
 -\frac{3991}{225}
 &
  -\frac{9824}{225}
  &  
  0   
  \cr
I^v_{41}  &   
\frac{1868  }{495}~\tau_{{a}, {b}} 
%
 &  
 \frac{47167}{9900}
 &  
 -\frac{20081}{4950}
 &  
 0   
 \cr
I^v_{42}  &   
\frac{1228 }{135} ~\tau_{{a}, {b}} + \frac{6524 }{297} ~\tau_{{a}, {c}} 
 &  
 \frac{257243}{14850}
 &  
 -\frac{16723}{2475}
 &  
 0   
 \cr
 I^v_{43}  &   
 \frac{9556  }{1485}~\tau_{{a}, {b}} 
%
 &  
 -\frac{16723}{14850}
 &  
 -\frac{2389}{825}
 &  
 0   
 \cr
 I^v_{44}  &   
 \frac{19112  }{1485}~\tau_{{a}, {b}}^2 
%
 &  
 -\frac{2047373}{133650}
 &
  \frac{212621}{7425}
  &  
  0   
  \cr
 I^v_{45}  &   
 \frac{10204}{1485}   
%
 &  
 \frac{323977}{29700}
 &  
 \frac{135203}{4950}
 &  
 0   
 \cr
 I^v_{46}  &   
 \frac{4778}{495}    
%
 &  
 \frac{16723}{3960}
 &  
 \frac{2389}{220}
 &
 0
    \cr
 I^v_{47}  &   
 -\frac{9556}{1485}   
%
 &  
 -\frac{31057}{2970}
 &  
 -\frac{38224}{1485}
 &  
 0   
 \cr
 I^v_{48}  &   
 -\frac{4778}{495}    
%
 &  
 -\frac{4778}{825}
 &  
 -\frac{33446}{2475}
 &  
 0   
 \cr
 I^v_{49}  &  
  \frac{9556}{1485}  
%
 &  
 \frac{31057}{4950}
 &  
 \frac{38224}{2475}
 &  
 0   
 \cr
 I^v_{50}  &   
 -\frac{16532  }{1485}~\tau_{{a}, {b}} + \frac{1403  }{135}~\tau_{{a}, {c}}    
 %
 &  
 -\frac{83233}{11880}
 &  
 -\frac{58961}{5940}
 & 
  0 
    \cr
I^v_{51}  &   
\frac{9556  }{1485}~\tau_{{a}, {b}}^2    
 %
 &  
 \frac{33309827}{2673000}
 &  
 -\frac{74059}{13500}
 &  
 0  
  \cr
I^v_{52}  &   
\frac{9556  }{1485}~\tau_{{a}, {b}}^2 
%
 &  
 -\frac{1048771}{148500}
 & 
  -\frac{1822807}{74250}
  &  
  -\frac{2389}{2970}   
  \cr
I^v_{53}  &   
-\frac{19112  }{1485}~\tau_{{a}, {b}} - \frac{19112  }{1485}~\tau_{{a}, {c}} 
%
 &  
 \frac{303403}{74250}
 &  
 \frac{1335451}{37125}
 &  
 0   
 \cr
I^v_{54}  &   
-\frac{17036  }{1485}~\tau_{{a}, {b}}    
%
 &  
 -\frac{8518}{1485}
 &  
 -\frac{34072}{1485}
 &  
 \frac{8518}{297}   
 \cr
 I^v_{55}  &   
 -\frac{19112}{1485}  
%
 &  
 \frac{962767}{29700}
 &  
 \frac{398963}{4950}
 &  
 0   
 \cr
 I^v_{56}  &   
 \frac{19112}{1485}   
%
 &  
 -\frac{16723}{1980}
 &  
 -\frac{2389}{110}
 &  
 0   
 \cr
 I^v_{57} &   
 -\frac{934}{99}  
%
 &  
 -\frac{91999}{4950}
 &  
 -\frac{38761}{825}
 &  
 0   
 \cr
I^v_{58}  &   
-\frac{2711  }{5940}~\tau_{{a}, {d}} + \frac{263  }{540}~\tau_{{b}, {c}}  
%
 &  
 \frac{89141}{19800}
 &  
 -\frac{11527}{2475}
 &  
 0
   \cr
 I^v_{59}  &   
 -\frac{2711}{2970}   
%
 &  
 -\frac{35243}{12375}
 &  
 -\frac{46087}{12375}
 &
   0   
   \cr
I^v_{60}  &   
-\frac{9556  }{1485}~\tau_{{a}, {b}} 
%
 &  
 -\frac{54947}{59400}
 &  
 -\frac{74059}{29700}
 &  
 0   
 \cr
I^v_{61}  &   
\frac{1868  }{495}~\tau_{{a}, {b}} 
%
 &  
 \frac{8873}{4125}
 &  
 -\frac{1868}{4125}
 &  
 0   
 \cr

 I^v_{62}  &   
 \frac{9556}{1485}    
%
 &  
 \frac{140951}{14850}
 &  
 \frac{174397}{7425}
 &  
 0   
 \cr
I^v_{63}  &   
\frac{1868  }{495}~\tau_{{a}, {b}}   
%
 &  
 \frac{34091}{12375}
 &  
 -\frac{31756}{12375}
 &  
 -\frac{934}{495}   
 \cr
 I^v_{64}  &   
 \frac{467  }{495}~\tau_{{a}, {b}}^2    
%
 &  
 -\frac{467}{600}
 &  
 \frac{17279}{4950}
 &  
 -\frac{467}{1980}   
 \cr
 I^v_{65}  &   
 \frac{14101}{1980}   
%
 &  
 \frac{183313}{6600}
 &  
 \frac{56404}{825}
 &  
 0   
 \cr
 I^v_{66}  &   
 -\frac{8588}{1485}    
%
 &  
 -\frac{916769}{74250}
 &  
 -\frac{382166}{12375}
 &  
 0   
 \cr
I^v_{67}  &   
\frac{1868  }{495}~\tau_{{a}, {b}}^2  
%
 &  
 \frac{1401}{550}
 &  
 \frac{3736}{825}
 &  
 \frac{467}{495}   
 \cr
 I^v_{68}  &   
 \frac{1868}{495}  ~\tau_{{a}, {b}}^2   
%
 &  
 \frac{467}{450}
 &  
 \frac{1868}{825}
 &  
 \frac{467}{165}   
 \cr
 I^v_{69}  &   
 \frac{4778}{1485}~\tau_{{a}, {b}}^2 
  %
 &  
 -\frac{3843901}{297000}
 &  
 \frac{231733}{148500}
 &  
 \frac{54947}{2970}   
 \cr
 \hline 
{\rm Total} &  
 -
 &  
 \frac{23}{2}
 &  
23
 &  
 \frac{23}{2}
 \end{array}
   \cr
\nonumber 
\eea
\end{table}


\section{Vacuum integrals and their expression in terms of master integrals}
\label{VacuumNumeratorsAppendix}

In the first column of \tab{NumeratorTable} we give the numerators of
the vacuum integrals in figs.~\ref{p1}--\ref{p3}, as they appear
in the expression for ${\widetilde {\cal V}}^{(4)}$ defined
in \eqns{TableNormalization}{V4def}.  Each integral can be reduced
to a linear combination of the three master integrals $V_1$, $V_2$ and
$V_8$.  The second, third and fourth columns of the table provide
the coefficients of $V_1$, $V_2$ and $V_8$, respectively, after this
reduction.  To obtain the coefficient of each vacuum integral $V_i$
in the final formula for the four-loop UV divergence in \eqn{FourLoopGpole},
we simply sum the numbers in each column labeled by a $V_i$, to obtain
\be
{\widetilde {\cal V}}^{(4)} = \frac{23}{2} \, ( V_1 + 2 V_2 + V_8 ) \,.
\label{V4tildefinal}
\ee
Inserting this value into \eqn{TableNormalization} yields the final
result~(\ref{FourLoopGpole}).

\end{document}